\documentclass[acmsmall]{acmart}

\usepackage{algorithm2e}
\usepackage{hyperref}
\usepackage{balance}
\usepackage{tabularx}
\usepackage{float}
\usepackage{multirow}
\usepackage{todonotes}
\usepackage{xltabular}

\newcolumntype{L}[1]{>{\raggedright\arraybackslash}m{#1}}
\newcolumntype{Y}{>{\raggedright\arraybackslash}X}

\definecolor{oxfordblue}{rgb}{0.0, 0.13, 0.28}
\definecolor{harvardcrimson}{rgb}{0.79, 0.0, 0.09}
\definecolor{dartmouthgreen}{rgb}{0.05, 0.5, 0.06}
\definecolor{princetonorange}{rgb}{1.0, 0.56, 0.0}
\definecolor{yaleblue}{rgb}{0.06, 0.3, 0.57}
\definecolor{usccardinal}{rgb}{0.6, 0.0, 0.0}
\definecolor{uclablue}{rgb}{0.33, 0.41, 0.58}
\definecolor{msugreen}{rgb}{0.09, 0.27, 0.23}
\definecolor{cornellred}{rgb}{0.7, 0.11, 0.11}
\definecolor{pomegranate}{RGB}{192, 57, 43}
\definecolor{anti-pomegranate}{RGB}{43,178,192}
\definecolor{alizarin}{RGB}{231, 76, 60}
\definecolor{anti-belize}{RGB}{185, 41, 56}
\definecolor{belize}{RGB}{41, 128, 185}
\definecolor{peter}{RGB}{52, 152, 219}
\definecolor{green}{RGB}{22, 160, 133}
\definecolor{anti-green}{RGB}{160,22,118}
\definecolor{turquoise}{RGB}{26, 188, 156}
\definecolor{pumpkin}{RGB}{211, 84, 0}
\definecolor{anti-pumpkin}{RGB}{0,22,211}
\definecolor{carrot}{RGB}{230, 126, 34}
\definecolor{wisteria}{RGB}{142, 68, 173}
\definecolor{anti-wisteria}{RGB}{99,173,68}
\definecolor{amethyst}{RGB}{155, 89, 182}
\definecolor{nephritis}{RGB}{39, 174, 96}
\definecolor{anti-nephritis}{RGB}{174,39,117}

\newcommand{\cmt}[1]{\ignorespaces}

\newcommand{\finalhighlight}[1]{{\textcolor{black}{#1}}}

\newcommand{\ylhighlight}[1]{{\textcolor{black}{#1}}}
\newcommand{\chaohighlight}[1]{{\textcolor{black}{#1}}}

\newcommand{\yuwen}[1]{{#1}}

\newcommand{\chao}[1]{{#1}}

\usepackage{mathtools}

\AtBeginDocument{%
  }

\setcopyright{acmlicensed}
\copyrightyear{2024}
\acmJournal{PACMHCI}
\acmYear{2024} \acmVolume{8} \acmNumber{CSCW1} \acmArticle{59} \acmMonth{4}\acmDOI{10.1145/3637336}

\begin{document}


\title[Exploring End-User Empowerment Interventions for Dark Patterns in UX]{From Awareness to Action: Exploring End-User Empowerment Interventions for Dark Patterns in UX}


\author{Yuwen Lu}
\authornote{Both authors contributed equally to this work.}
\affiliation{%
  \institution{University of Notre Dame}
  \city{Notre Dame}
  \state{IN}
  \country{USA}}
\email{ylu23@nd.edu}
\orcid{0000-0003-0845-5563}

\author{Chao Zhang}
\authornotemark[1]
\authornote{Work done as a visiting researcher at the University of Notre Dame.}
\affiliation{%
  \institution{Cornell University}
  \city{Ithaca}
  \state{NY}
  \country{USA}}
\email{cz468@cornell.edu}
\orcid{0000-0003-4286-8468}

\author{Yuewen Yang}
\authornotemark[2]
\affiliation{%
  \institution{Cornell Tech}
  \city{New York}
  \state{NY}
  \country{USA}}

\author{Yaxing Yao}
\affiliation{%
  \institution{Virginia Tech}
  \city{Blacksburg}
  \state{VA}
  \country{USA}}
\email{yaxing@vt.edu}

\author{Toby Jia-Jun Li}
\affiliation{%
  \institution{University of Notre Dame}
  \city{Notre Dame}
  \state{IN}
  \country{USA}}
\email{toby.j.li@nd.edu}
\renewcommand{\shortauthors}{Yuwen Lu et al.}

\begin{abstract}

The study of UX dark patterns, \textit{i.e.}, UI designs that seek to manipulate user behaviors, often for the benefit of online services, has drawn significant attention in the CHI and CSCW communities in recent years. To complement previous studies in addressing dark patterns from (1) the designer’s perspective on education and advocacy for ethical designs; and (2) the policymaker’s perspective on new regulations, we propose an end-user-empowerment intervention approach that helps users (1) raise the \textit{awareness} of dark patterns and understand their underlying design intents; (2) take \textit{actions} to counter the effects of dark patterns using a web augmentation approach. 
Through a two-phase co-design study, including 5 co-design workshops (\textit{N}=12) and a 2-week technology probe study (\textit{N}=15), we reported findings on the understanding of users' needs, preferences, and challenges in handling dark patterns and investigated the feedback and reactions to users' awareness of and action on dark patterns being empowered in a realistic in-situ setting.

\end{abstract}



\received{January 2023} 
\received[revised]{July 2023} 
\received[accepted]{November 2023}


\maketitle

\section{Introduction}

\textit{Dark patterns}~\cite{grayMappingLandscapeDark2023} are user interface design choices that lead to certain decisions users might not otherwise make, often for the purpose of benefiting an online service~\cite{narayanan_dark_2020,gray_dark_2018,grayPreliminaryOntologyDark2023}. Such dark patterns often result in user behaviors that are against the best interests of users e.g., \chao{pressured selling}, video binge-watching, giving up personal data, and installing applications that they do not need~\cite{digeronimo_dark_2020,mathur_2019_dark,chaudhary2022you}.
\chaohighlight{
Dark patterns vary in their complexity and impact.
Ranging from subtle manipulative nudges~\cite{hansen_nudge_2013} like non-consensual additions to shopping carts~\cite{mathur_what_2021} and prioritized options~\cite{nouwens_dark_2020}, to overt deception like alarming virus alerts, these patterns have led to adverse user impacts, raising public awareness and prompting regulation~\cite{united_states_senator_deb_fischer_senators_2019, warner_lawmakers_2021}.
}

\chaohighlight{Most current efforts in addressing dark patterns take (1) the designer’s perspective on education and advocacy for ethical designs~\cite{moser_impulse_2019,grasl_dark_2021,gray_dark_2018}; and (2) the policymaker’s perspective on new
regulations~\cite{leiser_dark_2020,markt_guidelines_2020,european_data_protection_board_guidelines_2020}. For example, a CHI 2021 workshop titled ``What can CHI do about dark patterns?~\cite{lukoff_what_2021}'' was held to discuss how designers can address dark patterns and what changes designers can advocate for through interactions with stakeholders. 
Another instance is an act named ``Deceptive Experiences To Online Users Reduction (DETOUR)'' introduced to prohibit large online platforms from using dark patterns~\cite{united_states_senator_deb_fischer_senators_2019, warner_lawmakers_2021}. 
However, these efforts often fell short of fully utilizing the autonomy of end users in self-protection~\cite{boush_deception_2016,ahujaConceptualizationsUserAutonomy2022}. 
End users have strong incentives and the desire to protect themselves from online threats, but often lack the capacity and associated support~\cite{smullen_managing_nodate,jin_exploring_2022,zou_examining_2020,lyngs_i_2020}.
Moreover, dark patterns are generative and shapeshifting, thus will continuously evolve, making it difficult to fully define and regulate through policies. 
Therefore, to complement previous efforts, we take on an \textbf{end-user-empowerment} orientation in this paper to explore the design of interventions for dark patterns considering the autonomy of end users.}

\chaohighlight{Guided by the Protection-Motivation Theory (PMT)~\cite{rogers_protection_1975}, we coined two types of intervention for our end-user-empowerment approach, targeting users' \textit{awareness} and \textit{action}. First, we enhance \textit{awareness} by increasing transparency about the presence and impacts of dark patterns. Second, we enable users to take \textit{action} against dark patterns, as previous studies have shown that awareness alone is not sufficient~\cite{bongard-blanchy_i_2021}. We employ a web augmentation approach, allowing users to select between pre-defined UI enhancements to dark patterns according to their preferences.}


\chaohighlight{
We also propose a Design-Behavior-Outcome framework, to map out the design space for UI enhancements in user \textit{action}. 
This framework situates individual intervention techniques (e.g., hiding, disabling, friction, etc.) from previous work~\cite{bongard-blanchy_i_2021,caraban_23_2019,lyngs_i_2020,kollnig_i_2021,boush_deception_2016,moser_impulse_2019,gould_special_2021,wang_field_2014} at different interaction phases between users and dark patterns. The resulting UI enhancements can change interface designs and user flows, or evoke users to reflect on the consequences caused by dark patterns.
}

\chao{To explore the design of our end-user-empowerment intervention for UX dark patterns, we conducted a two-phase co-design study. 
The first phase was five exploratory co-design workshops with 12 participants.
We investigated user needs, challenges, and preferences in handling UX dark patterns.} 
Through the workshops, we found \chao{that users have the desire to actively learn about dark patterns' impact, and their perceptions and coping mechanisms of dark patterns are individualized and dynamically changing. 
They also expect to be able to counteract dark patterns by changing interfaces, adjusting user flows, and reflecting on behavioral outcomes.}

\chao{Informed by the results of the first phase workshops, we further curated and deployed a technology probe study with 15 new participants for two weeks.
The probe study aims to contextualize users in their everyday experience, \ylhighlight{investigate their feedback towards our approach throughout 2 weeks}, and elicit more design implications for future end-user-empowerment interventions.
We materialized our awareness and action interventions as a probe named \textsc{Dark Pita}\footnote{\textsc{Dark Pita} is an acronym for \textbf{Dark} \textbf{P}attern \textbf{I}ntervention for \textbf{T}ransparency, and \textbf{A}ccountability.} in the form of a browser extension against a representative sample of dark patterns in popular online services.} The results showed that with our end-user empowerment approach, users gained transferable knowledge about dark patterns, felt empowered with autonomy over UIs, and chose UI enhancements to act against undesired dark patterns based on their dynamic, contextualized goals on different platforms.\looseness=-1


\chaohighlight{
Although the current version of \textsc{Dark Pita} is limited to handling a small sample of dark patterns with hand-crafted design enhancements, it exemplifies a new bottom-up end-user-empowerment approach. 
The study findings confirmed the effectiveness and presented useful design implications. 
The paper also outlines a research roadmap towards scaling up our approach with development in user behavior modeling, interface semantic understanding, and citizen science platforms. 
We end this paper with a discussion on how the end-user-empowerment approach connects to the ongoing efforts in policy-making and advocacy for design ethics.
}

In summary, this paper makes the following contributions.

\begin{enumerate}

\item A novel end-user-empowerment intervention approach for counteracting dark patterns in UX by enabling the end users of interfaces to recognize, understand, and take action upon dark patterns, including raising the transparency of dark patterns' presence and impacts; and modifying dark patterns by switching between UI enhancements according to their own personal preferences and goals.  

\item Findings and design implications from a two-phase co-design study consisting of 5 co-design workshops (\textit{N}=12) to explore users' underlying needs, preferences, and challenges in handling dark patterns; and a 2-week technology probe study (\textit{N}=15) to investigate users' feedback and reactions to their awareness of and action on dark patterns being empowered in an everyday setting.

\item An agenda for the research community to scale up this approach and deploy it in conjunction with ongoing efforts in crowd-sourced collective intelligence, citizen science, machine learning, policy making, and advocacy for design ethics. 

\end{enumerate}
\section{Background and Related Work}


\subsection{Studies of Dark Patterns}
\label{sec:studies_dark_pattern}

Brignull coined the term ``dark pattern'' (also known as ``deceptive design pattern'') which refers to ``a user interface that has been carefully crafted with an understanding of human psychology to trick users into doing things that they did not intend to''~\cite{gray_dark_2018}. Such dark patterns are prevalent---a previous study analyzed 240 popular mobile apps and found that 95\% of them contained at least one instance of dark patterns~\cite{digeronimo_dark_2020}. They commonly come in a variety of types across the web and mobile platforms~\cite{gunawan_comparative_2021} in different cultural context~\cite{hidakaLinguisticDeadEndsAlphabet2023}, exploiting users' attention~\cite{greenberg_dark_2014}, time~\cite{mildner_ethical_2021,chaudhary_are_2022}, money~\cite{mathur_2019_dark}, privacy~\cite{bosch_tales_2016} and autonomy in outright or subtle ways~\cite{conti_malicious_2010,toth_dark_2022}. 
At CHI 2023, a new SIG was formed to combat the growing issue of dark patterns in tech design through research, regulation, and interdisciplinary collaboration~\cite{grayDarkPatternsEmerging2023,grayEmergingTransdisciplinaryPerspectives2023}.

\ylhighlight{The prevalence and ``dark'' nature of dark patterns arose in a wide range of work published in the past years by HCI and CSCW academics. Brignull established a site\footnote{https://www.deceptive.design/} to collect examples of dark patterns and divided them into different types~\cite{brignull_deceptive_nodate}.
Gray et al.~\cite{gray_dark_2018} introduced ``dark patterns'' as an ethical phenomenon in design and identified five manipulative design strategies: nagging, obstruction, sneaking, interface interference, and forced action.
This foundational work has led researchers to uncover dark patterns on gaming~\cite{lewis_irresistible_2014,aagaard_game_2022,zagal_dark_2013}, robotics~\cite{laceyCutenessDarkPattern2019}, IoT devices~\cite{kowalczykUnderstandingDarkPatterns2023}, and social platforms~\cite{mildner_ethical_2021,mathur_endorsements_2018,mildnerEngagingGoverningStrategies2023,mildnerDefendingDarkArts2023,schaffnerUnderstandingAccountDeletion2022}, thereby establishing both generic~\cite{gray_dark_2018,conti_malicious_2010} and domain-specific~\cite{lewis_irresistible_2014,greenberg_dark_2014, mathur_2019_dark, chaudhary_are_2022, bosch_tales_2016} taxonomies of dark patterns. 
To unify these diverse taxonomies, Mathur et al.~\cite{mathur_what_2021} proposed six design attributes to characteristic dark patterns at a high level of generality. 
They described how dark patterns modify the disclosed information and underlying choice architecture for users, helping us to disclose manipulative mechanisms and provide targeted alternatives in our study.}\looseness=-1

\ylhighlight{Previous work also examined the designers' perspective regarding their intents and stakeholder values leading to ``dark'' designs~\cite{chivukula_dark_2018}. Moreover, studies on user attitude and perception have also expanded the dark pattern literature~\cite{chaudhary_are_2022,bongard-blanchy_i_2021,gray_enduser_2021}.
They investigated users' accounts of felt manipulation~\cite{gray_enduser_2021}, unintended behaviors~\cite{chaudhary_are_2022}, and perceived nuances between dark patterns and ``asshole design~\cite{gray_what_2020},'' foregrounding the need for users to have agency over their online experience~\cite{mathur_what_2021}.
Therefore, our work builds on previous efforts to address dark patterns (Section~\ref{section 2.2}) and understand user awareness (Section~\ref{section 2.3}), using co-design methods to explore a new intervention approach to \textit{empower users} against online manipulation. }\looseness=-1









\subsection{Efforts in Addressing Dark Patterns} \label{section 2.2}

\ylhighlight{Previous work investigated how designers, educators, and regulators can contribute to addressing dark patterns' adverse influence on end users~\cite{bongard-blanchy_i_2021,chivukula_dark_2018, maier2019dark,nouwens_dark_2020,gray_enduser_2021}. }

\chao{From the perspective of designers, a growing number of researchers called for the incorporation of ethics into the design process~\cite{soden_chi4evil_2019, narayanan_dark_2020}. 
Chivukula et al.~\cite{chivukula_dark_2018} revealed that designers often have dark and tacit intentions to persuade users with business purposes of satisfying stakeholders, even with sensitivity to user values.
\ylhighlight{Academics, therefore, have proposed design methods to foster better alignment with user values, such as value-centered design~\cite{gray_what_2020}.
From an education standpoint, Gray et al.~\cite{gray_dark_2018} encouraged UX professional organizations to build ethical education into the fabric of HCI/UX education and practice.  
Educators can also offer courses to deepen users' understanding of dark patterns~\cite{digeronimo_dark_2020},  train users to identify them~\cite{m_bhoot_towards_2020}, and increase their resistance through long-term boosts~\cite{hertwig_nudging_2017, wilson2022dark}. 
In terms of policymaking, efforts to investigate how dark patterns hurt user benefits (Section~\ref{sec:studies_dark_pattern}) have been raised as a space for new policies to be formed.
Regulators can implement economic incentives and regulatory interventions to force companies to reduce dark patterns in their services~\cite{leiser_dark_2020,markt_guidelines_2020,european_data_protection_board_guidelines_2020}.}
For example, recently published official reports from the European Union Commission~\cite{eu_commission_2022}, the European Data Protection Board (EDPB)~\cite{europeandataprotectionboardGuidelines2022Dark2022}, and the Federal Trade Commission (FTC)~\cite{federaltradecommisionBringingDarkPatterns2022} that specifically outline taxonomies of dark patterns, examples of violations, and opportunities for characterization and governance interventions.
}

\ylhighlight{However, most of these efforts overlooked the end users' autonomy of self-protection~\cite{boush_deception_2016,ahujaConceptualizationsUserAutonomy2022}. End users have strong incentives and the desire to protect themselves from threats in their online experiences, but often lack the capacity to do so~\cite{smullen_managing_nodate,jin_exploring_2022,zou_examining_2020,lyngs_i_2020}.} Meanwhile, dark patterns are shape-shifting and continuously evolving, making it hard to completely ban them with policies.

\ylhighlight{There is no one-size-fits-all solution. 
Previous studies have coined many intervention techniques to change individual types of dark patterns, such as enforcing consent~\cite{gunawanRedressDarkPatterns2022}, hiding or disabling~\cite{lyngs_i_2020,kollnig_i_2021}, adding friction~\cite{moser_impulse_2019}, and using ``bright patterns''~\cite{grasl_dark_2021}. 
Due to the diversity of dark patterns, these techniques can hardly be effective for all. 
To add an additional challenge, users' preferences for intervention techniques can change with their evolving understandings and perceptions~\cite{lyngs_i_2020,gray_enduser_2021}. 
Therefore, we need to better understand end users' expectations of interventions and their spontaneous approach to self-protection.}

\ylhighlight{In this work, inspired by previous studies of user awareness of dark patterns~\cite{chaudhary_are_2022,gray_enduser_2021,bongard-blanchy_i_2021,maierDarkDesignPatterns2020} and end-user web augmentation~\cite{stuerzlinger_user_2006,kim_stylette_2022,kumar_bricolage_2011,lim_ply_2018,zhang_fusion_2018}, we take a human-centered approach to support end users, by disclosing the presence and impact of dark patterns, and empowering users to ``fix'' the undesired ones with pre-defined UI alternatives.}



\subsection{User Perception of Dark Patterns}
\label{section 2.3}



\chao{Several researchers have conducted empirical studies to understand users' perception of dark patterns~\cite{chaudhary_are_2022,luguri_shining_2021,bongard-blanchy_i_2021,maierDarkDesignPatterns2020}. 
For example, Gray et al.~\cite{gray_enduser_2021} identified qualitatively supported insights to describe end users' experiences of being manipulated. 
They found a broad awareness from users that something is ``off'' or ``not correct,'' but still lacking the ability to precisely describe what drives the feeling of being manipulated~\cite{gray_enduser_2021}. 
Maier and Harr~\cite{maierDarkDesignPatterns2020} suggested users' perception of dark patterns goes through four stages---impression, assessment, balance, and acceptability. 
They have to get an impression of dark patterns, assess their convenience and manipulation, balance the trade-off, and then accept or reject dark patterns. 
\ylhighlight{However, the obscurity of design intents and the abuse of cognitive biases~\cite{mathur_2019_dark} make it difficult for users to comprehensively understand dark patterns (impression), and therefore hinder the subsequent assessment and balance processes. }
Therefore, we conducted the first-phase co-design workshops, seeking to answer what information users need to make up for the lack of transparency.
\ylhighlight{Based on the findings, we propose the \textit{awareness} intervention, aiming to empower the end users of an interface to recognize dark patterns (impression), understand the potential effects on their choice architecture and welfare (assessment), and balance the tension between user values and manipulation (balance).
}
Furthermore, Bongard-Blanchy et al~\cite{bongard-blanchy_i_2021}. discovered that the awareness of dark patterns does not necessarily lead to the ability to oppose adverse manipulative influences. 
A single ``transparency'' intervention may be insufficient to help users counteract dark patterns, implying the dual role of raising users' awareness and empowering them to take actions.
Therefore, we propose another intervention called \textit{action} which is based on an end user web augmentation approach.
}

\subsection{Web Augmentation}
\ylhighlight{Web augmentation~\cite{nouwensConsentOMaticAutomaticallyAnswering2022a} allows end users to customize existing web interfaces for personalized user experiences. GreaseMonkey\footnote{https://addons.mozilla.org/en-US/firefox/addon/greasemonkey/} is among the earliest browser extensions that manage user scripts to augment websites, many of which target adapting web UIs. Since GreaseMonkey requires users to write code scripts, it is mostly used by people with programming skills. Later on, many low-code or no-code web augmentation tools have been designed to lower the technical barrier~\cite{park_openhtml_2013,bolin_automation_2005,chang_webcrystal_2012,lim_ply_2018,zhang_fusion_2018}, allowing end users to change websites by direct interaction with UI elements~\cite{nebeling_xdbrowser_2016,nebeling_crowdadapt_2013,oppenlaender_crowdui_2020,kim_stylette_2022} or replacing components with defined alternatives~\cite{kumar_bricolage_2011,lee_designing_2010,fitzgerald_copystyler_2008}. This interaction paradigm makes it easier for end users without programming expertise due to its naturalness~\cite{myers2017making}.}

\ylhighlight{Many of these tools adopt a community-driven approach, where users share their web augmentations to be re-used by others (e.g. GreaseSpot\footnote{https://wiki.greasespot.net/} and Arc Boosts Gallery\footnote{https://arc.net/boosts}). However, these communities and their dynamics seem to be under-studied in CSCW, with only a few papers from adjacent research communities~\cite{nebeling2013crowdadapt, oppenlaender2020crowdui, alves2023gitui, firmenich2015user}.}

\ylhighlight{Previous work has investigated the use of web augmentation to address specific dark patterns. For example, Nouwens et al.~\cite{nouwensConsentOMaticAutomaticallyAnswering2022a} designed a browser extension, Consent-O-Matic, that automatically responds to consent pop-ups based on the user’s preferences. Kollnig et al.~\cite{kollnig_i_2021} proposed an approach named GreaseDroid, enabling Android users to remove dark patterns in mobile applications with ``patches''. While these two technical-centric studies extended the feasibility of dark pattern interventions through web/mobile augmentation, they did not fully investigate end users' needs for such interventions through a user-centered lens. In this work, our two-phase co-design study seeks to complement these technical UI augmentation work by trying to understand end users' needs, preferences, and expectations for interventions through UI augmentation. Our insights provide inspiration for community creators in designing alternatives to unethical design patterns in the future.}

\section{Co-Design Workshops}
\label{sec:co-design-workshop}
In the first phase of our study, we conducted 5 in-person exploratory co-design workshops\footnote{The protocol of workshops has been reviewed and approved by the IRB at our institution.} to achieve the following goals: \looseness=-1

\begin{enumerate}
    \item \ylhighlight{Exploring users' perceived disruptiveness and annoyance of different dark patterns in various usage contexts;}
    \item Learning the existing measures that users have developed or adopted, consciously or subconsciously, to cope with dark patterns;
    \item \ylhighlight{Investigating users' needs and expectations regarding dark pattern intervention techniques.}
\end{enumerate}


\subsection{Participants}
We recruited 12 participants (PA1--PA12; 5 men, 7 women) through word of mouth, email mailing list, and flyer distribution. Our participants represent diverse backgrounds in occupational domains (e.g., health, education, social assistance, and information services), Internet usage (ranging from 2--5 hours to 8+ hours per day), and knowledge of dark patterns (7 had heard of the concept, the remaining 5 had not). Detailed demographics can be found in Appendix~\ref{appendix:workshop_demographics}. We conducted 5 in-person workshops with 2--4 participants in each session. \ylhighlight{The groups were divided based on participants' time availability.} Each participant was compensated \$30 for their time.

\subsection{Workshop Activities}
Each workshop lasted 2 hours and started with a brief introduction to the concept and examples of dark patterns. The participants completed three activities together: a focus group discussion, a storyboard fill-in session, and a tangible website redesign activity.

\subsubsection{Focus Group Discussion}
 \ylhighlight{In a focus group, participants were first introduced to the concept of dark patterns, then reflected on and shared dark patterns examples they previously encountered in everyday lives. During the discussion, researchers provided feedback and clarifications to help them understand the boundary and varied ``darkness'' level of dark patterns~\cite{gray_dark_2018}. The participants were then asked to rank their examples by the level of perceived annoyance and disruptiveness and provide explanations. Researchers followed up with questions to find out the current strategies adopted by participants to address the impacts of dark patterns.}
 
 \chaohighlight{We designed this activity to help participants ground their understanding of dark patterns' prevalence and impact in their concrete personal experiences. Reflecting on dark patterns and the associated level of annoyance also acted as a stimulus, prompting participants to contemplate countermeasures in subsequent activities. The format was intentionally less structured, with an emphasis on encouraging speak-up and fostering a comfortable environment that would promote open dialogue in subsequent activities.}

\begin{figure}
\centering
\includegraphics[width=\linewidth]{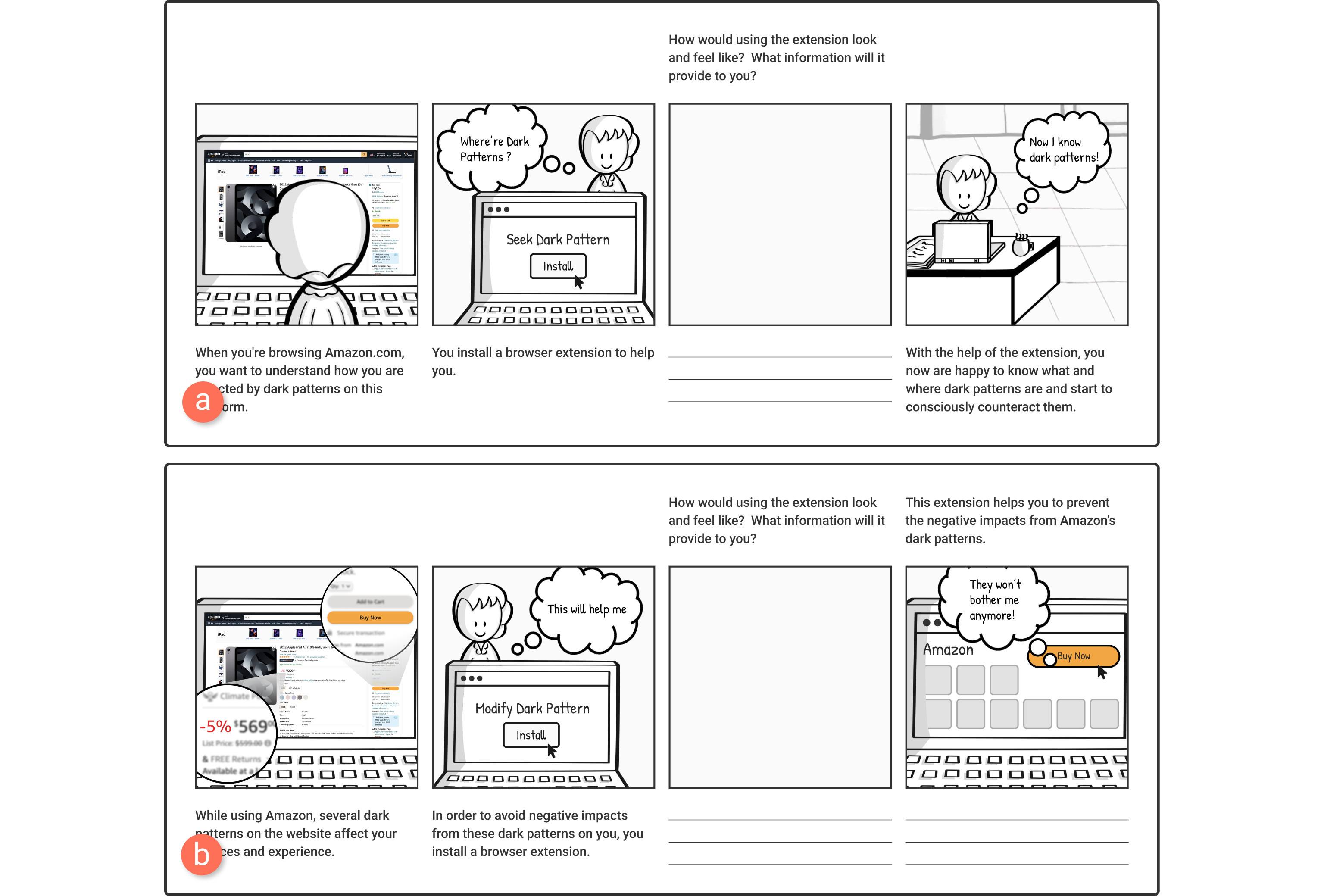}
\caption{Two examples of storyboards used in our co-design workshops.} \label{storyboard}
\end{figure}

\subsubsection{Storyboard Fill-in}
The second activity was a storyboard fill-in. Storyboards are commonly used HCI tools to visually communicate user experience scenarios to the audience~\cite{truong2006storyboarding, zhang_storybuddy_2022, sturdee2019sketching}. In co-design workshops, storyboards help contextualize participants and prompt them to think about their needs, goals, and constraints in the scenarios described~\cite{muller_participatory:_1993, bodker1999scenarios}. \chaohighlight{In our study, we adopted ``fill-in-the-blanks'' storyboards~\cite{wikstrom_exploring_2013,moraveji_comicboarding_2007} to explore the users' desired solutions to dark patterns. 
For each storyboard, we left one or two frames blank, encouraging participants to provide insights into their understanding of dark patterns and their preferred abilities to counter these patterns\footnote{Examples of participant-filled storyboards are available in the supplemental materials.}. 
}

\chaohighlight{
Specifically, we presented participants with 6 storyboards (Fig.~\ref{storyboard}) depicting scenarios in which a tool helped them mitigate the negative impacts of dark patterns. Of these, 3 focused on improving users’ awareness of dark patterns (Fig.~\ref{storyboard}a), and 3 (Fig.~\ref{storyboard}b) others on helping users take act against dark patterns. 
Each storyboard contained 4 frames, representing the background of the scenario, the tool used, how the tool helped, and the desired outcome respectively. 
We left the second or/and the third frames blank and asked participants to brainstorm their desired tools and their interactions.
}

\subsubsection{Tangible Website Redesign}
\ylhighlight{The last activity was a tangible user interface redesign. While storyboard fill-in focused on the design of intervention scenarios (e.g. the user flow of interventions), this activity targeted intervention at a lower, more granular level---the specific alternatives of dark patterns on UI design. We designed this activity to be tangible, in the form of paper prototyping instead of digital UI mockup modification, to encourage participants to make bold changes and think outside the box~\cite{snyder2003paper}. It also avoids the learning process for a new digital design tool.} 

We curated 13 representative dark pattern examples on 7 websites for online shopping, flight booking, video streaming, and social media as a diverse set of scenarios. These examples were selected from previous research literature~\cite{bongard-blanchy_i_2021, gray_dark_2018}, online discussions of dark patterns ~\cite{brignull_deceptive_nodate}, and the researchers' own experiences, with the goal of triggering discussions among participants. In each workshop, participants selected around 5 dark patterns that they were most concerned about to work on. \ylhighlight{For each dark pattern instance, we provided a printout of the website interface and a set of cut-out UI widgets from the same interface. During the activity, participants edited the cut-out widgets and drew new UI widgets using a variety of provided stationery\footnote{The provided stationery included but was not limited to pencils, colored pens, scissors, highlighters, and glue sticks}, and re-assembled them into a more desired design of the original website interface. The interfaces created by the participants were collected for analysis\footnote{Examples of participant-redesigned interfaces are available in the supplemental materials.}}.

\subsection{Workshop Findings}
\label{workshop_findings}
Following the open coding methods~\cite{braun2006using,lazar2017research}, two researchers conducted a thematic analysis of audio transcriptions for workshops and materials produced in the three activities. The researchers conducted three rounds of iterative labeling, in which descriptive labels on relevant transcript pieces were created, grouped, and generalized into higher-level themes. 
\ylhighlight{The analysis was discussion-based, with no necessity for inter-rater reliability due to the aim of discovering emergent themes~\cite{mcdonald2019reliability}}\looseness=-1. 

\ylhighlight{Comparative analysis was not the focus of the study, but a comparison of responses from tech industry users and other users showed no significant difference in their perceptions or coping mechanisms concerning dark patterns. This is in line with previous work~\cite{keleher2022well} which showed no significant differences between end users and experts in perceptions of dark patterns.}

Our workshop findings (WF) are described below in response to our workshop goals.


\paragraph{\textbf{WF1: Users would like to learn more about the impact of dark patterns}}
\label{WF1}

\ylhighlight{Participants expressed their desire to know more about the potential impacts of dark patterns. Although many were able to detect dark patterns, most participants only developed a vague assumption about the impacts, falling short of articulating the specifics. The particular mechanisms and impact remain as a ``blackbox'', confirming findings in~\cite{bongard-blanchy_i_2021,gray_enduser_2021}. }

Participants often asked about the detailed mechanisms and impact of dark patterns and felt the knowledge was useful. Importantly, clear knowledge of dark patterns' impact can help users choose services more consciously and potentially reduce the irritation of seeing dark patterns. \ylhighlight{PA6 and PA7 mentioned that for disguised ads on Instagram, \textit{``those are annoying at first, but once you know that (its impact) and come to expect it, it's like okay (less annoying) (PA6)''.} These findings complement existing research by demonstrating users' autonomy---they are not merely passive consumers of dark patterns. They are interested in actively learning, and the acquired knowledge can change their usage behavior and connections with online platforms.}

\paragraph{\textbf{WF2: Users' perceptions of dark patterns are personal and dynamically changing, which are formed based on user preferences, types of dark pattern instances, and usage contexts.}}
\label{WF2}
\ylhighlight{Despite the prevalent negative perception of manipulative UX design patterns, not all participants viewed them unfavorably. In fact, responses varied widely from negative to positive, echoing findings from studies on online behavioral advertising~\cite{ur_smart_2012}. Users often perceived a persuasive pattern as helpful when it aligns user goals with stakeholder profits. PA2 liked the autoplay feature on Netflix, even when understanding it used forced continuation, because \textit{``it is useful when I am away from my mouse''}. Similarly, for disguised ads on social media, PA12 expressed that \textit{``I won't block them. I usually don't engage or buy stuff... Maybe I'll see something in the future I like''.} We summarized 3 most common perceptions from participants: \textit{disruptiveness} (the user experience was disrupted by the dark pattern), \textit{indifference} (the dark pattern was neither harmful nor useful), and \textit{helpfulness} (the pattern was helpful in the current context).}


\ylhighlight{Users encountering a dark pattern typically evaluate its potential pros and cons subconsciously, influenced by factors like perceived convenience, potential consequences, and the pattern's apparent malicious intent. Accordingly, three factors---\textbf{the user, the dark pattern, and the usage context}---determine this perception. An example is the autoplay feature for the next episode on Netflix. PA1, PA10, and PA12 expressed their dislike of this feature, while PA2, PA5, and PA11 generally thought it was convenient. Furthermore, even the same user's perception can shift with different usage contexts and changing goals. PA11 found auto-play harmful when using Netflix during work because of the short break time, but useful when casually browsing after work just for fun. These findings are in parallel with results in~\cite{gray_enduser_2021} and enhance previous findings by highlighting the highly individualized and contextualized nature of users' perceptions of dark patterns.}

\paragraph{\textbf{WF3: Users develop varied coping mechanisms based on their different dark pattern perceptions.}}
\label{WF3}
\ylhighlight{This finding reveals more details regarding how end users react when facing dark patterns' perceived influences, in addition to the conclusions in~\cite{bongard-blanchy_i_2021}. For disruptive dark patterns, many users actively seek solutions to mitigate the impact. PA12 used a calendar to track the end of free subscription trials as a reminder to unsubscribe. On Instagram, PA6 developed a habit to avoid disguised ads when tapping through all stories. \textit{``Funnily enough, every time I watch a story, I have developed... an unconscious habit, I close out (by swiping down) and I click the next one. (PA6)''} If the user can successfully find a solution, it becomes a \textit{``muscle memory''} for them. For the Instagram habit, PA6 expressed that \textit{``I didn't know why I do that, but I guess that's a dark pattern and I am unconsciously adapted.''} PA1 also developed the habit of reaching their mouse before an episode ends on Netflix, to wiggle the cursor in time and avoid forced continuation.}

For dark patterns that users feel indifferent to, the most common strategy is ignoring them. PA6 mentioned that they gradually got used to ``confirmshaming'' dark patterns and ``just don't care anymore''. When asked about a disguised advertisement on a flight booking website during the redesign activity, PA8, PA9, and PA10 reported that they did not even notice it. They considered it ``too colorful'' to be relevant and therefore simply ignored it.


\subsection{Design Implications}
The thematic analysis results of our co-design workshops offered several design implications for dark pattern intervention techniques and user empowerment.

\paragraph{\textbf{DI1: Empower users with the ability to make changes on dark patterns.}}
\label{DI1}
When encountering disruptive dark patterns, users often feel manipulated but have no ability to resist (\textbf{\hyperref[WF1]{WF1}}). To help participants regain self-autonomy~\cite{mathur_what_2021}, we can empower end users with the ability to change the interfaces of dark patterns. It would help users take the initiative to mitigate the negative impact.

\paragraph{\textbf{DI2: Provide information on the potential consequences of dark patterns}}
\label{DI2}
During our workshops, participants expressed the need for information on the potential influences of dark patterns and envisioned a ranking of their severity. With such information, users can make better informed evaluations of the impact of a dark pattern on themselves (\textbf{\hyperref[WF2]{WF2}}) and develop their coping mechanism accordingly (\textbf{\hyperref[WF3]{WF3}}).

\paragraph{\textbf{DI3: Offer users multiple intervention options for each dark pattern.}}
\label{DI3}
\ylhighlight{Perceptions of dark patterns may shift with users, types of dark pattern instances, and usage contexts. Even for the same dark pattern, users may act differently (\textbf{\hyperref[WF3]{WF3}}). As a result, it is necessary to have multiple intervention options for users to choose from. In this way, users can have more flexibility in personalization and autonomy.
}

\paragraph{\textbf{DI4: Design dark pattern interventions with three strategies: interface design change, user flow adjustment, and behavioral outcome reflection.}}
\label{DI4}
\ylhighlight{In our redesign activity, participants proposed intervention techniques for dark patterns that can be categorized into three approaches: (1) modifying interface components and layouts to eliminate malicious design, (2) adjusting user flows to prevent users from falling into behavioral traps, and (3) evoking reflection by uncovering the outcomes of dark patterns for long-term self-change. Future intervention designs can take inspiration from these strategies and apply them in appropriate scenarios.}

For example, on a flight booking website, while the website highlighted the more expensive first-class and main-cabin options over the basic economy, PA8, PA9, and PA10 changed them to the same size and color to pursue a fair style. PA1, PA2, and PA3 designed an agent to provide an appropriate action guide with dark patterns on Amazon to help them save money. PA6 and PA7 wanted to know, in the long term, how many times dark patterns on a certain website affected their behavior, to reflect on their relationship with the platform.

We used these findings and design implications from our co-design workshop to guide our second co-design phase---a technology probe study.
\section{Technology Probe Study}

The co-design workshops (Section~\ref{sec:co-design-workshop}) served as a starting point for us to understand the existing relationship of users with dark patterns and their desired interventions. They looked at users' past daily experiences of encountering and coping with dark patterns with little or no external support. \ylhighlight{To further explore users' in-situ reactions toward ``fixing'' dark patterns on their own devices, we conducted a two-week deployment study of a technology probe.}

The technology probe method, proposed by Hutchinson et al.~\cite{hutchinson_technology_2003}, deploys ``simple, flexible technologies'' as probes in the real world with three goals: ``the \textit{social science goal} of collecting in-context information about the use and the users, the \textit{engineering goal} of testing the technology, and the \textit{design goal} of inspiring users and researchers to envision future technologies''~\cite{hutchinson_technology_2003}.
\chaohighlight{This method is widely used to examine the influence of new technologies on the daily experience of users as part of the co-design process~\cite{kaur_interpreting_2020,seymour_informing_2020}. It is worth mentioning that a technology probe study, while containing an engineering goal of field-testing a probe, is not equivalent to an evaluative study for the efficacy of well-developed systems.
This method does not seek to evaluate the probe's effectiveness on users' behavior change but to discover design implications and insights~\cite{hutchinson_technology_2003}. Thus we designed our technology probe, \textsc{Dark Pita},  with three research goals in mind: }
\label{sec:research_goals}
\begin{enumerate}
\item \textbf{Social science goal} is to understand end-user reactions, preferences, and desires in situ on \emph{awareness} of and \emph{action} for dark patterns in their online experiences.  
\item \textbf{Engineering goal} is to field test the technical feasibility of combining \emph{awareness} and \emph{action} as an end-user-empowerment intervention for dark patterns in realistic contexts of use. \looseness=-1

\item \textbf{Design goal} is to explore the design space of techniques, strategies, and interfaces for end-user-empowering interventions, with a specific focus on trade-offs and user constraints. 
\end{enumerate}

\begin{figure}
\centering
\includegraphics[width=\linewidth]{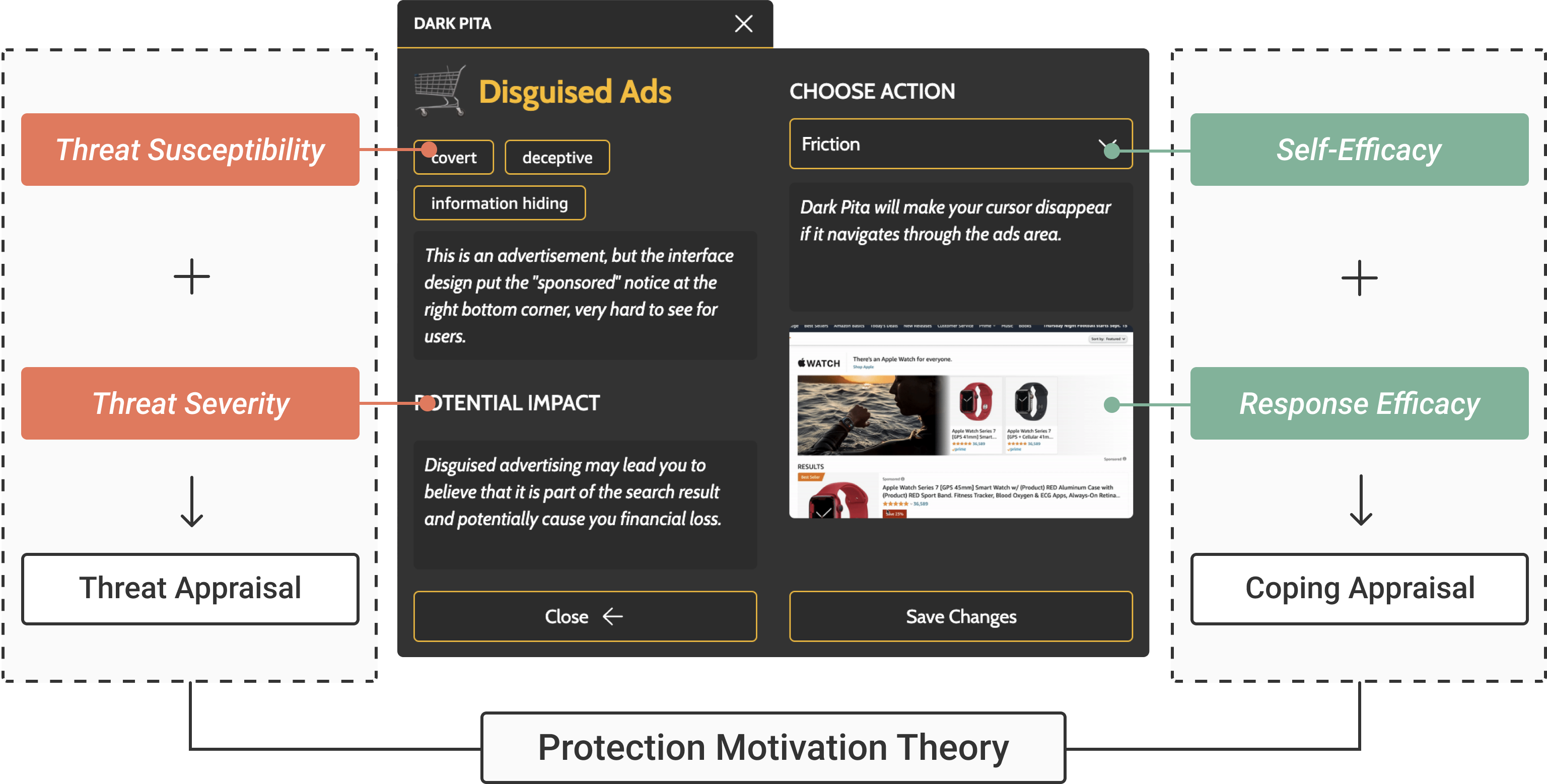}
\caption{The connections between \textsc{Dark Pita}'s design features and Protection Motivation Theory~\cite{rogers_protection_1975}.} \label{protection_motivation_theory}
\end{figure}

\subsection{Theoretical Grounding}
\label{sec:protection_motivation_theory}

\chaohighlight{
According to the design implications in the co-design workshops, the probe should help users to (1) raise awareness of dark patterns and
understand their underlying design intents (\textbf{\hyperref[DI2]{DI2}}) and (2) take actions to counter the effects of dark patterns in a web augmentation approach (\textbf{\hyperref[DI1]{DI1}}). 
The \emph{awareness} and \emph{action} mechanisms are naturally aligned with the \textit{\textbf{Protection Motivation Theory (PMT)}}~\cite{rogers_protection_1975}, which has been widely applied in behavior change design~\cite{van2019using, story2021design, prange2022secure, story2022increasing}.}

\chaohighlight{
PMT is commonly used to understand people's responses to triggers that appraise a potential threat~\cite{rogers_protection_1975}. 
It suggests to intervene in people's cognitive appraisal processes to motivate self-protection by articulating fear appeals. 
In PMT, two factors, \textit{threat appraisal} (how much people consider themselves at risk) and \textit{coping appraisal} (how effective people think their actions are against the risk), determine whether people would protect themselves~\cite{rogers_protection_1975}. This is in line with our workshop findings (\textbf{\textbf{\hyperref[WF2]{WF2}}} and \textbf{\textbf{\hyperref[WF3]{WF3}}}). Thus in our design, we used PMT as our theoretical basis and mapped our \textit{awareness} and \textit{action} mechanisms to \textit{threat} and \textit{coping appraisal} by disclosing the risk of dark patterns and guiding users to take effective measures against them (Figure~\ref{protection_motivation_theory}).
}

\subsection{The Probe: \textsc{Dark Pita}}

\chaohighlight{
To fulfill the research goals, we materialized our new PMT-based approach as a technology probe named \textsc{Dark Pita}, in the form of a browser extension that facilitates awareness and action for end-users against a small representative sample of UX dark patterns in several popular online services.} 

\chaohighlight{
In this section, we first describe an example scenario to demonstrate the user experience of interacting with \textsc{Dark Pita}. 
Then, we introduce the probe's main features in line with the five dimensions of probe design by Hutchinson et al.~\cite{hutchinson_technology_2003} and describe how we come up with UI enhancements for sampled dark pattern instances based on a new Design-Behavior-Outcome framework. 
Finally, we detail the technical implementation of \textsc{Dark Pita}.
}


\subsubsection{User Experience}

We selected five popular online services to support in \textsc{Dark Pita} (Amazon\footnote{https://www.amazon.com/}, Youtube\footnote{https://www.youtube.com/}, Netflix\footnote{https://www.netflix.com/}, Facebook\footnote{https://www.facebook.com/}, and Twitter\footnote{https://twitter.com/}). These samples represent different types of online services across task domains (i.e., online shopping, video streaming, and social media), containing diverse types of dark patterns (described in Section~\ref{sec:dark_pattern_instances}). \ylhighlight{They also possess substantial user bases, facilitating our recruitment process by accommodating a larger pool of potential participants.} In this section, we provide one example scenario of how a user may interact with \textsc{Dark Pita}. 



\begin{figure}
\centering
\includegraphics[width=\linewidth]{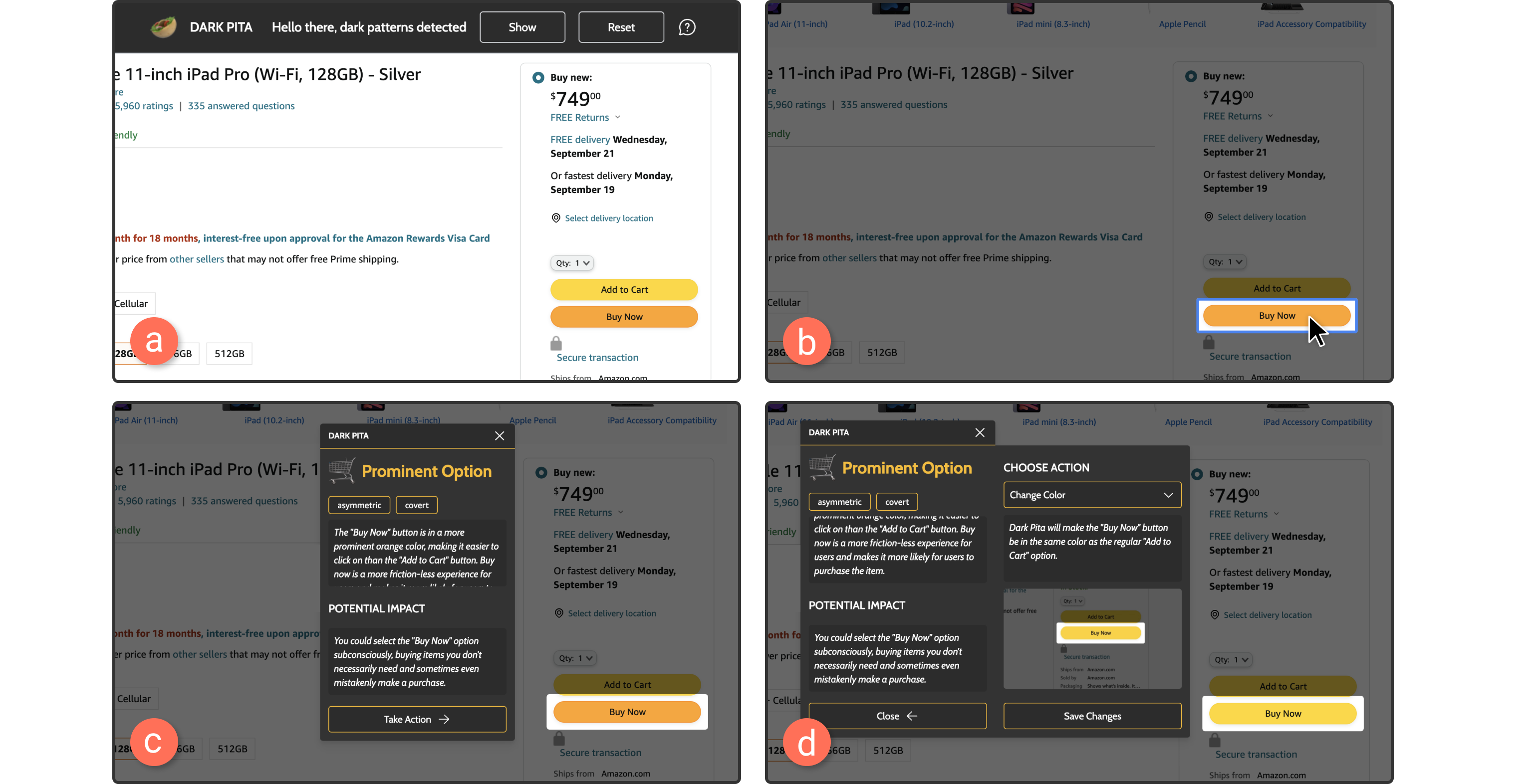}
\caption{An example scenario of how users might interact with \textsc{Dark Pita} on Amazon} \label{user_experience}
\end{figure}

Lisa is a frequent user of Amazon. When trying to check out on an item's page, she sees two buttons: ``Buy Now'' and ``Add to Cart''. The ``Buy Now'' button reduces the friction in checking out, improving Amazon's conversion rate; however, it can potentially cause users to buy unnecessary items they regret later~\cite{kuang2019user}. Here, Amazon designed ``Buy Now'' to be more visually prominent, making it easier to click on than ``Add to Cart''. \textsc{Dark Pita} notifies Lisa that dark patterns are detected on the page and highlights the ``Buy Now'' button (Fig.~\ref{user_experience}a and b). Then, she clicks on the highlighted area, and the \emph{awareness} panel reveals (Fig.~\ref{user_experience}c). It shows Lisa information about this dark pattern's manipulative mechanism and potential impacts, making her realize how the design can potentially trick her into directly checking out instead of adding the item to the cart. This way, Lisa will not be able to look at the total price of all items in the cart and reflect on the purchase before checking out, making her more likely to overspend on Amazon. 

\begin{figure}
\centering
\includegraphics[width=\linewidth]{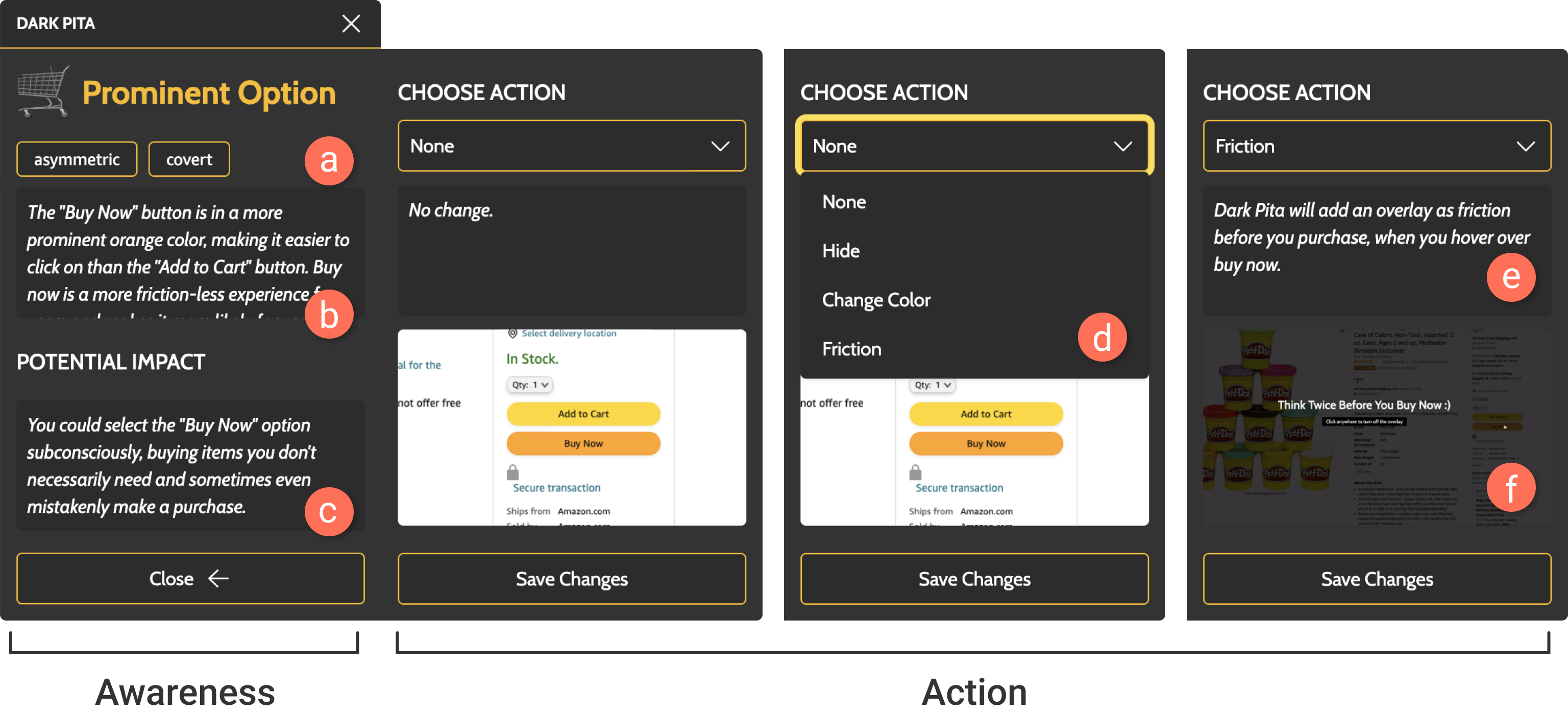}
\caption{\textsc{Dark Pita}'s \emph{awareness} and \emph{action} panels.} \label{design_features}
\end{figure}

To mitigate the effect of this dark pattern, Lisa opens the \emph{action} panel (Fig.~\ref{user_experience}d) and chooses a UI enhancement that changes the color of the ``Buy Now'' button to the same as the ``Add to Cart'' button (from several options available as shown in Fig.~\ref{design_features}d). With such experience, Lisa realizes that user interfaces can be styled differently to manipulate her decision-making process. Lisa feels that \textsc{Dark Pita} gives her more control and autonomy over these malicious interfaces. \looseness=-1

\subsubsection{Design Features}
Following the threat appraisal and coping appraisal processes of PMT (Section~\ref{sec:protection_motivation_theory}), \textsc{Dark Pita} consists of an \emph{awareness} panel and an \emph{action} panel (Fig.~\ref{protection_motivation_theory}). The \emph{awareness} panel brings the attention of end users by disclosing the manipulative mechanism of a dark pattern and the potential impact on user behavior (Fig.~\ref{design_features}). \ylhighlight{The \emph{action} panel empowers end users to take action to mitigate the negative impact by choosing from the multiple UI enhancements provided (Fig.~\ref{design_features})}. Overall, our probe can (1) detect and highlight dark patterns on websites, (2) disclose the manipulative mechanism (\textbf{\textbf{\hyperref[DI2]{DI2}}}), (3) articulate the potential impact (\textbf{\textbf{\hyperref[DI2]{DI2}}}), and allow users to (4) select UI enhancements for dark patterns (\textbf{\textbf{\hyperref[DI1]{DI1}}}, \textbf{\textbf{\hyperref[DI3]{DI3}}}, and \textbf{\textbf{\hyperref[DI4]{DI4}}}), and (5) preview the enhancement effect. Lastly, \textsc{Dark Pita} can (6) record participant interactions with the browser extension, the UI enhancements, and the supported dark patterns and allow participants to keep diary notes for study purposes.


\paragraph{\textbf{Dark pattern detection}} \ylhighlight{\textsc{Dark Pita} detects dark patterns when the user enters a website. Once the probe detects dark patterns (the detection technique is described in Section~\ref{sec:implementation}), a top banner appears, allowing the user to highlight all discovered dark patterns using the ``show'' button (Fig.~\ref{user_experience}b). When the user hovers their cursors over a highlighted dark pattern, a blue border suggests the clicking affordance for more information. \looseness=-1}

\paragraph{\textbf{Manipulative mechanism disclosure}}
\label{sec:design_feature2}

\ylhighlight{\textsc{Dark Pita} provides a brief explanation for each dark pattern (Fig.~\ref{design_features}b) and introduces the \textit{threat susceptibility} with ``dark'' attributes~\cite{mathur_what_2021} (Fig.~\ref{design_features}a). These attributes describe changes that the dark pattern imposes on the user's underlying choice architecture. Users can hover over each tag to get more details. For example, ``restrictive'' means that the dark pattern ``eliminates certain choices that should be available to you.'' }

\paragraph{\textbf{Potential impact articulation}} 
\label{sec:design_feature3}
For each dark pattern, \textsc{Dark Pita} explains its potential impact (\textit{threat severity}) through a normative perspective of individual welfare~\cite{mathur_what_2021} (Fig.~\ref{design_features}c). Specifically, we consider three types of individual welfare---financial loss, invasion of privacy, and cognitive burden, based on the framework proposed by Mathur et al.~\cite{mathur_what_2021}. \textsc{Dark Pita} identifies the theme to which the dark pattern belongs and elaborates on its impact. For example, it says ``You are likely to get distracted and watch videos that you never planned to. The automatic preview on hover grabs your attention and distracts you even further'' for the ``recommended videos'' dark pattern on Youtube.\looseness=-1

\paragraph{\textbf{User interface modification}}
If the user wants to take action to mitigate the potential impact of a dark pattern, they can press the ``Take Action'' button. The \emph{action} panel will appear on the left. \textsc{Dark Pita} provides 1--4 options of UI enhancements for each dark pattern to empower users to take protective actions (\textit{self-efficacy}) (Fig.~\ref{design_features}d). The user can select their desired enhancements and save the changes for their next visit to the site. We detail the design process of UI enhancements in Section~\ref{sec:ui_enhancement}.

\paragraph{\textbf{UI enhancement preview}}
For each enhancement, \textsc{Dark Pita} introduces its \textit{response efficacy} by explaining the effect (Fig.~\ref{design_features}e) and providing a preview (Fig.~\ref{design_features}f). The explanation informs the user about the intervention mechanism offered by the enhancement. It describes how this enhancement scaffolds the user to avoid harm to individual welfare. For example, the probe says ``\textsc{Dark Pita} will disable the preview function, which can protect you from being distracted.'' for a ``block preview'' enhancement. To demonstrate how an enhancement works, \textsc{Dark Pita} also displays a preview animation below the explanation. 

\paragraph{\textbf{Action logging and diary notes}}
\label{sec:action_diary_logging}
\ylhighlight{The probe records interactions of participants who have explicitly given their consent. All personal information contained in log entries is removed locally on the participant's device before being sent out. The log contains fine-grained interactions with the probe, such as timestamps, site information, panel openings, and saved UI enhancement choices. 
\textsc{Dark Pita} also provides a diary note panel for users to submit their reflections and expectations about the probe, dark patterns, and interventions. 
Participants can also attach screenshots that capture context information.}

Overall, \textsc{Dark Pita}'s features implement recommendations by Hutchinson et al.~\cite{hutchinson_technology_2003} that technology probes should be distinguished from regular design prototypes in five dimensions:

\begin{itemize}
\item \textbf{Functionality}: \textsc{Dark Pita} is simple enough to test only two key ideas: raising awareness of and taking action on dark patterns.

\item \textbf{Flexibility}: \textsc{Dark Pita} allows users to access its features on five sites in three task domains. It also provides multiple UI enhancements for each dark pattern, allowing users to modify dark patterns flexibly.

\item \textbf{Usability}: \textsc{Dark Pita} leverages a small sample of dark pattern instances to demonstrate its functionality, while leaving other dark patterns on these sites to evoke participants' reflections and design ideas. Usability was not a main concern in our deployment study.

\item \textbf{Logging}: \textsc{Dark Pita} implemented a comprehensive logging mechanism of participant interactions. It allows participants to record their situational thoughts and keep diary notes.

\item \textbf{Design phase}: \textsc{Dark Pita} proposes an end-user-initiated intervention paradigm in complement to existing solutions to dark patterns led by designers, educators, policymakers, etc. It is used in early design and aims to influence the future design of interventions and user empowerment.
\end{itemize}

\subsubsection{UI Enhancements}
\label{sec:ui_enhancement}
\chaohighlight{
We first sampled 13 instances of dark patterns across five popular sites. Then, we designed 1--4 UI enhancements for each instance (31 in total) using selected intervention techniques from previous studies~\cite{bongard-blanchy_i_2021,caraban_23_2019,lyngs_i_2020,kollnig_i_2021,boush_deception_2016,moser_impulse_2019,gould_special_2021,wang_field_2014} and co-design workshops (\textbf{\textbf{\hyperref[DI4]{DI4}}}) based on a Design-Behavior-Outcome framework.}

\begin{table}[!htbp]
\centering
\small
\caption{Instances of dark pattern with ``dark'' attributes~\cite{mathur_what_2021}. We designed 1--4 UI enhancements for each instance, employing different intervention strategies.}
\label{instance_attribute}
\begin{tabularx}{\textwidth}{lL{4cm}L{4cm}}
\toprule
\textbf{Name}          & \textbf{Attributes} & \textbf{Interventions} \\ \midrule
\yuwen{Prominent} ``Buy Now'' Button & Asymmetric, Covert & Hiding, Fairness, Friction \\ [8pt]
Disguised Ads & Covert, Deceptive, Information Hiding & Hiding, Friction, Counterfactual Thinking, Information Disclosure \\ [8pt]
\yuwen{Fake Discounts} & Information Hiding & Hiding, Action Guide, Counterfactual Thinking, Information Disclosure \\ [8pt]
\yuwen{Limited Time Recommendation} & Asymmetric, Counterfactual Thinking, Disparate Treatment, Information Hiding & Hiding, Counterfactual Thinking, Reflection \\ [8pt]
\yuwen{Video Autoplay} & Asymmetric, Disparate Treatment, Covert             & Hiding, Disabling, Reflection \\ [8pt]
\yuwen{Hiding Dislike Count} & Information Hiding & Information Disclosure \\ [8pt]
\yuwen{Auto Recommendations} & Asymmetric, Disparate Treatment, Covert & Hiding, Disabling, Reflection \\ [8pt]
\yuwen{Hiding Total Episode Time} & Restrictive, Covert, Information Hiding & Reflection \\ [8pt]
Automatic Preview & Restrictive, Covert & Disabling \\ [8pt]
\yuwen{Fake Trending Content} & Disparate Treatment, Covert & Hiding \\ [8pt]
\yuwen{Disguised Suggested Tweets} & Information Hiding, Covert & Information Disclosure, Friction \\ [8pt]
\yuwen{Sneaking Short Videos Into Feed} & Covert, Asymmetric & Hiding, Counterfactual Thinking, Friction \\ [8pt]
\yuwen{Disguised Sponsorship} & Covert, Information Hiding & Hiding, Information Disclosure \\ 
\bottomrule
\end{tabularx}
\end{table}

\paragraph{\textbf{Dark pattern samples}}

\label{sec:dark_pattern_instances}
\ylhighlight{Given that millions of dark patterns exist on the internet~\cite{digeronimo_dark_2020}, it is infeasible to cover all instances in our design probe. To build a simple and flexible probe~\cite{hutchinson_technology_2003} as a demonstration of our approach, we sampled a representative group of dark patterns. Also to ensure our participants can use the probe frequently during the study period,  we selected 13 instances from five popular websites across different service categories (Amazon, YouTube, Netflix, Facebook, and Twitter).} Based on Mathur et al.'s approach~\cite{mathur_2019_dark}, three researchers determined each instance's ``dark'' attributes individually and discussed to resolve conflicts and reach a consensus. Detailed \yuwen{category information and descriptions of the selected dark patterns are included in Appendix~\ref{appendix:dark_pattern_instance}. The research team balanced the dark pattern instances with different attributes and from different categories to form a representative sample, while some dark pattern attributes or categories are unavoidably more common than others.}

\begin{table}[!htbp]
\centering
\small
\caption{The intervention strategies used in the design of UI enhancements.}
\label{intervention}
\begin{tabularx}{\textwidth}{llX}
\toprule
\textbf{Name} & \textbf{Dimension} & \textbf{Definition}                                        \\ \midrule
Hiding & Design & Hide the dark pattern \\
Fairness & Design & Eliminate visual prominence of one option over others \\
Information Disclosure & Design & Exposing the hidden information from the dark pattern \\
Counterfactual Thinking & Behavior & Trigger users to be suspicious of the dark pattern \\
Disabling & Behavior & Disable the functionality of the dark pattern \\
Action Guide & Behavior            & Provide helpful behavioral guidance to deal with the dark pattern \\
Friction & Behavior & Add additional steps to the interaction flow of the dark pattern \\
Reflection & Outcome & Provoke users to reflect on the outcomes from the dark pattern \\ \bottomrule
\end{tabularx}
\end{table}

\begin{figure}
\centering
\includegraphics[width=\linewidth]{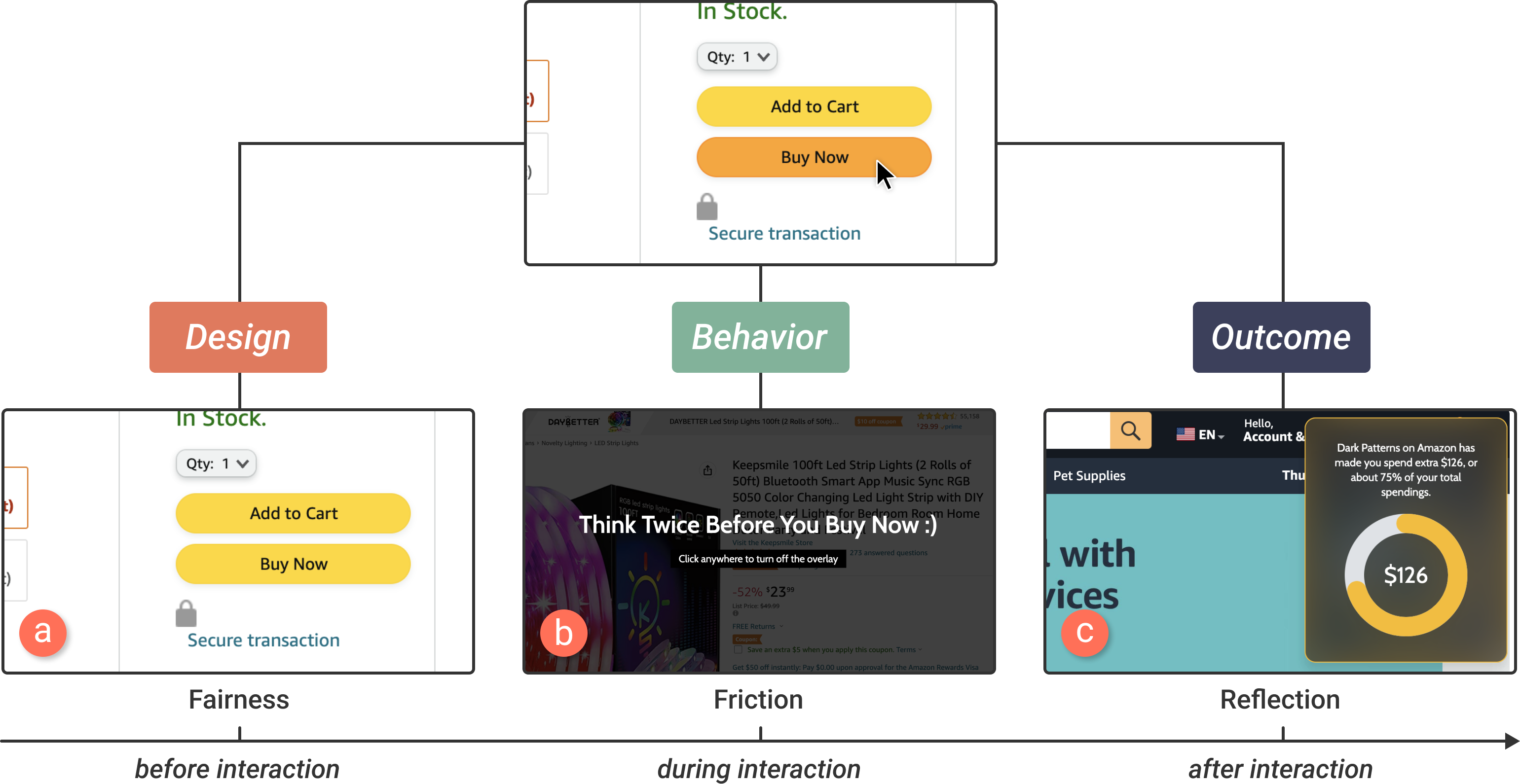}
\caption{The Design-Behavior-Outcome framework illustrated with examples of UI enhancements.} \label{dbo_framework}
\end{figure}

\chaohighlight{
\paragraph{\textbf{Design-Behavior-Outcome framework}}
\label{sec:dbo_framework}
Previous studies have explored various intervention techniques to counteract the influence of dark patterns on user behaviors~\cite{bongard-blanchy_i_2021}. These techniques can act in different phases of user interaction with dark patterns. \ylhighlight{Based on \textbf{\textbf{\hyperref[DI4]{DI4}}} from our co-design workshops, we propose a Design-Behavior-Outcome framework to situate different intervention techniques in their corresponding interaction stages to design appropriate UI enhancements for end users (Fig.~\ref{dbo_framework}).}
\ylhighlight{This framework can inspire future intervention technique designs \textit{before, during, and after} a user's interaction with dark patterns.}
}

\begin{itemize}
    \item \textbf{Design}: \textit{Design} interventions change the visual style of the interfaces or the information displayed \textit{before} the user interacts with the dark pattern.
    \ylhighlight{\item \textbf{Behavior}: \textit{Behavior} interventions directly guide, modify, or constrain users' behavior \textit{during} the interaction flow with the dark pattern.}
    \ylhighlight{\item \textbf{Outcome}: \textit{Outcome} interventions explain the possible consequences of provoking user reflection \textit{after} interacting with the dark pattern. }
\end{itemize}

\chaohighlight{
Specifically, we first selected 7 intervention strategies from previous literature on dark patterns~\cite{bongard-blanchy_i_2021} and technology-mediated nudging~\cite{caraban_23_2019}, including \textit{detection}~\cite{m_bhoot_towards_2020,mathur_2019_dark}, \textit{warning consequences}~\cite{lyngs_i_2020}, \textit{hiding}~\cite{lyngs_i_2020}, \textit{disabling}~\cite{kollnig_i_2021}, \textit{counterfactual thinking}~\cite{boush_deception_2016}, \textit{friction}~\cite{moser_impulse_2019,gould_special_2021}, and \textit{reflection}~\cite{wang_field_2014}. 
\textit{Detection} and \textit{warning consequences} were excluded because we have implemented them in the \textit{awareness} features of \textsc{Dark Pita}. 
As a complement, we derived 3 participant-designed intervention strategies from our co-design workshops, that is, \textit{fairness}, \textit{information disclosure}, and \textit{action guide} (\textbf{DI4}). 
Overall, we selected 8 techniques and situated them in our proposed Design-Behavior-Outcome framework (Table~\ref{intervention}). \looseness=-1 
}

\paragraph{\textbf{Intervention design}}
\label{sec:intervention_selection}

For each instance, we designed 1--4 UI enhancements (31 in total, shown in Table~\ref{intervention}).
\ylhighlight{In Fig.~\ref{dbo_framework}, we illustrate how we designed 3 enhancements against Amazon's ``Buy Now''. 
By using a more prominent color and hiding the cart item subtotal from users, the ``Buy Now'' button potentially promotes users' impulsive buying behavior~\cite{sin2022dark}. 
To target the visual \textit{design} of this dark pattern before interaction, we designed an enhancement to change the ``Buy Now'' button's color to the same as the alternative option (the \textit{fairness} strategy, Fig.~\ref{dbo_framework}a). 
During interaction, the enhancement using \textit{friction} opens a popup alert when the user hovers over ``Buy Now'' (Fig.~\ref{dbo_framework}b). 
Another alternative enhancement \textit{after} interaction, utilizing the \textit{reflection} strategy, provides a summary on how much dark patterns on Amazon have potentially led to unintentional shopping (Fig.~\ref{dbo_framework}c). A similar suite of UI enhancements can be used for similar dark patterns such as Subscribe \& Save or Buy Now \& Pay Later.}
Details of UI enhancements can be found in the Appendix~\ref{appendix:ui_enhancement}.

\subsubsection{Implementation}
\label{sec:implementation}
The \textsc{Dark Pita} probe is a Chrome browser extension. It was implemented using the Vue\footnote{https://vuejs.org/} framework and the Chrome extension API\footnote{https://developer.chrome.com/docs/extensions/reference/}. For dark pattern detection, it uses a rule-based method by matching the attributes of HTML elements (e.g., \texttt{id}, \texttt{aria-label}, \texttt{data-uia}) using manually authored regular expressions or finding unique section title strings in the \texttt{innerHTML} attribute of HTML elements. This approach allowed \textsc{Dark Pita} to detect all instances of the same dark patterns regardless of the content of the page. For UI enhancements, we implemented them by programming the browser extension to automatically add, remove, and/or modify the corresponding DOM elements when the target website is loaded. Once the user selects a UI enhancement for a detected dark pattern, \textsc{Dark Pita} immediately executes the corresponding script to modify the UI of the website. Additionally, we use the Chrome storage API\footnote{https://developer.chrome.com/docs/extensions/reference/storage/} to store user configurations of UI enhancements. Every time a user opens a new instance of the target web page, the extension automatically retrieves the user's saved configurations and applies their previous UI enhancement setup. \looseness=-1









\subsection{Study Participants}
After implementing the probe, we recruited participants through online advertising and word-of-mouth. \ylhighlight{None of the previous co-design workshop participants were included in order to mitigate the geographic biases in in-person activity recruitment. We conducted purposive sampling to ensure the diversity of participant demographics (e.g., age, gender), technology literacy, occupation, Chrome usage, and familiarity with dark patterns. In total, 17 participants were recruited, and 15 (PB1--PB15; 9 males, 6 females) completed the study. One of the two dropouts did not find time to use our probe after the entry interview. We were unable to get in touch with the other dropout after the entry interview.} Therefore, we excluded the data from these two participants from the analysis. Detailed demographic information about the 15 participants can be found in Appendix~\ref{appendix:deployment_demographics}.

\subsection{Study Protocol}
\ylhighlight{The two-week technology probe deployment study\footnote{The study protocol has been approved by the IRB at our institution.} began with a semi-structured entry interview for study introduction and \textsc{Dark Pita} installation. We also discussed participants' perceptions, attitudes, and behaviors toward dark patterns in past experiences to help them better understand the concept and contextualize our study. During the study, participants were asked to use \textsc{Dark Pita} on routine websites according to their everyday habits. We obtained consent from all participants prior to the study. We understand that experiences with dark patterns can be annoying, so we ensured that each participant understood their right to leave the study at any point if they wish.}

During the two weeks of use, we encouraged participants to log their thoughts and feelings as they interacted \textsc{Dark Pita}. We provided several heuristic questions to guide participants. For example, they can talk about a specific dark pattern, answering questions such as ``How does the dark pattern affect your online experience?'' and ``Are there any other intervention designs that you can think of?'', or they can report any reflections or issues they encounter when using the probe. 
\ylhighlight{As an incentive, we offered a \$2 (USD) reward for each submitted note (up to \$16) and encouraged each participant to submit at least one note every two days. Our probe also recorded detailed log data as described in Section~\ref{sec:action_diary_logging}.}

\ylhighlight{In the middle of the study, a semi-structured 30-minute check-in interview was conducted. The main goals of this interview were to (1) clarify users' questions and concerns from the first week of use; (2) understand their usage behavior with our probe; and (3) remind them of the study procedures, including probe usages and diary note submissions.} \ylhighlight{At the end of the study, we conducted a semi-structured one-hour exit interview with each participant to (1) collect information about participants' experiences with \textsc{Dark Pita} and explore the rationale behind interesting user behaviors or diary notes; (2) understand changes in user behaviors, perceptions, and attitudes toward dark patterns; and (3) find out users' feedback for interventions in \textsc{Dark Pita}}. 

All interviews were conducted online through Zoom and recorded with the consent of the participants. Each participant was compensated \$100 for using our probe and joining the three interviews. Additional compensation for sending diary notes is also provided. Appendix~\ref{appendix:deploymen_protocol} contains the protocols of our entry, middle, and exit interviews.


\subsection{Data Analysis Methods}
\ylhighlight{Three researchers conducted open coding and thematic analysis~\cite{braun2006using,lazar2017research} of interview transcriptions and diary notes.} Throughout the analysis, they went through three rounds of labeling and engaged in constant discussions to identify code, merge themes, and resolve conflicts. Similar to the co-design workshop analysis process (Section~\ref{workshop_findings}), our goal was to discover emergent themes, and the analysis process was discussion-based, so the inter-rater reliability is not necessary~\cite{mcdonald2019reliability}. \finalhighlight{The themes we derived during our data analysis were included in Appendix \ref{sec:workshop_analysis_themes}.} Our analysis paid attention to users' reactions and changes throughout the 2-week period, with the three research goals in mind (Section~\ref{sec:research_goals}). 

\section{Results}

\begin{figure}
\centering
\includegraphics[width=\linewidth]{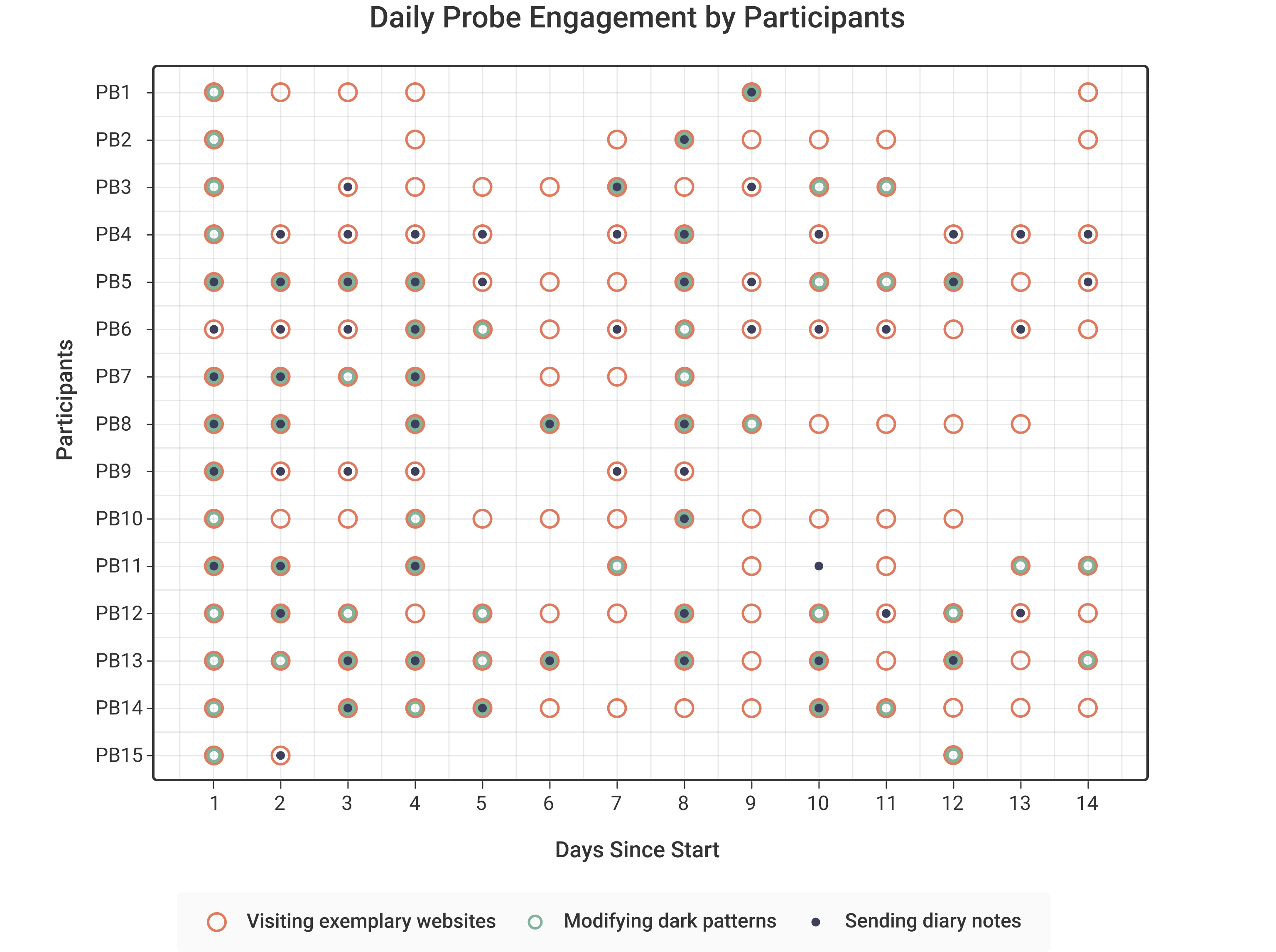}
\caption{The daily probe engagement of our participants. Red and green circles denote participants who visited sites containing dark pattern instances or modified dark pattern instances with UI enhancements at least once on that day, respectively. The blue dots denote participants who sent diary notes at least once on that day. The touchpoints (mid-study check-in interviews) were on Day 8.}
\label{daily_logs}
\end{figure}



\subsection{Probe Engagement}
The engagement validates the feasibility of our probe (\textit{engineering goal}). In total, 48,368 (\emph{mean}=3224.53, \emph{std}=2836.14, \emph{min}=352, \emph{max}=9112) action logging entries and 115 (\emph{mean}=7.67, \emph{std}=4.84, \emph{min}=1, \emph{max}=17) diary note entries were created. The participants visited a total of 13,611 (\emph{mean}=907.4, \emph{std}=819.88, \emph{min}=58, \emph{max}=3220) distinct web pages on their browsers instrumented with \textsc{Dark Pita}, where our probe was triggered 4,834 (\emph{mean}=322.27, \emph{std}=316.65, \emph{min}=28, \emph{max}=1188) times. \finalhighlight{10 participants (66.7\%) visited all three types of online services, 4 participants (26.7\%) visited video streaming platforms and social media platforms, and 1 participant only visited video streaming platforms over the course of the study.} The participants set up UI enhancements for dark pattern instances 280 (\emph{mean}=18.67, \emph{std}=8.93, \emph{min}=1, \emph{max}=34) times. During the two weeks, these UI enhancements were triggered 14,355 times (\emph{mean}=957, \emph{std}=1178.48, \emph{min}=2, \emph{max}=4621) in total. For each UI enhancement, the average number of times that it was triggered was 463.06. 2 UI enhancements (i.e., \textit{counterfactual thinking} (Appendix~\ref{appendix:ui_enhancement}) for the discount price on Amazon and \textit{reflection} for the remaining time on Netflix) were not successfully used due to technical problems caused by website updates on Amazon and Netflix. Fig.~\ref{daily_logs} shows the daily engagement of participants with our probe. Overall, the logs indicate that most participants were actively engaged with our probe. They visited example websites that contain instances of dark patterns, modified the interfaces of these websites with UI enhancements to mitigate the impact of dark patterns, and submitted diary entries. 









\subsection{Key Insights}

Our findings from the entry, mid-study, and exit interviews with 15 participants, together with their 115 diary entries, provide key insights (KIs) for understanding the reactions and needs of end users on \emph{awareness} of and \emph{action} for dark patterns (\emph{social science goal}) and demonstrate the usefulness of our end-user-empowerment intervention approach (\emph{engineering goal}).


\paragraph{\textbf{KI1: Providing information about specific instances of dark patterns allows users to gain transferable knowledge}} 

\label{KI1}
\finalhighlight{The users found the information presented about dark patterns on the \emph{awareness} panel to be educational.} PB12 found that attribute tags are ``great information'' and helped him better understand the design rationale of designers hidden behind dark patterns. PB16 added that \textsc{Dark Pita} made him know \ylhighlight{\textit{``what are the certain things that kind of triggered me into going down a hole''}}. Equipped with such knowledge, the participants developed new perspectives on online services. PB3, PB6, PB6, and PB12 mentioned \textsc{Dark Pita} helped them more explicitly see a large number of dark pattern instances on Twitter, which reduced their level of trust in the platform. PB6 became more critical of disguised ads on some platforms, and PB10 behaved more cautiously to avoid dark patterns. By seeing the dislike count on YouTube again (enabled by a UI enhancement of \textsc{Dark Pita}) (Appendix~\ref{appendix:ui_enhancement}), PB7 was able to gain a more comprehensive view of the videos they watch. \ylhighlight{This finding showed that improved awareness through information disclosure on dark patterns can change users' perception of digital platforms, extending previous survey results of end users' mistrust~\cite{gray_enduser_2021}.}

Importantly, users were able to transfer their newly learned knowledge of dark patterns to other platforms \textsc{Dark Pita} did not yet support. For example, PB12 would investigate \ylhighlight{\textit{``how the similar designs (of dark patterns) might be applied to other interfaces that I use''}. PB2, PB5, PB9, and PB12 started to think about dark patterns in mobile apps and desired to use \textsc{Dark Pita} on phones. The probe even inspired PB9 to ponder ``textual'' dark patterns (e.g., misleading and manipulative language on interfaces). To summarize, these findings demonstrated that providing information about dark patterns can not only raise users' awareness, but also inspire them to transfer the learned knowledge to dark patterns on other platforms.}

\paragraph{\textbf{KI2: The capability to modify existing interfaces boosts the user perception of empowerment and autonomy}} 
\label{KI2}

\finalhighlight{In our deployment study, 7 participants (46.7\%) explicitly mentioned the feeling of empowerment of being able to change interfaces, as they were no longer just passive consumers of decisions made by designers.} PB3 mentioned that \textit{``the most empowering was being able to highlight algorithm-recommended content on Twitter... it provided a level of consciousness (during browsing)''.} Previous work on dark patterns in HCI and CSCW has primarily regarded end users as passive receivers of these deceptive practices~\cite{gray_enduser_2021, bongard-blanchy_i_2021}. These findings show the benefit of our end-user-empowerment approach, treating users as engaged actors in mitigating dark patterns.

Notably, participants emphasized the importance of having this support from a third-party tool. \ylhighlight{PB5, PB6, and PB12 mentioned that some platforms also allow users to make interface changes; for example, on Facebook, users can remove disguised ads and select why they are not interested. While this allows the user to hide the ad, it serves the company's business interest; some users realized this and chose not to use it. PB12 mentioned that users and companies have contradictory goals: users want to see fewer targeted ads, while the company wants to make ads more personalized to generate more revenue. On the contrary, a third-party tool like \textsc{Dark Pita} presents no conflict of interest with users, leading to enhanced user trust.}\looseness=-1

\paragraph{\textbf{KI3: The dynamic goals and usage contexts of users when using online services determine their desired UI enhancements.}} 
\label{KI3}


\ylhighlight{Users have diverse goals online: for example, PB10 wants to reduce their time on Twitter, while PB3 and PB5 do not mind spending more time on Twitter browsing. This is in line with our workshop finding on users' perceptions of dark patterns (\textbf{\hyperref[WF2]{WF2}}). In the deployment study, these differences in goals shaped users' choices of UI enhancements.}

\ylhighlight{Even a single user's goals can change with different usage scenarios, which in turn modifies their choice of UI enhancements. For instance, PB2, PB3, and PB14 separately reported that they did not mind the ``video autoplay on hover'' feature on YouTube homepage but disabled it on individual video-watching pages. This was because the feature was helpful for previewing content on the homepage, but distracting and time-wasting when watching individual videos. PB14 also turns the focus mode on YouTube on and off between \textit{``when I want to focus... and when I just want to have fun or just relax... I think they get changed based on what I wanted to do at the time.''}} \looseness=-1

\ylhighlight{Users' goals often reflect and shape their personal relationship with a service platform. During our study, the ability to modify dark patterns reminded users of their long-term goals against impulsive behaviors.} In the exit interview, PB12 mentioned that seeing so many dark patterns explicitly marked on Twitter made him want to use the platform less. They described the feeling as ``someone nagging you should stop doing this''. \ylhighlight{Although sometimes annoying, they still found it helpful for their long-term goal of reducing Twitter usage.} This is also related to studies on self-control such as ~\cite{baumeister2002yielding, hofmann2009impulse}.


\ylhighlight{Previous discussions mostly view the ``darkness'' level of deceptive patterns as an objective attribute~\cite{gray_dark_2018}; however, this insight extends such narrative by showing the \textit{personal} and \textit{dynamic} nature of such ``darkness''.} End users' goals and usage contexts are dynamic and individualized. Therefore, a one-size-fits-all approach or a designer- or policymaker-initiated approach cannot fully accommodate them. By offering users the choice of multiple UI enhancements, our approach enables them to customize their intervention for a dark pattern based on individual contextual preferences. \looseness=-1
    
\subsection{Design Implications}
\label{sec:deployment_design_implications}

Our findings offer design implications (DI) for future techniques, strategies, and interfaces of end-user-empowerment interventions for dark patterns (\emph{design goal}).

\paragraph{\textbf{DI5: Design non-intrusive and less-interrupting UI enhancements}}
\label{DF5}

Users prefer non-intrusive and less-interrupting UI enhancements. Visually non-intrusive UI enhancements that do not interfere with users’ normal browsing experiences were greatly appreciated during our study. For example, PB15 shared that the highlighted disguised ads with thick red borders (Appendix~\ref{appendix:ui_enhancement}) became annoying, so he chose to directly hide the ads (Appendix~\ref{appendix:ui_enhancement}) instead. On the contrary, blocking the previews of recommended videos (Appendix~\ref{appendix:ui_enhancement}) is a ``gentle'' method to remove distracting content while avoiding users' fear of missing out on information~\cite{lyngs_i_2020}.


\paragraph{\textbf{DI6: Provide fine-grained control over modification of dark patterns}}
\label{DF6}

Participants expected future interventions to give them more fine-grained control over dark patterns. For example, PB13 wanted to ``control the quantity'' of promoted content on Twitter and leave approximately one-third on their feed instead of removing all of them. Similarly, PB2 envisioned a filter to ``control content'' based on personal interests, i.e., automatically identifying information potentially beneficial to her and removing the rest.

\paragraph{\textbf{DI7: Improve transparency and provide global control for the UI enhancements.}}
\label{DI7}

In addition to the ``design transparency'' of dark patterns, our participants also wanted transparency in the design of UI enhancements. They expected to see the intentions and mechanisms of these UI enhancements, so they could understand how these work against dark patterns and \yuwen{select the ones that fit their needs}. 
For example, PB5 and PB15 suggested providing explanations on how \textsc{Dark Pita} calculates the time or money spent on reflection (Appendix~\ref{appendix:ui_enhancement}). PB15 also mentioned that they wanted a global control panel and dashboard for all active UI enhancements so that they could quickly get explanations of them, view their status, and change their configurations.

\paragraph{\textbf{DI8: Contemplate with the boundary between UI enhancements and dark patterns.}}
\label{DF8}

UI enhancements usually involve a certain degree of persuasive design or nudge techniques themselves, which is similar to dark patterns. \ylhighlight{Based on what we have learned from our studies, it is important to align the goals of the user and the goals of the intervention tools (\textbf{\hyperref[KI2]{KI2}} and \textbf{\hyperref[KI3]{KI3}}) to protect the welfare of users (e.g. privacy data)~\cite{mathur_what_2021}. In addition, according to Hansen and Jespersen's framework~\cite{hansen_nudge_2013}, the dividing line between manipulative and beneficial nudges is transparency (i.e., if the user can perceive the intentions and means behind the nudge) (\textbf{\hyperref[DI7]{DI7}}). Such potential alignments should be clearly explained to users to help them make informed decisions about adopting UI enhancements. With carefully set boundary between UI enhancements and dark patterns through goal alignment and transparency, we can meaningfully prevent further manipulation against end users' will. }







\section{Scaling Up: A Research Agenda}
\finalhighlight{Our findings highlight the potential of an end-user-empowerment approach in helping users understand, intervene, and make informed decisions about dark patterns based on their specific needs, goals, and context. By disclosing information and enabling actions against dark patterns, users gained an increased sense of autonomy in online experiences (KI1). Our proposed design-behavior-outcome framework maps out design opportunities for future dark pattern interventions. Through our 2-phase studies, we revealed that end users desire non-intrusive (DI6), personalized (KI2), and dynamic (KI3). Future research needs to carefully consider the distinct preferences of the target user groups and the context of use of digital services (DI8).}

Although our two-week technology probe study illustrated the usefulness and technical feasibility of this approach, scalability remains a challenge. \ylhighlight{Our manual process for 31 UI enhancements on 5 websites is adequate for a small-scale probe, but cannot practically cover a significant selection of millions of dark patterns for real-world impact~\cite{digeronimo_dark_2020}. This scalability challenge is two-fold: on one hand, new dark patterns emerge quickly, and it requires considerable maintenance overhead to keep up-to-date for all sites; on the other, designing multiple user-desired UI enhancements for each dark pattern requires significant effort (illustrated in Section~\ref{sec:deployment_design_implications}).}

To help a large audience effectively mitigate dark patterns' impacts in real-world settings, a new approach is needed to scale up this effort. Here, we propose several possible future directions and discuss relevant efforts in adjacent research areas. 

\subsection{A Crowd-Sourced Collective Intelligence Approach}
\label{sec:crowd_sourced_approach}

\finalhighlight{A crowd-sourced, collective intelligence approach can be an effective way to tackle scalability issues~\cite{allen2018design}. This could involve community contributions for identifying dark patterns, their impacts, and potential UI enhancements.} The aggregated data can expand the capabilities of tools like \textsc{Dark Pita}, as well as provide training data for future machine learning (ML) models that detect dark patterns, predict user behaviors, and generate interventions, as we will discuss in Section~\ref{sec:ml-approach}. This crowd-sourced approach can involve multiple stakeholders:

\begin{enumerate}
    
    \item \ylhighlight{\textit{End users} can identify dark patterns in their daily experiences, report their behaviors in response to dark patterns, and express their desired changes.} Public release and wide adoption of tools such as \textsc{Dark Pita} can provide a platform for soliciting such information.

    \item \textit{Designers} who are \yuwen{motivated to contribute can provide meta-information of their design intentions behind UX designs} along with these features. 


    \item \textit{Third-party developers} can develop new detectors and UI enhancements for instances of dark patterns and contribute them to a unified repository or ``community wiki'' for public use.\looseness=-1
    
\end{enumerate}

\subsection{A Citizen Science Approach}
\label{sec:citizen_science_approach}

\finalhighlight{A citizen science approach~\cite{bonney2014next} can enhance transparency by gathering data on the design processes that result in dark patterns.
UX practitioners commonly use A/B tests~\cite{siroker2015b} to examine the effect of dark patterns (e.g., discouraging subscription cancellation). This approach seeks to involve the public in contributing the hypotheses, protocols, and outcomes of these A/B tests to reveal the hidden design intents behind dark patterns. This approach can improve design transparency, similar to how pre-registration of experiments and data transparency contribute to the open science movement~\cite{nosek2015promoting,nosek2018preregistration}. Meanwhile, guardrails must be put in place once this information becomes public: if a study shows the business benefits of including dark patterns, they should not be blindly misused by other companies and designers.}

\ylhighlight{Implementing the citizen science approach would involve (1) a consistent format to report the relevant experiment information; (2) a community repository where the information can be aggregated, organized, and shared; and (3) optionally, a platform or a set of tools for conducting UX experiments that make it easier to share the experiment information. The citizen science approach can also engage multiple stakeholders:}

\begin{enumerate}
    
\item \textit{UX practitioners} who are ethically-minded can participate by sharing the hypotheses, protocols, and outcomes of these experiments.

\item \textit{Third-party researchers} can audit the shared results by replicating experiments using the information provided.

\item \ylhighlight{\textit{Policy makers and community activists} can mandate or advocate for companies' adoption of this approach, which we will discuss in Section~\ref{sec:design_policy_effort_coordinate}.}

\end{enumerate}

\subsection{A Machine Learning Approach}
\label{sec:ml-approach}
\finalhighlight{With the latest advances in the computational UI understanding~\cite{jiang2022computational,deka_rico:_2017,huang_swire_2019,li_screen2vec:_2021,wen2023empowering} and user behavior modeling~\cite{langley1999user,zhou_atrank_2018,yin_temporal_2014,zheran2018reinforcement, li2018kite}, machine learning (ML) techniques to model UX dark patterns show great promise in scaling up the effort in dark pattern intervention. \ylhighlight{Early explorations in this area such as AidUI~\cite{mansur2023aidui} have demonstrated the impressive performance of ML models in automated dark pattern recognition.} Previous efforts to automate the detection of dark patterns~\cite{donnellyBePatternWorld2022a} in consent banners with ML~\cite{soe2022automated} and reverse engineering~\cite{nouwensConsentOMaticAutomaticallyAnswering2022a} have also shown promising results in this area. Specifically, ML models have the potential to (1) identify instances of dark patterns and categorize them, (2) predict the consequences/user behaviors under the influence of these dark patterns; and (3) generate UI enhancements with different dark pattern intervention strategies.} 

However, the lack of large datasets on dark pattern designs, user behavior under the influence of dark patterns, and users' preferred action against dark patterns are major barriers to the ML approach. \ylhighlight{In addition to the \textsc{ContextDP} dataset proposed in AidUI~\cite{mansur2023aidui}, such datasets may also be constructed from existing curated lists of dark patterns, e.g., Deceptive Design Hall of Shame\footnote{https://www.deceptive.design/hall-of-shame/all}, from website crawling, or from data collected using our proposed crowd-sourced (Section~\ref{sec:crowd_sourced_approach}) and citizen science (Section~\ref{sec:citizen_science_approach}) approaches}. 

\finalhighlight{It is vital to acknowledge the potential abuse of ML techniques in efficiently creating dark patterns in interfaces. We implore researchers and practitioners using ML in design to be vigilant, adhere to previous empirical results of dark patterns~\cite{gray_what_2020, gray_enduser_2021, mathur_what_2021}, and ensure their designs align with user goals, to mitigate misuse. The development of such ML models must consider the intended users' goals and ethical values to guarantee their widespread utility and benefit~\cite{shneiderman2022human}.}

\subsection{Coordinating Efforts with Design Ethics Advocacy and Policy Making}
\label{sec:design_policy_effort_coordinate}
\ylhighlight{Our end-user-empowerment approach complements designer-centered ethical practices and policy-focused regulation against dark patterns.} We propose several directions that coordinate these efforts to scale up the impact.

\finalhighlight{
Designer-focused efforts can strengthen the proposed crowd-sourced (Section~\ref{sec:crowd_sourced_approach}) and citizen science (Section~\ref{sec:citizen_science_approach}) approaches. As discussed, designers play an important role in both approaches; designer education and advocacy are crucial to boosting their participation and engagement~\cite{gray_dark_2018}.}

\finalhighlight{The citizen science method~\cite{bonney2014next} helps with a key issue in policy making: defining dark patterns for regulation is challenging as a comprehensive definition is currently lacking~\cite{gray_dark_2018, mathur_2019_dark}.  Citizen science promotes ``design transparency''~\cite{eiband2018bringing} in policy making, such as mandating study preregistration and sharing of A/B testing experiments. If a mandate is not yet practical, we can also take gradual steps, such as issuing ``design transparency'' or ``ethical design'' badges to companies or organizations that comply with the requirement.}



\subsection{The Power Imbalance between End Users and Designers}
Our end-user-empowerment approach has implications for addressing the power imbalance between end users and designers in interfaces. \ylhighlight{Today, designers usually have dictating power over interface design. Even if users can modify interfaces, the possible configurations are often pre-defined by designers. Our design probe \textsc{Dark Pita} and our end-user-empowerment approach attempt to shift this power imbalance through \emph{awareness} and \emph{action}.}

\ylhighlight{Together with advancements in ML for dark patterns (Section~\ref{sec:ml-approach}), new community-based approaches~\cite{kollnig_i_2021} will further empower end users against designers' ``interface dictatorship''. 
New communities such as Arc Boost Gallery\footnote{https://arc.net/boosts} provide great opportunities for future investigation. For example, CSCW academics can investigate the common community structures, dynamics, and member values to gain insights for sustaining web augmentation communities around dark patterns. Existing research on dark pattern Reddit communities~\cite{gray_what_2020} and CSCW research on online communities~\cite{dym2018online, fiesler2017growing} have built solid foundations for such explorations. Meanwhile, we hope our Design-Behavior-Outcome framework cast light on the design space for intervention techniques and could guide future community creators in coming up with useful solutions. }

To address this ``tug-of-war'' between designers and users, we can also achieve end-user empowerment by ``de-powering'' designers, a more radical and aggressive approach. \ylhighlight{Research on malleable interfaces~\cite{nichols_2009_description,nichols_2002_remote} has shown feasibility in generating UIs automatically based on the specifications of service functionalities, user preferences, and usage context.} In this way, the role of designers will be limited to describing the specifications of a system, with little power over the visual presentations of information and the interaction mechanisms. This would prevent the creation of many dark patterns in the first place.

\section{Limitations and Future Work}

This presented work has several limitations. First, given the diverse range of dark pattern taxonomies~\cite{mathur_what_2021}, it is difficult to comprehensively cover all types of dark patterns in one work. Our sampled dark pattern instances were limited to three genres of online services: online shopping, video streaming, and social media. However, dark patterns can exist in a wide variety of platforms or task domains. In addition, although we ensured diversity in the sample of dark pattern instances we used, our manual curation process allowed us to reach only a limited sample. These limitations are in line with our above-mentioned scalability challenge; we hope that future studies can scale up our efforts with the research agenda we have discussed.

Also, although our probe \textsc{Dark Pita} only supports web browsers on computers, dark patterns also exist in mobile applications~\cite{digeronimo_dark_2020,gunawan_comparative_2021}. In fact, many participants in our deployment study expressed their desire to use \textsc{Dark Pita} on smartphones: They feel more vulnerable and less alert to dark patterns on mobile devices, given the often casual usage context. Future work can explore the expansion of our end-user-empowerment approach to mobile platforms. Previous work such as~\cite{li_sugilite:_2017,li2019pumice} has shown the technical feasibility of a similar approach on Android with the Accessibility API, while the stricter developer permissions on iOS remain a challenge.

\finalhighlight{The use of in-person co-design workshops, while facilitating more effective and smooth interactions, could have biased our participant pool towards individuals close to the workshop locations. This geographically constrained recruitment approach might limit the diversity of experiences and perspectives contributing to our co-design process (reflected in \hyperref[appendix:workshop_demographics]{Appendix A.1}). To offset this potential bias, we conducted the probe study online, recruiting participants from a wider range of backgrounds (shown in \hyperref[appendix:deployment_demographics]{Appendix A.5}). 
In addition, in-person interaction with participants might make them hesitant to share negative opinions, which could introduce bias to our results. Future studies can improve by recruiting more geographically diverse participants to explore their perspectives based on different cultural and socio-economic backgrounds.}

\ylhighlight{The scale and primary qualitative nature of our two-phase study inherently limit the representativeness of our findings. Despite our best efforts to cultivate a diverse participant pool, our conclusions mainly reflect the perspectives of our specific sample group, failing to fully capture views from homogeneous groups (e.g., experts and non-experts). We recommend future research to consider larger-scale studies for a more comprehensive exploration of this topic.} \looseness=-1

Despite the promising qualitative results reported in the probe deployment study, the limited duration and scale of our technology probe study did not allow us to track long-term user behaviors to quantitatively examine the efficacy of our approach in behavioral changes. \ylhighlight{Consequently, the feedback we gathered could be affected by novelty effect~\cite{sung2009robots}}. We plan to further develop our probe into a fully functional system and conduct larger-scale longer-term field experiments to measure the impacts on user individual welfare (e.g., financial loss) and autonomy~\cite{mathur_what_2021}. Furthermore, to measure collective welfare~\cite{mathur_what_2021} and collect community intelligence, we also plan to release \textsc{Dark Pita} to the general public.










\section{Conclusion}
Through a series of co-design workshops and a 2-week deployment study of a technology probe, we proposed and tested an end-user-empowerment approach for dark pattern intervention. We discussed implications for the design of future interventions to support users' awareness and actions against undesired UX dark patterns. Our approach presents opportunities in coordinating with the ongoing efforts in addressing dark patterns from the perspectives of designers, educators, and policymakers. We laid out a research agenda to scale up this approach by utilizing developments in crowd-sourced collective intelligence, citizen science platforms, computational UI techniques, and user behavior modeling to guide future work in this domain.
\begin{acks}

We extend our sincere appreciation to our participants for their contributions to this project. We thank all anonymous reviewers for their feedback. This work is supported in part by the National Science Foundation (Grant No. CMMI-2326378).

\end{acks}

\balance
\bibliographystyle{ACM-Reference-Format}
\bibliography{references}


\begin{thebibliography}{136}


\ifx \showCODEN    \undefined \def \showCODEN     #1{\unskip}     \fi
\ifx \showDOI      \undefined \def \showDOI       #1{#1}\fi
\ifx \showISBNx    \undefined \def \showISBNx     #1{\unskip}     \fi
\ifx \showISBNxiii \undefined \def \showISBNxiii  #1{\unskip}     \fi
\ifx \showISSN     \undefined \def \showISSN      #1{\unskip}     \fi
\ifx \showLCCN     \undefined \def \showLCCN      #1{\unskip}     \fi
\ifx \shownote     \undefined \def \shownote      #1{#1}          \fi
\ifx \showarticletitle \undefined \def \showarticletitle #1{#1}   \fi
\ifx \showURL      \undefined \def \showURL       {\relax}        \fi
\providecommand\bibfield[2]{#2}
\providecommand\bibinfo[2]{#2}
\providecommand\natexlab[1]{#1}
\providecommand\showeprint[2][]{arXiv:#2}

\bibitem[Aagaard et~al\mbox{.}(2022)]%
        {aagaard_game_2022}
\bibfield{author}{\bibinfo{person}{Jacob Aagaard}, \bibinfo{person}{Miria Emma~Clausen Knudsen}, \bibinfo{person}{Per Bækgaard}, {and} \bibinfo{person}{Kevin Doherty}.} \bibinfo{year}{2022}\natexlab{}.
\newblock \showarticletitle{A {Game} of {Dark} {Patterns}: {Designing} {Healthy}, {Highly}-{Engaging} {Mobile} {Games}}. In \bibinfo{booktitle}{\emph{Extended {Abstracts} of the 2022 {CHI} {Conference} on {Human} {Factors} in {Computing} {Systems}}} \emph{(\bibinfo{series}{{CHI} {EA} '22})}. \bibinfo{publisher}{Association for Computing Machinery}, \bibinfo{address}{New York, NY, USA}, \bibinfo{pages}{1--8}.
\newblock
\showISBNx{978-1-4503-9156-6}
\urldef\tempurl%
\url{https://doi.org/10.1145/3491101.3519837}
\showDOI{\tempurl}


\bibitem[Ahuja and Kumar(2022)]%
        {ahujaConceptualizationsUserAutonomy2022}
\bibfield{author}{\bibinfo{person}{Sanju Ahuja} {and} \bibinfo{person}{Jyoti Kumar}.} \bibinfo{year}{2022}\natexlab{}.
\newblock \showarticletitle{Conceptualizations of User Autonomy within the Normative Evaluation of Dark Patterns}.
\newblock \bibinfo{journal}{\emph{Ethics and Information Technology}} \bibinfo{volume}{24}, \bibinfo{number}{4} (\bibinfo{date}{Dec.} \bibinfo{year}{2022}), \bibinfo{pages}{52}.
\newblock
\showISSN{1572-8439}
\urldef\tempurl%
\url{https://doi.org/10.1007/s10676-022-09672-9}
\showDOI{\tempurl}


\bibitem[Allen et~al\mbox{.}(2018)]%
        {allen2018design}
\bibfield{author}{\bibinfo{person}{BJ Allen}, \bibinfo{person}{Deepa Chandrasekaran}, {and} \bibinfo{person}{Suman Basuroy}.} \bibinfo{year}{2018}\natexlab{}.
\newblock \showarticletitle{Design crowdsourcing: The impact on new product performance of sourcing design solutions from the “crowd”}.
\newblock \bibinfo{journal}{\emph{Journal of Marketing}} \bibinfo{volume}{82}, \bibinfo{number}{2} (\bibinfo{year}{2018}), \bibinfo{pages}{106--123}.
\newblock


\bibitem[Alves et~al\mbox{.}(2023)]%
        {alves2023gitui}
\bibfield{author}{\bibinfo{person}{S{\'e}rgio Alves}, \bibinfo{person}{Ricardo Costa}, \bibinfo{person}{Kyle Montague}, {and} \bibinfo{person}{Tiago Guerreiro}.} \bibinfo{year}{2023}\natexlab{}.
\newblock \showarticletitle{GitUI: A Community-Based Platform to Democratize User Interfaces}. In \bibinfo{booktitle}{\emph{Extended Abstracts of the 2023 CHI Conference on Human Factors in Computing Systems}}. \bibinfo{publisher}{Association for Computing Machinery}, \bibinfo{address}{New York, NY, USA}, \bibinfo{pages}{1--6}.
\newblock


\bibitem[Amazon(2022)]%
        {amazon_pricing_explanation}
\bibfield{author}{\bibinfo{person}{Amazon}.} \bibinfo{year}{2022}\natexlab{}.
\newblock \bibinfo{title}{Items on Amazon may display a List Price, Was Price, or other strike-through pricing or saving information on the product detail page.}
\newblock
\newblock
\urldef\tempurl%
\url{https://www.amazon.com/gp/help/customer/display.html?nodeId=GQ6B6RH72AX8D2TD&ref_=dp_hp}
\showURL{%
\tempurl}


\bibitem[Baumeister(2002)]%
        {baumeister2002yielding}
\bibfield{author}{\bibinfo{person}{Roy~F Baumeister}.} \bibinfo{year}{2002}\natexlab{}.
\newblock \showarticletitle{Yielding to temptation: Self-control failure, impulsive purchasing, and consumer behavior}.
\newblock \bibinfo{journal}{\emph{Journal of consumer Research}} \bibinfo{volume}{28}, \bibinfo{number}{4} (\bibinfo{year}{2002}), \bibinfo{pages}{670--676}.
\newblock


\bibitem[Board(2020)]%
        {european_data_protection_board_guidelines_2020}
\bibfield{author}{\bibinfo{person}{European Data~Protection Board}.} \bibinfo{year}{2020}\natexlab{}.
\newblock \bibinfo{title}{Guidelines 05/2020 on consent under {Regulation} 2016/679}.
\newblock
\newblock
\urldef\tempurl%
\url{https://edpb.europa.eu/our-work-tools/our-documents/guidelines/guidelines-052020-consent-under-regulation-2016679_en}
\showURL{%
\tempurl}


\bibitem[Board(2022)]%
        {europeandataprotectionboardGuidelines2022Dark2022}
\bibfield{author}{\bibinfo{person}{European Data~Protection Board}.} \bibinfo{year}{2022}\natexlab{}.
\newblock \bibinfo{booktitle}{\emph{Guidelines 3/2022 on Dark Patterns in Social Media Platform Interfaces: How to Recognise and Avoid Them}}.
\newblock \bibinfo{type}{{T}echnical {R}eport}. \bibinfo{institution}{European Data Protection Board}, \bibinfo{address}{Brussels, Belgique}.
\newblock


\bibitem[Bodker(1999)]%
        {bodker1999scenarios}
\bibfield{author}{\bibinfo{person}{Susanne Bodker}.} \bibinfo{year}{1999}\natexlab{}.
\newblock \showarticletitle{Scenarios in user-centred design-setting the stage for reflection and action}. In \bibinfo{booktitle}{\emph{Proceedings of the 32nd Annual Hawaii International Conference on Systems Sciences. 1999. HICSS-32. Abstracts and CD-ROM of Full Papers}}. \bibinfo{publisher}{IEEE}, \bibinfo{address}{Maui, HI, USA}, \bibinfo{pages}{11--pp}.
\newblock


\bibitem[Bolin et~al\mbox{.}(2005)]%
        {bolin_automation_2005}
\bibfield{author}{\bibinfo{person}{Michael Bolin}, \bibinfo{person}{Matthew Webber}, \bibinfo{person}{Philip Rha}, \bibinfo{person}{Tom Wilson}, {and} \bibinfo{person}{Robert~C. Miller}.} \bibinfo{year}{2005}\natexlab{}.
\newblock \showarticletitle{Automation and customization of rendered web pages}. In \bibinfo{booktitle}{\emph{Proceedings of the 18th annual {ACM} symposium on {User} interface software and technology}} \emph{(\bibinfo{series}{{UIST} '05})}. \bibinfo{publisher}{Association for Computing Machinery}, \bibinfo{address}{New York, NY, USA}, \bibinfo{pages}{163--172}.
\newblock
\showISBNx{978-1-59593-271-6}
\urldef\tempurl%
\url{https://doi.org/10.1145/1095034.1095062}
\showDOI{\tempurl}


\bibitem[Bongard-Blanchy et~al\mbox{.}(2021)]%
        {bongard-blanchy_i_2021}
\bibfield{author}{\bibinfo{person}{Kerstin Bongard-Blanchy}, \bibinfo{person}{Arianna Rossi}, \bibinfo{person}{Salvador Rivas}, \bibinfo{person}{Sophie Doublet}, \bibinfo{person}{Vincent Koenig}, {and} \bibinfo{person}{Gabriele Lenzini}.} \bibinfo{year}{2021}\natexlab{}.
\newblock \showarticletitle{”{I} am {Definitely} {Manipulated}, {Even} {When} {I} am {Aware} of it. {It}’s {Ridiculous}!” - {Dark} {Patterns} from the {End}-{User} {Perspective}}. In \bibinfo{booktitle}{\emph{Designing {Interactive} {Systems} {Conference} 2021}} \emph{(\bibinfo{series}{{DIS} '21})}. \bibinfo{publisher}{Association for Computing Machinery}, \bibinfo{address}{New York, NY, USA}, \bibinfo{pages}{763--776}.
\newblock
\showISBNx{978-1-4503-8476-6}
\urldef\tempurl%
\url{https://doi.org/10.1145/3461778.3462086}
\showDOI{\tempurl}


\bibitem[Bonney et~al\mbox{.}(2014)]%
        {bonney2014next}
\bibfield{author}{\bibinfo{person}{Rick Bonney}, \bibinfo{person}{Jennifer~L Shirk}, \bibinfo{person}{Tina~B Phillips}, \bibinfo{person}{Andrea Wiggins}, \bibinfo{person}{Heidi~L Ballard}, \bibinfo{person}{Abraham~J Miller-Rushing}, {and} \bibinfo{person}{Julia~K Parrish}.} \bibinfo{year}{2014}\natexlab{}.
\newblock \showarticletitle{Next steps for citizen science}.
\newblock \bibinfo{journal}{\emph{Science}} \bibinfo{volume}{343}, \bibinfo{number}{6178} (\bibinfo{year}{2014}), \bibinfo{pages}{1436--1437}.
\newblock


\bibitem[Boush et~al\mbox{.}(2016)]%
        {boush_deception_2016}
\bibfield{author}{\bibinfo{person}{David~M. Boush}, \bibinfo{person}{Marian Friestad}, {and} \bibinfo{person}{Peter Wright}.} \bibinfo{year}{2016}\natexlab{}.
\newblock \bibinfo{booktitle}{\emph{Deception {In} {The} {Marketplace}: {The} {Psychology} of {Deceptive} {Persuasion} and {Consumer} {Self}-{Protection}}}.
\newblock \bibinfo{publisher}{Routledge}, \bibinfo{address}{New York}.
\newblock
\showISBNx{978-0-203-80552-7}
\urldef\tempurl%
\url{https://doi.org/10.4324/9780203805527}
\showDOI{\tempurl}


\bibitem[Braun and Clarke(2006)]%
        {braun2006using}
\bibfield{author}{\bibinfo{person}{Virginia Braun} {and} \bibinfo{person}{Victoria Clarke}.} \bibinfo{year}{2006}\natexlab{}.
\newblock \showarticletitle{Using thematic analysis in psychology}.
\newblock \bibinfo{journal}{\emph{Qualitative research in psychology}} \bibinfo{volume}{3}, \bibinfo{number}{2} (\bibinfo{year}{2006}), \bibinfo{pages}{77--101}.
\newblock


\bibitem[Brignull(2010)]%
        {brignull_deceptive_nodate}
\bibfield{author}{\bibinfo{person}{Harry Brignull}.} \bibinfo{year}{2010}\natexlab{}.
\newblock \bibinfo{title}{Deceptive design - {Types} of deceptive design}.
\newblock
\newblock
\urldef\tempurl%
\url{https://www.deceptive.design/types}
\showURL{%
\tempurl}


\bibitem[Bösch et~al\mbox{.}(2016)]%
        {bosch_tales_2016}
\bibfield{author}{\bibinfo{person}{Christoph Bösch}, \bibinfo{person}{Benjamin Erb}, \bibinfo{person}{Frank Kargl}, \bibinfo{person}{Henning Kopp}, {and} \bibinfo{person}{Stefan Pfattheicher}.} \bibinfo{year}{2016}\natexlab{}.
\newblock \showarticletitle{Tales from the {Dark} {Side}: {Privacy} {Dark} {Strategies} and {Privacy} {Dark} {Patterns}}.
\newblock \bibinfo{journal}{\emph{Proceedings on Privacy Enhancing Technologies}} \bibinfo{volume}{2016}, \bibinfo{number}{4} (\bibinfo{date}{Oct.} \bibinfo{year}{2016}), \bibinfo{pages}{237--254}.
\newblock
\urldef\tempurl%
\url{https://doi.org/10.1515/popets-2016-0038}
\showDOI{\tempurl}


\bibitem[Caraban et~al\mbox{.}(2019)]%
        {caraban_23_2019}
\bibfield{author}{\bibinfo{person}{Ana Caraban}, \bibinfo{person}{Evangelos Karapanos}, \bibinfo{person}{Daniel Gonçalves}, {and} \bibinfo{person}{Pedro Campos}.} \bibinfo{year}{2019}\natexlab{}.
\newblock \showarticletitle{23 {Ways} to {Nudge}: {A} {Review} of {Technology}-{Mediated} {Nudging} in {Human}-{Computer} {Interaction}}. In \bibinfo{booktitle}{\emph{Proceedings of the 2019 {CHI} {Conference} on {Human} {Factors} in {Computing} {Systems}}}. \bibinfo{publisher}{ACM}, \bibinfo{address}{Glasgow Scotland Uk}, \bibinfo{pages}{1--15}.
\newblock
\showISBNx{978-1-4503-5970-2}
\urldef\tempurl%
\url{https://doi.org/10.1145/3290605.3300733}
\showDOI{\tempurl}


\bibitem[Chang and Myers(2012)]%
        {chang_webcrystal_2012}
\bibfield{author}{\bibinfo{person}{Kerry Shih-Ping Chang} {and} \bibinfo{person}{Brad~A. Myers}.} \bibinfo{year}{2012}\natexlab{}.
\newblock \showarticletitle{{WebCrystal}: understanding and reusing examples in web authoring}. In \bibinfo{booktitle}{\emph{Proceedings of the {SIGCHI} {Conference} on {Human} {Factors} in {Computing} {Systems}}} \emph{(\bibinfo{series}{{CHI} '12})}. \bibinfo{publisher}{Association for Computing Machinery}, \bibinfo{address}{New York, NY, USA}, \bibinfo{pages}{3205--3214}.
\newblock
\showISBNx{978-1-4503-1015-4}
\urldef\tempurl%
\url{https://doi.org/10.1145/2207676.2208740}
\showDOI{\tempurl}


\bibitem[Chaudhary et~al\mbox{.}(2022a)]%
        {chaudhary2022you}
\bibfield{author}{\bibinfo{person}{Akash Chaudhary}, \bibinfo{person}{Jaivrat Saroha}, \bibinfo{person}{Kyzyl Monteiro}, \bibinfo{person}{Angus~G. Forbes}, {and} \bibinfo{person}{Aman Parnami}.} \bibinfo{year}{2022}\natexlab{a}.
\newblock \showarticletitle{“{Are} {You} {Still} {Watching}?”: {Exploring} {Unintended} {User} {Behaviors} and {Dark} {Patterns} on {Video} {Streaming} {Platforms}}. In \bibinfo{booktitle}{\emph{Designing {Interactive} {Systems} {Conference}}} \emph{(\bibinfo{series}{{DIS} '22})}. \bibinfo{publisher}{Association for Computing Machinery}, \bibinfo{address}{New York, NY, USA}, \bibinfo{pages}{776--791}.
\newblock
\showISBNx{978-1-4503-9358-4}
\urldef\tempurl%
\url{https://doi.org/10.1145/3532106.3533562}
\showDOI{\tempurl}


\bibitem[Chaudhary et~al\mbox{.}(2022b)]%
        {chaudhary_are_2022}
\bibfield{author}{\bibinfo{person}{Akash Chaudhary}, \bibinfo{person}{Jaivrat Saroha}, \bibinfo{person}{Kyzyl Monteiro}, \bibinfo{person}{Angus~G. Forbes}, {and} \bibinfo{person}{Aman Parnami}.} \bibinfo{year}{2022}\natexlab{b}.
\newblock \showarticletitle{“{Are} {You} {Still} {Watching}?”: {Exploring} {Unintended} {User} {Behaviors} and {Dark} {Patterns} on {Video} {Streaming} {Platforms}}. In \bibinfo{booktitle}{\emph{Designing {Interactive} {Systems} {Conference}}} \emph{(\bibinfo{series}{{DIS} '22})}. \bibinfo{publisher}{Association for Computing Machinery}, \bibinfo{address}{New York, NY, USA}, \bibinfo{pages}{776--791}.
\newblock
\showISBNx{978-1-4503-9358-4}
\urldef\tempurl%
\url{https://doi.org/10.1145/3532106.3533562}
\showDOI{\tempurl}


\bibitem[Chivukula et~al\mbox{.}(2018)]%
        {chivukula_dark_2018}
\bibfield{author}{\bibinfo{person}{Shruthi~Sai Chivukula}, \bibinfo{person}{Jason Brier}, {and} \bibinfo{person}{Colin~M. Gray}.} \bibinfo{year}{2018}\natexlab{}.
\newblock \showarticletitle{Dark Intentions or Persuasion? UX Designers' Activation of Stakeholder and User Values}. In \bibinfo{booktitle}{\emph{Proceedings of the 2018 ACM Conference Companion Publication on Designing Interactive Systems}} (Hong Kong, China) \emph{(\bibinfo{series}{DIS '18 Companion})}. \bibinfo{publisher}{Association for Computing Machinery}, \bibinfo{address}{New York, NY, USA}, \bibinfo{pages}{87–91}.
\newblock
\showISBNx{9781450356312}
\urldef\tempurl%
\url{https://doi.org/10.1145/3197391.3205417}
\showDOI{\tempurl}


\bibitem[Commision(2022)]%
        {federaltradecommisionBringingDarkPatterns2022}
\bibfield{author}{\bibinfo{person}{Federal~Trade Commision}.} \bibinfo{year}{2022}\natexlab{}.
\newblock \bibinfo{booktitle}{\emph{Bringing Dark Patterns to Light}}.
\newblock \bibinfo{type}{{T}echnical {R}eport}. \bibinfo{institution}{Federal Trade Commision}. \bibinfo{pages}{48} pages.
\newblock


\bibitem[Commission et~al\mbox{.}(2022)]%
        {eu_commission_2022}
\bibfield{author}{\bibinfo{person}{European Commission}, \bibinfo{person}{Directorate-General for Justice}, \bibinfo{person}{Consumers}, \bibinfo{person}{F Lupiáñez-Villanueva}, \bibinfo{person}{A Boluda}, \bibinfo{person}{F Bogliacino}, \bibinfo{person}{G Liva}, \bibinfo{person}{L Lechardoy}, {and} \bibinfo{person}{T Rodríguez de~las Heras~Ballell}.} \bibinfo{year}{2022}\natexlab{}.
\newblock \bibinfo{booktitle}{\emph{Behavioural study on unfair commercial practices in the digital environment : dark patterns and manipulative personalisation : final report}}.
\newblock \bibinfo{publisher}{Publications Office of the European Union}, \bibinfo{address}{Brussels, Belgium}.
\newblock
\urldef\tempurl%
\url{https://doi.org/10.2838/859030}
\showDOI{\tempurl}


\bibitem[Conti and Sobiesk(2010)]%
        {conti_malicious_2010}
\bibfield{author}{\bibinfo{person}{Gregory Conti} {and} \bibinfo{person}{Edward Sobiesk}.} \bibinfo{year}{2010}\natexlab{}.
\newblock \showarticletitle{Malicious interface design: exploiting the user}. In \bibinfo{booktitle}{\emph{Proceedings of the 19th international conference on {World} wide web}} \emph{(\bibinfo{series}{{WWW} '10})}. \bibinfo{publisher}{Association for Computing Machinery}, \bibinfo{address}{New York, NY, USA}, \bibinfo{pages}{271--280}.
\newblock
\showISBNx{978-1-60558-799-8}
\urldef\tempurl%
\url{https://doi.org/10.1145/1772690.1772719}
\showDOI{\tempurl}


\bibitem[Deka et~al\mbox{.}(2017)]%
        {deka_rico:_2017}
\bibfield{author}{\bibinfo{person}{Biplab Deka}, \bibinfo{person}{Zifeng Huang}, \bibinfo{person}{Chad Franzen}, \bibinfo{person}{Joshua Hibschman}, \bibinfo{person}{Daniel Afergan}, \bibinfo{person}{Yang Li}, \bibinfo{person}{Jeffrey Nichols}, {and} \bibinfo{person}{Ranjitha Kumar}.} \bibinfo{year}{2017}\natexlab{}.
\newblock \showarticletitle{Rico: {A} {Mobile} {App} {Dataset} for {Building} {Data}-{Driven} {Design} {Applications}}. In \bibinfo{booktitle}{\emph{Proceedings of the 30th {Annual} {ACM} {Symposium} on {User} {Interface} {Software} and {Technology}}} \emph{(\bibinfo{series}{{UIST} '17})}. \bibinfo{publisher}{ACM}, \bibinfo{address}{New York, NY, USA}, \bibinfo{pages}{845--854}.
\newblock
\showISBNx{978-1-4503-4981-9}
\urldef\tempurl%
\url{https://doi.org/10.1145/3126594.3126651}
\showDOI{\tempurl}


\bibitem[Di~Geronimo et~al\mbox{.}(2020)]%
        {digeronimo_dark_2020}
\bibfield{author}{\bibinfo{person}{Linda Di~Geronimo}, \bibinfo{person}{Larissa Braz}, \bibinfo{person}{Enrico Fregnan}, \bibinfo{person}{Fabio Palomba}, {and} \bibinfo{person}{Alberto Bacchelli}.} \bibinfo{year}{2020}\natexlab{}.
\newblock \bibinfo{booktitle}{\emph{UI Dark Patterns and Where to Find Them: A Study on Mobile Applications and User Perception}}.
\newblock \bibinfo{publisher}{Association for Computing Machinery}, \bibinfo{address}{New York, NY, USA}, \bibinfo{pages}{1–14}.
\newblock
\showISBNx{9781450367080}
\urldef\tempurl%
\url{https://doi.org/10.1145/3313831.3376600}
\showURL{%
\tempurl}


\bibitem[Donnelly et~al\mbox{.}(2022)]%
        {donnellyBePatternWorld2022a}
\bibfield{author}{\bibinfo{person}{Jordan Donnelly}, \bibinfo{person}{Alan Downley}, \bibinfo{person}{Yunpeng Liu}, \bibinfo{person}{Yufei Su}, \bibinfo{person}{Quanwei Sun}, \bibinfo{person}{Lan Zeng}, \bibinfo{person}{Andrea Curley}, \bibinfo{person}{Damian Gordon}, \bibinfo{person}{Paul Kelly}, \bibinfo{person}{Dympna O'Sullivan}, {and} \bibinfo{person}{Anna Becevel}.} \bibinfo{year}{2022}\natexlab{}.
\newblock \showarticletitle{``Be a Pattern for the World'': The Development of a Dark Patterns Detection Tool to Prevent Online User Loss}. In \bibinfo{booktitle}{\emph{Proceedings of Ethicomp, 20th International Conference on the Ethical and Social Issues in Information and Communication Technologies}}. \bibinfo{publisher}{University of Turku}, \bibinfo{address}{Turku, Finland}, \bibinfo{pages}{578--582}.
\newblock
\urldef\tempurl%
\url{https://doi.org/10.21427/2Y2Q-6323}
\showDOI{\tempurl}


\bibitem[Dym et~al\mbox{.}(2018)]%
        {dym2018online}
\bibfield{author}{\bibinfo{person}{Brianna Dym}, \bibinfo{person}{Cecilia Aragon}, \bibinfo{person}{Julia Bullard}, \bibinfo{person}{Ruby Davis}, {and} \bibinfo{person}{Casey Fiesler}.} \bibinfo{year}{2018}\natexlab{}.
\newblock \showarticletitle{Online fandom: Boldly going where few CSCW researchers have gone before}. In \bibinfo{booktitle}{\emph{Companion of the 2018 ACM Conference on Computer Supported Cooperative Work and Social Computing}}. \bibinfo{publisher}{Association for Computing Machinery}, \bibinfo{address}{New York, NY, USA}, \bibinfo{pages}{121--124}.
\newblock


\bibitem[Eiband et~al\mbox{.}(2018)]%
        {eiband2018bringing}
\bibfield{author}{\bibinfo{person}{Malin Eiband}, \bibinfo{person}{Hanna Schneider}, \bibinfo{person}{Mark Bilandzic}, \bibinfo{person}{Julian Fazekas-Con}, \bibinfo{person}{Mareike Haug}, {and} \bibinfo{person}{Heinrich Hussmann}.} \bibinfo{year}{2018}\natexlab{}.
\newblock \showarticletitle{Bringing transparency design into practice}. In \bibinfo{booktitle}{\emph{23rd international conference on intelligent user interfaces}}. \bibinfo{pages}{211--223}.
\newblock


\bibitem[Fiesler et~al\mbox{.}(2017)]%
        {fiesler2017growing}
\bibfield{author}{\bibinfo{person}{Casey Fiesler}, \bibinfo{person}{Shannon Morrison}, \bibinfo{person}{R~Benjamin Shapiro}, {and} \bibinfo{person}{Amy~S Bruckman}.} \bibinfo{year}{2017}\natexlab{}.
\newblock \showarticletitle{Growing their own: Legitimate peripheral participation for computational learning in an online fandom community}. In \bibinfo{booktitle}{\emph{Proceedings of the 2017 ACM conference on computer supported cooperative work and social computing}}. \bibinfo{publisher}{Association for Computing Machinery}, \bibinfo{address}{New York, NY, USA}, \bibinfo{pages}{1375--1386}.
\newblock


\bibitem[Firmenich et~al\mbox{.}(2015)]%
        {firmenich2015user}
\bibfield{author}{\bibinfo{person}{Diego Firmenich}, \bibinfo{person}{Sergio Firmenich}, \bibinfo{person}{Gustavo Rossi}, \bibinfo{person}{Marco Winckler}, {and} \bibinfo{person}{Damiano Distante}.} \bibinfo{year}{2015}\natexlab{}.
\newblock \showarticletitle{User interface adaptation using web augmentation techniques: towards a negotiated approach}. In \bibinfo{booktitle}{\emph{Engineering the Web in the Big Data Era: 15th International Conference, ICWE 2015, Rotterdam, The Netherlands, June 23-26, 2015, Proceedings 15}}. \bibinfo{publisher}{Springer}, \bibinfo{address}{New York, NY, USA}, \bibinfo{pages}{147--164}.
\newblock


\bibitem[Fischer(2019)]%
        {united_states_senator_deb_fischer_senators_2019}
\bibfield{author}{\bibinfo{person}{United States Senator~Deb Fischer}.} \bibinfo{year}{2019}\natexlab{}.
\newblock \bibinfo{title}{Senators {Introduce} {Bipartisan} {Legislation} to {Ban} {Manipulative} '{Dark} {Patterns}'}.
\newblock
\newblock
\urldef\tempurl%
\url{http://www.fischer.senate.gov/public/index.cfm/2019/4/senators-introduce-bipartisan-legislation-to-ban-manipulative-dark-patterns}
\showURL{%
\tempurl}


\bibitem[Fitzgerald(2008)]%
        {fitzgerald_copystyler_2008}
\bibfield{author}{\bibinfo{person}{Michael~J. Fitzgerald}.} \bibinfo{year}{2008}\natexlab{}.
\newblock \emph{\bibinfo{title}{{CopyStyler} : {Web} design by example}}.
\newblock Thesis. \bibinfo{school}{Massachusetts Institute of Technology}.
\newblock
\urldef\tempurl%
\url{https://dspace.mit.edu/handle/1721.1/46032}
\showURL{%
\tempurl}
\newblock
\shownote{Accepted: 2009-06-30T17:06:01Z Journal Abbreviation: Web design by example}.


\bibitem[Gould et~al\mbox{.}(2021)]%
        {gould_special_2021}
\bibfield{author}{\bibinfo{person}{Sandy J.~J. Gould}, \bibinfo{person}{Lewis~L. Chuang}, \bibinfo{person}{Ioanna Iacovides}, \bibinfo{person}{Diego Garaialde}, \bibinfo{person}{Marta~E. Cecchinato}, \bibinfo{person}{Benjamin~R. Cowan}, {and} \bibinfo{person}{Anna~L. Cox}.} \bibinfo{year}{2021}\natexlab{}.
\newblock \showarticletitle{A {Special} {Interest} {Group} on {Designed} and {Engineered} {Friction} in {Interaction}}. In \bibinfo{booktitle}{\emph{Extended {Abstracts} of the 2021 {CHI} {Conference} on {Human} {Factors} in {Computing} {Systems}}} \emph{(\bibinfo{series}{{CHI} {EA} '21})}. \bibinfo{publisher}{Association for Computing Machinery}, \bibinfo{address}{New York, NY, USA}, \bibinfo{pages}{1--4}.
\newblock
\showISBNx{978-1-4503-8095-9}
\urldef\tempurl%
\url{https://doi.org/10.1145/3411763.3450404}
\showDOI{\tempurl}


\bibitem[Gray et~al\mbox{.}(2021)]%
        {gray_enduser_2021}
\bibfield{author}{\bibinfo{person}{Colin~M. Gray}, \bibinfo{person}{Jingle Chen}, \bibinfo{person}{Shruthi~Sai Chivukula}, {and} \bibinfo{person}{Liyang Qu}.} \bibinfo{year}{2021}\natexlab{}.
\newblock \showarticletitle{End User Accounts of Dark Patterns as Felt Manipulation}.
\newblock \bibinfo{journal}{\emph{Proc. ACM Hum.-Comput. Interact.}} \bibinfo{volume}{5}, \bibinfo{number}{CSCW2}, Article \bibinfo{articleno}{372} (\bibinfo{date}{oct} \bibinfo{year}{2021}), \bibinfo{numpages}{25}~pages.
\newblock
\urldef\tempurl%
\url{https://doi.org/10.1145/3479516}
\showDOI{\tempurl}


\bibitem[Gray et~al\mbox{.}(2023a)]%
        {grayEmergingTransdisciplinaryPerspectives2023}
\bibfield{author}{\bibinfo{person}{Colin~M. Gray}, \bibinfo{person}{Shruthi~Sai Chivukula}, \bibinfo{person}{Kerstin {Bongard-Blanchy}}, \bibinfo{person}{Arunesh Mathur}, \bibinfo{person}{Johanna~T. Gunawan}, {and} \bibinfo{person}{Brennan Schaffner}.} \bibinfo{year}{2023}\natexlab{a}.
\newblock \showarticletitle{Emerging Transdisciplinary Perspectives to Confront Dark Patterns}. In \bibinfo{booktitle}{\emph{Extended Abstracts of the 2023 CHI Conference on Human Factors in Computing Systems}} \emph{(\bibinfo{series}{CHI EA '23})}. \bibinfo{publisher}{Association for Computing Machinery}, \bibinfo{address}{New York, NY, USA}, \bibinfo{pages}{1--4}.
\newblock
\showISBNx{978-1-4503-9422-2}
\urldef\tempurl%
\url{https://doi.org/10.1145/3544549.3583745}
\showDOI{\tempurl}


\bibitem[Gray et~al\mbox{.}(2020)]%
        {gray_what_2020}
\bibfield{author}{\bibinfo{person}{Colin~M. Gray}, \bibinfo{person}{Shruthi~Sai Chivukula}, {and} \bibinfo{person}{Ahreum Lee}.} \bibinfo{year}{2020}\natexlab{}.
\newblock \showarticletitle{What {Kind} of {Work} {Do} "{Asshole} {Designers}" {Create}? {Describing} {Properties} of {Ethical} {Concern} on {Reddit}}. In \bibinfo{booktitle}{\emph{Proceedings of the 2020 {ACM} {Designing} {Interactive} {Systems} {Conference}}}. \bibinfo{publisher}{ACM}, \bibinfo{address}{Eindhoven Netherlands}, \bibinfo{pages}{61--73}.
\newblock
\showISBNx{978-1-4503-6974-9}
\urldef\tempurl%
\url{https://doi.org/10.1145/3357236.3395486}
\showDOI{\tempurl}


\bibitem[Gray et~al\mbox{.}(2018)]%
        {gray_dark_2018}
\bibfield{author}{\bibinfo{person}{Colin~M. Gray}, \bibinfo{person}{Yubo Kou}, \bibinfo{person}{Bryan Battles}, \bibinfo{person}{Joseph Hoggatt}, {and} \bibinfo{person}{Austin~L. Toombs}.} \bibinfo{year}{2018}\natexlab{}.
\newblock \showarticletitle{The Dark (Patterns) Side of UX Design}. In \bibinfo{booktitle}{\emph{Proceedings of the 2018 CHI Conference on Human Factors in Computing Systems}} (Montreal QC, Canada) \emph{(\bibinfo{series}{CHI '18})}. \bibinfo{publisher}{Association for Computing Machinery}, \bibinfo{address}{New York, NY, USA}, \bibinfo{pages}{1–14}.
\newblock
\showISBNx{9781450356206}
\urldef\tempurl%
\url{https://doi.org/10.1145/3173574.3174108}
\showDOI{\tempurl}


\bibitem[Gray et~al\mbox{.}(2023b)]%
        {grayMappingLandscapeDark2023}
\bibfield{author}{\bibinfo{person}{Colin~M. Gray}, \bibinfo{person}{Lorena Sanchez~Chamorro}, \bibinfo{person}{Ike Obi}, {and} \bibinfo{person}{Ja-Nae Duane}.} \bibinfo{year}{2023}\natexlab{b}.
\newblock \showarticletitle{Mapping the Landscape of Dark Patterns Scholarship: A Systematic Literature Review}. In \bibinfo{booktitle}{\emph{Companion Publication of the 2023 ACM Designing Interactive Systems Conference}} \emph{(\bibinfo{series}{DIS '23 Companion})}. \bibinfo{publisher}{Association for Computing Machinery}, \bibinfo{address}{New York, NY, USA}, \bibinfo{pages}{188--193}.
\newblock
\showISBNx{978-1-4503-9898-5}
\urldef\tempurl%
\url{https://doi.org/10.1145/3563703.3596635}
\showDOI{\tempurl}


\bibitem[Gray et~al\mbox{.}(2023c)]%
        {grayPreliminaryOntologyDark2023}
\bibfield{author}{\bibinfo{person}{Colin~M. Gray}, \bibinfo{person}{Cristiana Santos}, {and} \bibinfo{person}{Nataliia Bielova}.} \bibinfo{year}{2023}\natexlab{c}.
\newblock \showarticletitle{Towards a Preliminary Ontology of Dark Patterns Knowledge}. In \bibinfo{booktitle}{\emph{Extended Abstracts of the 2023 CHI Conference on Human Factors in Computing Systems}} \emph{(\bibinfo{series}{CHI EA '23})}. \bibinfo{publisher}{Association for Computing Machinery}, \bibinfo{address}{New York, NY, USA}, \bibinfo{pages}{1--9}.
\newblock
\showISBNx{978-1-4503-9422-2}
\urldef\tempurl%
\url{https://doi.org/10.1145/3544549.3585676}
\showDOI{\tempurl}


\bibitem[Gray et~al\mbox{.}(2023d)]%
        {grayDarkPatternsEmerging2023}
\bibfield{author}{\bibinfo{person}{Colin~M. Gray}, \bibinfo{person}{Cristiana~Teixeira Santos}, \bibinfo{person}{Nicole Tong}, \bibinfo{person}{Thomas Mildner}, \bibinfo{person}{Arianna Rossi}, \bibinfo{person}{Johanna~T. Gunawan}, {and} \bibinfo{person}{Caroline Sinders}.} \bibinfo{year}{2023}\natexlab{d}.
\newblock \showarticletitle{Dark Patterns and the Emerging Threats of Deceptive Design Practices}. In \bibinfo{booktitle}{\emph{Extended Abstracts of the 2023 CHI Conference on Human Factors in Computing Systems}} \emph{(\bibinfo{series}{CHI EA '23})}. \bibinfo{publisher}{Association for Computing Machinery}, \bibinfo{address}{New York, NY, USA}, \bibinfo{pages}{1--4}.
\newblock
\showISBNx{978-1-4503-9422-2}
\urldef\tempurl%
\url{https://doi.org/10.1145/3544549.3583173}
\showDOI{\tempurl}


\bibitem[Graßl et~al\mbox{.}(2021)]%
        {grasl_dark_2021}
\bibfield{author}{\bibinfo{person}{Paul Graßl}, \bibinfo{person}{Hanna Schraffenberger}, \bibinfo{person}{Frederik~Zuiderveen Borgesius}, {and} \bibinfo{person}{Moniek Buijzen}.} \bibinfo{year}{2021}\natexlab{}.
\newblock \showarticletitle{Dark and {Bright} {Patterns} in {Cookie} {Consent} {Requests}}.
\newblock \bibinfo{journal}{\emph{Journal of Digital Social Research}} \bibinfo{volume}{3}, \bibinfo{number}{1} (\bibinfo{date}{Feb.} \bibinfo{year}{2021}), \bibinfo{pages}{1--38}.
\newblock
\showISSN{2003-1998}
\urldef\tempurl%
\url{https://doi.org/10.33621/jdsr.v3i1.54}
\showDOI{\tempurl}
\newblock
\shownote{Number: 1}.


\bibitem[Greenberg et~al\mbox{.}(2014)]%
        {greenberg_dark_2014}
\bibfield{author}{\bibinfo{person}{Saul Greenberg}, \bibinfo{person}{Sebastian Boring}, \bibinfo{person}{Jo Vermeulen}, {and} \bibinfo{person}{Jakub Dostal}.} \bibinfo{year}{2014}\natexlab{}.
\newblock \showarticletitle{Dark patterns in proxemic interactions: a critical perspective}. In \bibinfo{booktitle}{\emph{Proceedings of the 2014 conference on {Designing} interactive systems}} \emph{(\bibinfo{series}{{DIS} '14})}. \bibinfo{publisher}{Association for Computing Machinery}, \bibinfo{address}{New York, NY, USA}, \bibinfo{pages}{523--532}.
\newblock
\showISBNx{978-1-4503-2902-6}
\urldef\tempurl%
\url{https://doi.org/10.1145/2598510.2598541}
\showDOI{\tempurl}


\bibitem[Gunawan et~al\mbox{.}(2021)]%
        {gunawan_comparative_2021}
\bibfield{author}{\bibinfo{person}{Johanna Gunawan}, \bibinfo{person}{Amogh Pradeep}, \bibinfo{person}{David Choffnes}, \bibinfo{person}{Woodrow Hartzog}, {and} \bibinfo{person}{Christo Wilson}.} \bibinfo{year}{2021}\natexlab{}.
\newblock \showarticletitle{A {Comparative} {Study} of {Dark} {Patterns} {Across} {Web} and {Mobile} {Modalities}}.
\newblock \bibinfo{journal}{\emph{Proceedings of the ACM on Human-Computer Interaction}} \bibinfo{volume}{5}, \bibinfo{number}{CSCW2} (\bibinfo{date}{Oct.} \bibinfo{year}{2021}), \bibinfo{pages}{377:1--377:29}.
\newblock
\urldef\tempurl%
\url{https://doi.org/10.1145/3479521}
\showDOI{\tempurl}


\bibitem[Gunawan et~al\mbox{.}(2022)]%
        {gunawanRedressDarkPatterns2022}
\bibfield{author}{\bibinfo{person}{Johanna Gunawan}, \bibinfo{person}{Cristiana Santos}, {and} \bibinfo{person}{Irene Kamara}.} \bibinfo{year}{2022}\natexlab{}.
\newblock \showarticletitle{Redress for Dark Patterns Privacy Harms? A Case Study on Consent Interactions}. In \bibinfo{booktitle}{\emph{Proceedings of the 2022 Symposium on Computer Science and Law}} \emph{(\bibinfo{series}{CSLAW '22})}. \bibinfo{publisher}{Association for Computing Machinery}, \bibinfo{address}{New York, NY, USA}, \bibinfo{pages}{181--194}.
\newblock
\showISBNx{978-1-4503-9234-1}
\urldef\tempurl%
\url{https://doi.org/10.1145/3511265.3550448}
\showDOI{\tempurl}


\bibitem[Hansen and Jespersen(2013)]%
        {hansen_nudge_2013}
\bibfield{author}{\bibinfo{person}{Pelle~Guldborg Hansen} {and} \bibinfo{person}{Andreas~Maaløe Jespersen}.} \bibinfo{year}{2013}\natexlab{}.
\newblock \showarticletitle{Nudge and the {Manipulation} of {Choice}: {A} {Framework} for the {Responsible} {Use} of the {Nudge} {Approach} to {Behaviour} {Change} in {Public} {Policy}}.
\newblock \bibinfo{journal}{\emph{European Journal of Risk Regulation}} \bibinfo{volume}{4}, \bibinfo{number}{1} (\bibinfo{date}{March} \bibinfo{year}{2013}), \bibinfo{pages}{3--28}.
\newblock
\showISSN{1867-299X, 2190-8249}
\urldef\tempurl%
\url{https://doi.org/10.1017/S1867299X00002762}
\showDOI{\tempurl}


\bibitem[Hertwig and Grüne-Yanoff(2017)]%
        {hertwig_nudging_2017}
\bibfield{author}{\bibinfo{person}{Ralph Hertwig} {and} \bibinfo{person}{Till Grüne-Yanoff}.} \bibinfo{year}{2017}\natexlab{}.
\newblock \showarticletitle{Nudging and {Boosting}: {Steering} or {Empowering} {Good} {Decisions}}.
\newblock \bibinfo{journal}{\emph{Perspectives on Psychological Science}} \bibinfo{volume}{12}, \bibinfo{number}{6} (\bibinfo{date}{Nov.} \bibinfo{year}{2017}), \bibinfo{pages}{973--986}.
\newblock
\showISSN{1745-6916}
\urldef\tempurl%
\url{https://doi.org/10.1177/1745691617702496}
\showDOI{\tempurl}
\newblock
\shownote{Publisher: SAGE Publications Inc}.


\bibitem[Hidaka et~al\mbox{.}(2023)]%
        {hidakaLinguisticDeadEndsAlphabet2023}
\bibfield{author}{\bibinfo{person}{Shun Hidaka}, \bibinfo{person}{Sota Kobuki}, \bibinfo{person}{Mizuki Watanabe}, {and} \bibinfo{person}{Katie Seaborn}.} \bibinfo{year}{2023}\natexlab{}.
\newblock \showarticletitle{Linguistic Dead-Ends and Alphabet Soup: Finding Dark Patterns in Japanese Apps}. In \bibinfo{booktitle}{\emph{Proceedings of the 2023 CHI Conference on Human Factors in Computing Systems}} \emph{(\bibinfo{series}{CHI '23})}. \bibinfo{publisher}{Association for Computing Machinery}, \bibinfo{address}{New York, NY, USA}, \bibinfo{pages}{1--13}.
\newblock
\showISBNx{978-1-4503-9421-5}
\urldef\tempurl%
\url{https://doi.org/10.1145/3544548.3580942}
\showDOI{\tempurl}


\bibitem[Hofmann et~al\mbox{.}(2009)]%
        {hofmann2009impulse}
\bibfield{author}{\bibinfo{person}{Wilhelm Hofmann}, \bibinfo{person}{Malte Friese}, {and} \bibinfo{person}{Fritz Strack}.} \bibinfo{year}{2009}\natexlab{}.
\newblock \showarticletitle{Impulse and self-control from a dual-systems perspective}.
\newblock \bibinfo{journal}{\emph{Perspectives on psychological science}} \bibinfo{volume}{4}, \bibinfo{number}{2} (\bibinfo{year}{2009}), \bibinfo{pages}{162--176}.
\newblock


\bibitem[Huang et~al\mbox{.}(2019)]%
        {huang_swire_2019}
\bibfield{author}{\bibinfo{person}{Forrest Huang}, \bibinfo{person}{John~F. Canny}, {and} \bibinfo{person}{Jeffrey Nichols}.} \bibinfo{year}{2019}\natexlab{}.
\newblock \showarticletitle{Swire: Sketch-Based User Interface Retrieval}. In \bibinfo{booktitle}{\emph{Proceedings of the 2019 CHI Conference on Human Factors in Computing Systems}} (Glasgow, Scotland Uk) \emph{(\bibinfo{series}{CHI '19})}. \bibinfo{publisher}{ACM}, \bibinfo{address}{New York, NY, USA}, \bibinfo{pages}{1–10}.
\newblock
\showISBNx{9781450359702}
\urldef\tempurl%
\url{https://doi.org/10.1145/3290605.3300334}
\showDOI{\tempurl}


\bibitem[Hutchinson et~al\mbox{.}(2003)]%
        {hutchinson_technology_2003}
\bibfield{author}{\bibinfo{person}{Hilary Hutchinson}, \bibinfo{person}{Wendy Mackay}, \bibinfo{person}{Bo Westerlund}, \bibinfo{person}{Benjamin~B. Bederson}, \bibinfo{person}{Allison Druin}, \bibinfo{person}{Catherine Plaisant}, \bibinfo{person}{Michel Beaudouin-Lafon}, \bibinfo{person}{Stéphane Conversy}, \bibinfo{person}{Helen Evans}, \bibinfo{person}{Heiko Hansen}, \bibinfo{person}{Nicolas Roussel}, {and} \bibinfo{person}{Björn Eiderbäck}.} \bibinfo{year}{2003}\natexlab{}.
\newblock \showarticletitle{Technology probes: inspiring design for and with families}. In \bibinfo{booktitle}{\emph{Proceedings of the {SIGCHI} {Conference} on {Human} {Factors} in {Computing} {Systems}}} \emph{(\bibinfo{series}{{CHI} '03})}. \bibinfo{publisher}{Association for Computing Machinery}, \bibinfo{address}{New York, NY, USA}, \bibinfo{pages}{17--24}.
\newblock
\showISBNx{978-1-58113-630-2}
\urldef\tempurl%
\url{https://doi.org/10.1145/642611.642616}
\showDOI{\tempurl}


\bibitem[Jiang et~al\mbox{.}(2022)]%
        {jiang2022computational}
\bibfield{author}{\bibinfo{person}{Yue Jiang}, \bibinfo{person}{Yuwen Lu}, \bibinfo{person}{Jeffrey Nichols}, \bibinfo{person}{Wolfgang Stuerzlinger}, \bibinfo{person}{Chun Yu}, \bibinfo{person}{Christof Lutteroth}, \bibinfo{person}{Yang Li}, \bibinfo{person}{Ranjitha Kumar}, {and} \bibinfo{person}{Toby Jia-Jun Li}.} \bibinfo{year}{2022}\natexlab{}.
\newblock \showarticletitle{Computational {Approaches} for {Understanding}, {Generating}, and {Adapting} {User} {Interfaces}}. In \bibinfo{booktitle}{\emph{Extended {Abstracts} of the 2022 {CHI} {Conference} on {Human} {Factors} in {Computing} {Systems}}} \emph{(\bibinfo{series}{{CHI} {EA} '22})}. \bibinfo{publisher}{Association for Computing Machinery}, \bibinfo{address}{New York, NY, USA}, \bibinfo{pages}{1--6}.
\newblock
\showISBNx{978-1-4503-9156-6}
\urldef\tempurl%
\url{https://doi.org/10.1145/3491101.3504030}
\showDOI{\tempurl}


\bibitem[Jin et~al\mbox{.}(2022)]%
        {jin_exploring_2022}
\bibfield{author}{\bibinfo{person}{Haojian Jin}, \bibinfo{person}{Boyuan Guo}, \bibinfo{person}{Rituparna Roychoudhury}, \bibinfo{person}{Yaxing Yao}, \bibinfo{person}{Swarun Kumar}, \bibinfo{person}{Yuvraj Agarwal}, {and} \bibinfo{person}{Jason~I. Hong}.} \bibinfo{year}{2022}\natexlab{}.
\newblock \showarticletitle{Exploring the {Needs} of {Users} for {Supporting} {Privacy}-{Protective} {Behaviors} in {Smart} {Homes}}. In \bibinfo{booktitle}{\emph{Proceedings of the 2022 {CHI} {Conference} on {Human} {Factors} in {Computing} {Systems}}} \emph{(\bibinfo{series}{{CHI} '22})}. \bibinfo{publisher}{Association for Computing Machinery}, \bibinfo{address}{New York, NY, USA}, \bibinfo{pages}{1--19}.
\newblock
\showISBNx{978-1-4503-9157-3}
\urldef\tempurl%
\url{https://doi.org/10.1145/3491102.3517602}
\showDOI{\tempurl}


\bibitem[Kaur et~al\mbox{.}(2020)]%
        {kaur_interpreting_2020}
\bibfield{author}{\bibinfo{person}{Harmanpreet Kaur}, \bibinfo{person}{Harsha Nori}, \bibinfo{person}{Samuel Jenkins}, \bibinfo{person}{Rich Caruana}, \bibinfo{person}{Hanna Wallach}, {and} \bibinfo{person}{Jennifer Wortman~Vaughan}.} \bibinfo{year}{2020}\natexlab{}.
\newblock \showarticletitle{Interpreting {Interpretability}: {Understanding} {Data} {Scientists}' {Use} of {Interpretability} {Tools} for {Machine} {Learning}}. In \bibinfo{booktitle}{\emph{Proceedings of the 2020 {CHI} {Conference} on {Human} {Factors} in {Computing} {Systems}}} \emph{(\bibinfo{series}{{CHI} '20})}. \bibinfo{publisher}{Association for Computing Machinery}, \bibinfo{address}{New York, NY, USA}, \bibinfo{pages}{1--14}.
\newblock
\showISBNx{978-1-4503-6708-0}
\urldef\tempurl%
\url{https://doi.org/10.1145/3313831.3376219}
\showDOI{\tempurl}


\bibitem[Keleher et~al\mbox{.}(2022)]%
        {keleher2022well}
\bibfield{author}{\bibinfo{person}{Maxwell Keleher}, \bibinfo{person}{Fiona Westin}, \bibinfo{person}{Preethi Nagabandi}, {and} \bibinfo{person}{Sonia Chiasson}.} \bibinfo{year}{2022}\natexlab{}.
\newblock \showarticletitle{How Well Do Experts Understand End-Users’ Perceptions of Manipulative Patterns?}. In \bibinfo{booktitle}{\emph{Nordic Human-Computer Interaction Conference}}. \bibinfo{publisher}{Association for Computing Machinery}, \bibinfo{address}{New York, NY, USA}, \bibinfo{pages}{1--21}.
\newblock


\bibitem[Kim et~al\mbox{.}(2022)]%
        {kim_stylette_2022}
\bibfield{author}{\bibinfo{person}{Tae~Soo Kim}, \bibinfo{person}{DaEun Choi}, \bibinfo{person}{Yoonseo Choi}, {and} \bibinfo{person}{Juho Kim}.} \bibinfo{year}{2022}\natexlab{}.
\newblock \showarticletitle{Stylette: {Styling} the {Web} with {Natural} {Language}}. In \bibinfo{booktitle}{\emph{{CHI} {Conference} on {Human} {Factors} in {Computing} {Systems}}} \emph{(\bibinfo{series}{{CHI} '22})}. \bibinfo{publisher}{Association for Computing Machinery}, \bibinfo{address}{New York, NY, USA}, \bibinfo{pages}{1--17}.
\newblock
\showISBNx{978-1-4503-9157-3}
\urldef\tempurl%
\url{https://doi.org/10.1145/3491102.3501931}
\showDOI{\tempurl}


\bibitem[Kollnig et~al\mbox{.}(2021)]%
        {kollnig_i_2021}
\bibfield{author}{\bibinfo{person}{Konrad Kollnig}, \bibinfo{person}{Siddhartha Datta}, {and} \bibinfo{person}{Max Van~Kleek}.} \bibinfo{year}{2021}\natexlab{}.
\newblock \showarticletitle{I {Want} {My} {App} {That} {Way}: {Reclaiming} {Sovereignty} {Over} {Personal} {Devices}}. In \bibinfo{booktitle}{\emph{Extended {Abstracts} of the 2021 {CHI} {Conference} on {Human} {Factors} in {Computing} {Systems}}} \emph{(\bibinfo{series}{{CHI} {EA} '21})}. \bibinfo{publisher}{Association for Computing Machinery}, \bibinfo{address}{New York, NY, USA}, \bibinfo{pages}{1--8}.
\newblock
\showISBNx{978-1-4503-8095-9}
\urldef\tempurl%
\url{https://doi.org/10.1145/3411763.3451632}
\showDOI{\tempurl}


\bibitem[Kowalczyk et~al\mbox{.}(2023)]%
        {kowalczykUnderstandingDarkPatterns2023}
\bibfield{author}{\bibinfo{person}{Monica Kowalczyk}, \bibinfo{person}{Johanna~T. Gunawan}, \bibinfo{person}{David Choffnes}, \bibinfo{person}{Daniel~J Dubois}, \bibinfo{person}{Woodrow Hartzog}, {and} \bibinfo{person}{Christo Wilson}.} \bibinfo{year}{2023}\natexlab{}.
\newblock \showarticletitle{Understanding Dark Patterns in Home IoT Devices}. In \bibinfo{booktitle}{\emph{Proceedings of the 2023 CHI Conference on Human Factors in Computing Systems}} \emph{(\bibinfo{series}{CHI '23})}. \bibinfo{publisher}{Association for Computing Machinery}, \bibinfo{address}{New York, NY, USA}, \bibinfo{pages}{1--27}.
\newblock
\showISBNx{978-1-4503-9421-5}
\urldef\tempurl%
\url{https://doi.org/10.1145/3544548.3581432}
\showDOI{\tempurl}


\bibitem[Kuang and Fabricant(2019)]%
        {kuang2019user}
\bibfield{author}{\bibinfo{person}{Cliff Kuang} {and} \bibinfo{person}{Robert Fabricant}.} \bibinfo{year}{2019}\natexlab{}.
\newblock \bibinfo{booktitle}{\emph{User friendly: How the hidden rules of design are changing the way we live, work \& play}}.
\newblock \bibinfo{publisher}{Random House}, \bibinfo{address}{New York}.
\newblock


\bibitem[Kumar et~al\mbox{.}(2011)]%
        {kumar_bricolage_2011}
\bibfield{author}{\bibinfo{person}{Ranjitha Kumar}, \bibinfo{person}{Jerry~O. Talton}, \bibinfo{person}{Salman Ahmad}, {and} \bibinfo{person}{Scott~R. Klemmer}.} \bibinfo{year}{2011}\natexlab{}.
\newblock \showarticletitle{Bricolage: example-based retargeting for web design}. In \bibinfo{booktitle}{\emph{Proceedings of the {SIGCHI} {Conference} on {Human} {Factors} in {Computing} {Systems}}} \emph{(\bibinfo{series}{{CHI} '11})}. \bibinfo{publisher}{Association for Computing Machinery}, \bibinfo{address}{New York, NY, USA}, \bibinfo{pages}{2197--2206}.
\newblock
\showISBNx{978-1-4503-0228-9}
\urldef\tempurl%
\url{https://doi.org/10.1145/1978942.1979262}
\showDOI{\tempurl}


\bibitem[Lacey and Caudwell(2019)]%
        {laceyCutenessDarkPattern2019}
\bibfield{author}{\bibinfo{person}{Cherie Lacey} {and} \bibinfo{person}{Catherine Caudwell}.} \bibinfo{year}{2019}\natexlab{}.
\newblock \showarticletitle{Cuteness as a ‘Dark Pattern’ in Home Robots}. In \bibinfo{booktitle}{\emph{2019 14th ACM/IEEE International Conference on Human-Robot Interaction (HRI)}} (2019-03). \bibinfo{publisher}{IEEE}, \bibinfo{address}{Daegu, South Korea}, \bibinfo{pages}{374--381}.
\newblock
\showISSN{2167-2148}
\urldef\tempurl%
\url{https://doi.org/10.1109/HRI.2019.8673274}
\showDOI{\tempurl}


\bibitem[Langley(1999)]%
        {langley1999user}
\bibfield{author}{\bibinfo{person}{Pat Langley}.} \bibinfo{year}{1999}\natexlab{}.
\newblock \showarticletitle{User modeling in adaptive interface}.
\newblock In \bibinfo{booktitle}{\emph{UM99 User Modeling}}. \bibinfo{publisher}{Springer}, \bibinfo{address}{Udine, Italy}, \bibinfo{pages}{357--370}.
\newblock


\bibitem[Lazar et~al\mbox{.}(2017)]%
        {lazar2017research}
\bibfield{author}{\bibinfo{person}{Jonathan Lazar}, \bibinfo{person}{Jinjuan~Heidi Feng}, {and} \bibinfo{person}{Harry Hochheiser}.} \bibinfo{year}{2017}\natexlab{}.
\newblock \bibinfo{booktitle}{\emph{Research methods in human-computer interaction}}.
\newblock \bibinfo{publisher}{Morgan Kaufmann}, \bibinfo{address}{Chichester, West Sussex, U.K}.
\newblock


\bibitem[Lee et~al\mbox{.}(2010)]%
        {lee_designing_2010}
\bibfield{author}{\bibinfo{person}{Brian Lee}, \bibinfo{person}{Savil Srivastava}, \bibinfo{person}{Ranjitha Kumar}, \bibinfo{person}{Ronen Brafman}, {and} \bibinfo{person}{Scott~R. Klemmer}.} \bibinfo{year}{2010}\natexlab{}.
\newblock \showarticletitle{Designing with interactive example galleries}. In \bibinfo{booktitle}{\emph{Proceedings of the {SIGCHI} {Conference} on {Human} {Factors} in {Computing} {Systems}}} \emph{(\bibinfo{series}{{CHI} '10})}. \bibinfo{publisher}{Association for Computing Machinery}, \bibinfo{address}{New York, NY, USA}, \bibinfo{pages}{2257--2266}.
\newblock
\showISBNx{978-1-60558-929-9}
\urldef\tempurl%
\url{https://doi.org/10.1145/1753326.1753667}
\showDOI{\tempurl}


\bibitem[Leiser(2020)]%
        {leiser_dark_2020}
\bibfield{author}{\bibinfo{person}{M.~R. Leiser}.} \bibinfo{year}{2020}\natexlab{}.
\newblock \bibinfo{booktitle}{\emph{'{Dark} {Patterns}': {The} {Case} for {Regulatory} {Pluralism}}}.
\newblock \bibinfo{type}{{SSRN} {Scholarly} {Paper}} 3625637. \bibinfo{institution}{Social Science Research Network}, \bibinfo{address}{Rochester, NY}.
\newblock
\urldef\tempurl%
\url{https://doi.org/10.2139/ssrn.3625637}
\showDOI{\tempurl}


\bibitem[Lewis(2014)]%
        {lewis_irresistible_2014}
\bibfield{author}{\bibinfo{person}{Chris Lewis}.} \bibinfo{year}{2014}\natexlab{}.
\newblock \bibinfo{booktitle}{\emph{Irresistible {Apps}: {Motivational} {Design} {Patterns} for {Apps}, {Games}, and {Web}-based {Communities}} (\bibinfo{edition}{1st} ed.)}.
\newblock \bibinfo{publisher}{Apress}, \bibinfo{address}{USA}.
\newblock
\showISBNx{978-1-4302-6421-7}


\bibitem[Li et~al\mbox{.}(2017)]%
        {li_sugilite:_2017}
\bibfield{author}{\bibinfo{person}{Toby Jia-Jun Li}, \bibinfo{person}{Amos Azaria}, {and} \bibinfo{person}{Brad~A. Myers}.} \bibinfo{year}{2017}\natexlab{}.
\newblock \showarticletitle{{SUGILITE}: {Creating} {Multimodal} {Smartphone} {Automation} by {Demonstration}}. In \bibinfo{booktitle}{\emph{Proceedings of the 2017 {CHI} {Conference} on {Human} {Factors} in {Computing} {Systems}}} \emph{(\bibinfo{series}{{CHI} '17})}. \bibinfo{publisher}{ACM}, \bibinfo{address}{New York, NY, USA}, \bibinfo{pages}{6038--6049}.
\newblock
\showISBNx{978-1-4503-4655-9}
\urldef\tempurl%
\url{https://doi.org/10.1145/3025453.3025483}
\showDOI{\tempurl}


\bibitem[Li et~al\mbox{.}(2021)]%
        {li_screen2vec:_2021}
\bibfield{author}{\bibinfo{person}{Toby Jia-Jun Li}, \bibinfo{person}{Lindsay Popowski}, \bibinfo{person}{Tom Mitchell}, {and} \bibinfo{person}{Brad~A Myers}.} \bibinfo{year}{2021}\natexlab{}.
\newblock \showarticletitle{{Screen2Vec}: {Semantic} {Embedding} of {GUI} {Screens} and {GUI} {Components}}. In \bibinfo{booktitle}{\emph{Proceedings of the 2021 {CHI} {Conference} on {Human} {Factors} in {Computing} {Systems}}} \emph{(\bibinfo{series}{{CHI} '21})}. \bibinfo{publisher}{Association for Computing Machinery}, \bibinfo{address}{New York, NY, USA}, \bibinfo{pages}{1--15}.
\newblock
\showISBNx{978-1-4503-8096-6}
\urldef\tempurl%
\url{https://doi.org/10.1145/3411764.3445049}
\showDOI{\tempurl}


\bibitem[Li et~al\mbox{.}(2019)]%
        {li2019pumice}
\bibfield{author}{\bibinfo{person}{Toby Jia-Jun Li}, \bibinfo{person}{Marissa Radensky}, \bibinfo{person}{Justin Jia}, \bibinfo{person}{Kirielle Singarajah}, \bibinfo{person}{Tom~M. Mitchell}, {and} \bibinfo{person}{Brad~A. Myers}.} \bibinfo{year}{2019}\natexlab{}.
\newblock \showarticletitle{PUMICE: A Multi-Modal Agent That Learns Concepts and Conditionals from Natural Language and Demonstrations}. In \bibinfo{booktitle}{\emph{Proceedings of the 32nd Annual ACM Symposium on User Interface Software and Technology}} (New Orleans, LA, USA) \emph{(\bibinfo{series}{UIST '19})}. \bibinfo{publisher}{Association for Computing Machinery}, \bibinfo{address}{New York, NY, USA}, \bibinfo{pages}{577–589}.
\newblock
\showISBNx{9781450368162}
\urldef\tempurl%
\url{https://doi.org/10.1145/3332165.3347899}
\showDOI{\tempurl}


\bibitem[Li and Riva(2018)]%
        {li2018kite}
\bibfield{author}{\bibinfo{person}{Toby Jia-Jun Li} {and} \bibinfo{person}{Oriana Riva}.} \bibinfo{year}{2018}\natexlab{}.
\newblock \showarticletitle{Kite: {Building} {Conversational} {Bots} from {Mobile} {Apps}}. In \bibinfo{booktitle}{\emph{Proceedings of the 16th {Annual} {International} {Conference} on {Mobile} {Systems}, {Applications}, and {Services}}} \emph{(\bibinfo{series}{{MobiSys} '18})}. \bibinfo{publisher}{Association for Computing Machinery}, \bibinfo{address}{New York, NY, USA}, \bibinfo{pages}{96--109}.
\newblock
\showISBNx{978-1-4503-5720-3}
\urldef\tempurl%
\url{https://doi.org/10.1145/3210240.3210339}
\showDOI{\tempurl}


\bibitem[Lim et~al\mbox{.}(2018)]%
        {lim_ply_2018}
\bibfield{author}{\bibinfo{person}{Sarah Lim}, \bibinfo{person}{Joshua Hibschman}, \bibinfo{person}{Haoqi Zhang}, {and} \bibinfo{person}{Eleanor O'Rourke}.} \bibinfo{year}{2018}\natexlab{}.
\newblock \showarticletitle{Ply: {A} {Visual} {Web} {Inspector} for {Learning} from {Professional} {Webpages}}. In \bibinfo{booktitle}{\emph{Proceedings of the 31st {Annual} {ACM} {Symposium} on {User} {Interface} {Software} and {Technology}}} \emph{(\bibinfo{series}{{UIST} '18})}. \bibinfo{publisher}{Association for Computing Machinery}, \bibinfo{address}{New York, NY, USA}, \bibinfo{pages}{991--1002}.
\newblock
\showISBNx{978-1-4503-5948-1}
\urldef\tempurl%
\url{https://doi.org/10.1145/3242587.3242660}
\showDOI{\tempurl}


\bibitem[Liu et~al\mbox{.}(2018)]%
        {zheran2018reinforcement}
\bibfield{author}{\bibinfo{person}{Evan~Zheran Liu}, \bibinfo{person}{Kelvin Guu}, \bibinfo{person}{Panupong Pasupat}, \bibinfo{person}{Tianlin Shi}, {and} \bibinfo{person}{Percy Liang}.} \bibinfo{year}{2018}\natexlab{}.
\newblock \bibinfo{title}{Reinforcement {Learning} on {Web} {Interfaces} {Using} {Workflow}-{Guided} {Exploration}}.
\newblock
\newblock
\urldef\tempurl%
\url{http://arxiv.org/abs/1802.08802}
\showURL{%
\tempurl}
\newblock
\shownote{arXiv:1802.08802 [cs]}.


\bibitem[Luguri and Strahilevitz(2021)]%
        {luguri_shining_2021}
\bibfield{author}{\bibinfo{person}{Jamie Luguri} {and} \bibinfo{person}{Lior~Jacob Strahilevitz}.} \bibinfo{year}{2021}\natexlab{}.
\newblock \showarticletitle{Shining a {Light} on {Dark} {Patterns}}.
\newblock \bibinfo{journal}{\emph{Journal of Legal Analysis}} \bibinfo{volume}{13}, \bibinfo{number}{1} (\bibinfo{date}{Jan.} \bibinfo{year}{2021}), \bibinfo{pages}{43--109}.
\newblock
\showISSN{2161-7201}
\urldef\tempurl%
\url{https://doi.org/10.1093/jla/laaa006}
\showDOI{\tempurl}


\bibitem[Lukoff et~al\mbox{.}(2021)]%
        {lukoff_what_2021}
\bibfield{author}{\bibinfo{person}{Kai Lukoff}, \bibinfo{person}{Alexis Hiniker}, \bibinfo{person}{Colin~M. Gray}, \bibinfo{person}{Arunesh Mathur}, {and} \bibinfo{person}{Shruthi~Sai Chivukula}.} \bibinfo{year}{2021}\natexlab{}.
\newblock \showarticletitle{What Can CHI Do About Dark Patterns?}. In \bibinfo{booktitle}{\emph{Extended Abstracts of the 2021 CHI Conference on Human Factors in Computing Systems}} (Yokohama, Japan) \emph{(\bibinfo{series}{CHI EA '21})}. \bibinfo{publisher}{Association for Computing Machinery}, \bibinfo{address}{New York, NY, USA}, Article \bibinfo{articleno}{102}, \bibinfo{numpages}{6}~pages.
\newblock
\showISBNx{9781450380959}
\urldef\tempurl%
\url{https://doi.org/10.1145/3411763.3441360}
\showDOI{\tempurl}


\bibitem[Lyngs et~al\mbox{.}(2020)]%
        {lyngs_i_2020}
\bibfield{author}{\bibinfo{person}{Ulrik Lyngs}, \bibinfo{person}{Kai Lukoff}, \bibinfo{person}{Petr Slovak}, \bibinfo{person}{William Seymour}, \bibinfo{person}{Helena Webb}, \bibinfo{person}{Marina Jirotka}, \bibinfo{person}{Jun Zhao}, \bibinfo{person}{Max Van~Kleek}, {and} \bibinfo{person}{Nigel Shadbolt}.} \bibinfo{year}{2020}\natexlab{}.
\newblock \showarticletitle{'{I} {Just} {Want} to {Hack} {Myself} to {Not} {Get} {Distracted}': {Evaluating} {Design} {Interventions} for {Self}-{Control} on {Facebook}}. In \bibinfo{booktitle}{\emph{Proceedings of the 2020 {CHI} {Conference} on {Human} {Factors} in {Computing} {Systems}}}. \bibinfo{publisher}{Association for Computing Machinery}, \bibinfo{address}{New York, NY, USA}, \bibinfo{pages}{1--15}.
\newblock
\showISBNx{978-1-4503-6708-0}
\urldef\tempurl%
\url{https://doi.org/10.1145/3313831.3376672}
\showURL{%
\tempurl}


\bibitem[M.~Bhoot et~al\mbox{.}(2020)]%
        {m_bhoot_towards_2020}
\bibfield{author}{\bibinfo{person}{Aditi M.~Bhoot}, \bibinfo{person}{Mayuri A.~Shinde}, {and} \bibinfo{person}{Wricha P.~Mishra}.} \bibinfo{year}{2020}\natexlab{}.
\newblock \showarticletitle{Towards the {Identification} of {Dark} {Patterns}: {An} {Analysis} {Based} on {End}-{User} {Reactions}}. In \bibinfo{booktitle}{\emph{{IndiaHCI} '20: {Proceedings} of the 11th {Indian} {Conference} on {Human}-{Computer} {Interaction}}} \emph{(\bibinfo{series}{{IndiaHCI} 2020})}. \bibinfo{publisher}{Association for Computing Machinery}, \bibinfo{address}{New York, NY, USA}, \bibinfo{pages}{24--33}.
\newblock
\showISBNx{978-1-4503-8944-0}
\urldef\tempurl%
\url{https://doi.org/10.1145/3429290.3429293}
\showDOI{\tempurl}


\bibitem[Maier(2019)]%
        {maier2019dark}
\bibfield{author}{\bibinfo{person}{Maximilian Maier}.} \bibinfo{year}{2019}\natexlab{}.
\newblock \bibinfo{title}{Dark patterns--An end user perspective}.
\newblock
\newblock


\bibitem[Maier and Harr(2020)]%
        {maierDarkDesignPatterns2020}
\bibfield{author}{\bibinfo{person}{Maximilian Maier} {and} \bibinfo{person}{Rikard Harr}.} \bibinfo{year}{2020}\natexlab{}.
\newblock \showarticletitle{Dark Design Patterns: An End-User Perspective}.
\newblock \bibinfo{journal}{\emph{Human Technology}} \bibinfo{volume}{16}, \bibinfo{number}{2} (\bibinfo{year}{2020}), \bibinfo{pages}{170--199}.
\newblock
\showISSN{17956889}
\urldef\tempurl%
\url{https://doi.org/10.17011/ht/urn.202008245641}
\showDOI{\tempurl}


\bibitem[Mansur et~al\mbox{.}(2023)]%
        {mansur2023aidui}
\bibfield{author}{\bibinfo{person}{SM Mansur}, \bibinfo{person}{Sabiha Salma}, \bibinfo{person}{Damilola Awofisayo}, {and} \bibinfo{person}{Kevin Moran}.} \bibinfo{year}{2023}\natexlab{}.
\newblock \bibinfo{title}{AidUI: Toward Automated Recognition of Dark Patterns in User Interfaces}.
\newblock
\newblock


\bibitem[Markt(2020)]%
        {markt_guidelines_2020}
\bibfield{author}{\bibinfo{person}{Autoriteit Consument~\& Markt}.} \bibinfo{year}{2020}\natexlab{}.
\newblock \bibinfo{title}{Guidelines on the protection of the online consumer {\textbar} {ACM}.nl}.
\newblock
\newblock
\urldef\tempurl%
\url{https://www.acm.nl/en/publications/guidelines-protection-online-consumer}
\showURL{%
\tempurl}
\newblock
\shownote{Last Modified: 2020-11-23T14:40:37+0100}.


\bibitem[Mathur et~al\mbox{.}(2019)]%
        {mathur_2019_dark}
\bibfield{author}{\bibinfo{person}{Arunesh Mathur}, \bibinfo{person}{Gunes Acar}, \bibinfo{person}{Michael~J. Friedman}, \bibinfo{person}{Eli Lucherini}, \bibinfo{person}{Jonathan Mayer}, \bibinfo{person}{Marshini Chetty}, {and} \bibinfo{person}{Arvind Narayanan}.} \bibinfo{year}{2019}\natexlab{}.
\newblock \showarticletitle{Dark Patterns at Scale: Findings from a Crawl of 11K Shopping Websites}.
\newblock \bibinfo{journal}{\emph{Proc. ACM Hum.-Comput. Interact.}} \bibinfo{volume}{3}, \bibinfo{number}{CSCW}, Article \bibinfo{articleno}{81} (\bibinfo{date}{nov} \bibinfo{year}{2019}), \bibinfo{numpages}{32}~pages.
\newblock
\urldef\tempurl%
\url{https://doi.org/10.1145/3359183}
\showDOI{\tempurl}


\bibitem[Mathur et~al\mbox{.}(2021)]%
        {mathur_what_2021}
\bibfield{author}{\bibinfo{person}{Arunesh Mathur}, \bibinfo{person}{Mihir Kshirsagar}, {and} \bibinfo{person}{Jonathan Mayer}.} \bibinfo{year}{2021}\natexlab{}.
\newblock \showarticletitle{What Makes a Dark Pattern... Dark? Design Attributes, Normative Considerations, and Measurement Methods}. In \bibinfo{booktitle}{\emph{Proceedings of the 2021 CHI Conference on Human Factors in Computing Systems}} (Yokohama, Japan) \emph{(\bibinfo{series}{CHI '21})}. \bibinfo{publisher}{Association for Computing Machinery}, \bibinfo{address}{New York, NY, USA}, Article \bibinfo{articleno}{360}, \bibinfo{numpages}{18}~pages.
\newblock
\showISBNx{9781450380966}
\urldef\tempurl%
\url{https://doi.org/10.1145/3411764.3445610}
\showDOI{\tempurl}


\bibitem[Mathur et~al\mbox{.}(2018)]%
        {mathur_endorsements_2018}
\bibfield{author}{\bibinfo{person}{Arunesh Mathur}, \bibinfo{person}{Arvind Narayanan}, {and} \bibinfo{person}{Marshini Chetty}.} \bibinfo{year}{2018}\natexlab{}.
\newblock \showarticletitle{Endorsements on {Social} {Media}: {An} {Empirical} {Study} of {Affiliate} {Marketing} {Disclosures} on {YouTube} and {Pinterest}}.
\newblock \bibinfo{journal}{\emph{Proceedings of the ACM on Human-Computer Interaction}} \bibinfo{volume}{2}, \bibinfo{number}{CSCW} (\bibinfo{date}{Nov.} \bibinfo{year}{2018}), \bibinfo{pages}{1--26}.
\newblock
\showISSN{2573-0142}
\urldef\tempurl%
\url{https://doi.org/10.1145/3274388}
\showDOI{\tempurl}


\bibitem[McDonald et~al\mbox{.}(2019)]%
        {mcdonald2019reliability}
\bibfield{author}{\bibinfo{person}{Nora McDonald}, \bibinfo{person}{Sarita Schoenebeck}, {and} \bibinfo{person}{Andrea Forte}.} \bibinfo{year}{2019}\natexlab{}.
\newblock \showarticletitle{Reliability and {Inter}-rater {Reliability} in {Qualitative} {Research}: {Norms} and {Guidelines} for {CSCW} and {HCI} {Practice}}.
\newblock \bibinfo{journal}{\emph{Proceedings of the ACM on Human-Computer Interaction}} \bibinfo{volume}{3}, \bibinfo{number}{CSCW} (\bibinfo{date}{Nov.} \bibinfo{year}{2019}), \bibinfo{pages}{72:1--72:23}.
\newblock
\urldef\tempurl%
\url{https://doi.org/10.1145/3359174}
\showDOI{\tempurl}


\bibitem[Mildner et~al\mbox{.}(2023a)]%
        {mildnerDefendingDarkArts2023}
\bibfield{author}{\bibinfo{person}{Thomas Mildner}, \bibinfo{person}{Merle Freye}, \bibinfo{person}{Gian-Luca Savino}, \bibinfo{person}{Philip~R. Doyle}, \bibinfo{person}{Benjamin~R. Cowan}, {and} \bibinfo{person}{Rainer Malaka}.} \bibinfo{year}{2023}\natexlab{a}.
\newblock \showarticletitle{Defending Against the Dark Arts: Recognising Dark Patterns in Social Media}. In \bibinfo{booktitle}{\emph{Proceedings of the 2023 ACM Designing Interactive Systems Conference}} \emph{(\bibinfo{series}{DIS '23})}. \bibinfo{publisher}{Association for Computing Machinery}, \bibinfo{address}{New York, NY, USA}, \bibinfo{pages}{2362--2374}.
\newblock
\showISBNx{978-1-4503-9893-0}
\urldef\tempurl%
\url{https://doi.org/10.1145/3563657.3595964}
\showDOI{\tempurl}


\bibitem[Mildner and Savino(2021)]%
        {mildner_ethical_2021}
\bibfield{author}{\bibinfo{person}{Thomas Mildner} {and} \bibinfo{person}{Gian-Luca Savino}.} \bibinfo{year}{2021}\natexlab{}.
\newblock \showarticletitle{Ethical {User} {Interfaces}: {Exploring} the {Effects} of {Dark} {Patterns} on {Facebook}}. In \bibinfo{booktitle}{\emph{Extended {Abstracts} of the 2021 {CHI} {Conference} on {Human} {Factors} in {Computing} {Systems}}} \emph{(\bibinfo{series}{{CHI} {EA} '21})}. \bibinfo{publisher}{Association for Computing Machinery}, \bibinfo{address}{New York, NY, USA}, \bibinfo{pages}{1--7}.
\newblock
\showISBNx{978-1-4503-8095-9}
\urldef\tempurl%
\url{https://doi.org/10.1145/3411763.3451659}
\showDOI{\tempurl}


\bibitem[Mildner et~al\mbox{.}(2023b)]%
        {mildnerEngagingGoverningStrategies2023}
\bibfield{author}{\bibinfo{person}{Thomas Mildner}, \bibinfo{person}{Gian-Luca Savino}, \bibinfo{person}{Philip~R. Doyle}, \bibinfo{person}{Benjamin~R. Cowan}, {and} \bibinfo{person}{Rainer Malaka}.} \bibinfo{year}{2023}\natexlab{b}.
\newblock \showarticletitle{About Engaging and Governing Strategies: A Thematic Analysis of Dark Patterns in Social Networking Services}. In \bibinfo{booktitle}{\emph{Proceedings of the 2023 CHI Conference on Human Factors in Computing Systems}} \emph{(\bibinfo{series}{CHI '23})}. \bibinfo{publisher}{Association for Computing Machinery}, \bibinfo{address}{New York, NY, USA}, \bibinfo{pages}{1--15}.
\newblock
\showISBNx{978-1-4503-9421-5}
\urldef\tempurl%
\url{https://doi.org/10.1145/3544548.3580695}
\showDOI{\tempurl}


\bibitem[Moraveji et~al\mbox{.}(2007)]%
        {moraveji_comicboarding_2007}
\bibfield{author}{\bibinfo{person}{Neema Moraveji}, \bibinfo{person}{Jason Li}, \bibinfo{person}{Jiarong Ding}, \bibinfo{person}{Patrick O'Kelley}, {and} \bibinfo{person}{Suze Woolf}.} \bibinfo{year}{2007}\natexlab{}.
\newblock \showarticletitle{Comicboarding: using comics as proxies for participatory design with children}. In \bibinfo{booktitle}{\emph{Proceedings of the {SIGCHI} {Conference} on {Human} {Factors} in {Computing} {Systems}}} \emph{(\bibinfo{series}{{CHI} '07})}. \bibinfo{publisher}{Association for Computing Machinery}, \bibinfo{address}{New York, NY, USA}, \bibinfo{pages}{1371--1374}.
\newblock
\showISBNx{978-1-59593-593-9}
\urldef\tempurl%
\url{https://doi.org/10.1145/1240624.1240832}
\showDOI{\tempurl}


\bibitem[Moser et~al\mbox{.}(2019)]%
        {moser_impulse_2019}
\bibfield{author}{\bibinfo{person}{Carol Moser}, \bibinfo{person}{Sarita~Y. Schoenebeck}, {and} \bibinfo{person}{Paul Resnick}.} \bibinfo{year}{2019}\natexlab{}.
\newblock \showarticletitle{Impulse {Buying}: {Design} {Practices} and {Consumer} {Needs}}. In \bibinfo{booktitle}{\emph{Proceedings of the 2019 {CHI} {Conference} on {Human} {Factors} in {Computing} {Systems}}} \emph{(\bibinfo{series}{{CHI} '19})}. \bibinfo{publisher}{Association for Computing Machinery}, \bibinfo{address}{New York, NY, USA}, \bibinfo{pages}{1--15}.
\newblock
\showISBNx{978-1-4503-5970-2}
\urldef\tempurl%
\url{https://doi.org/10.1145/3290605.3300472}
\showDOI{\tempurl}


\bibitem[Muller and Kuhn(1993)]%
        {muller_participatory:_1993}
\bibfield{author}{\bibinfo{person}{Michael~J. Muller} {and} \bibinfo{person}{Sarah Kuhn}.} \bibinfo{year}{1993}\natexlab{}.
\newblock \showarticletitle{Participatory Design}.
\newblock \bibinfo{journal}{\emph{Commun. ACM}} \bibinfo{volume}{36}, \bibinfo{number}{6} (\bibinfo{date}{June} \bibinfo{year}{1993}), \bibinfo{pages}{24–28}.
\newblock
\showISSN{0001-0782}
\urldef\tempurl%
\url{https://doi.org/10.1145/153571.255960}
\showDOI{\tempurl}


\bibitem[Myers et~al\mbox{.}(2017)]%
        {myers2017making}
\bibfield{author}{\bibinfo{person}{Brad~A Myers}, \bibinfo{person}{Andrew~J Ko}, \bibinfo{person}{Chris Scaffidi}, \bibinfo{person}{Stephen Oney}, \bibinfo{person}{YoungSeok Yoon}, \bibinfo{person}{Kerry Chang}, \bibinfo{person}{Mary~Beth Kery}, {and} \bibinfo{person}{Toby Jia-Jun Li}.} \bibinfo{year}{2017}\natexlab{}.
\newblock \showarticletitle{Making end user development more natural}.
\newblock In \bibinfo{booktitle}{\emph{New perspectives in end-user development}}. \bibinfo{publisher}{Springer}, \bibinfo{address}{New York, NY}, \bibinfo{pages}{1--22}.
\newblock


\bibitem[Narayanan et~al\mbox{.}(2020)]%
        {narayanan_dark_2020}
\bibfield{author}{\bibinfo{person}{Arvind Narayanan}, \bibinfo{person}{Arunesh Mathur}, \bibinfo{person}{Marshini Chetty}, {and} \bibinfo{person}{Mihir Kshirsagar}.} \bibinfo{year}{2020}\natexlab{}.
\newblock \showarticletitle{Dark Patterns: Past, Present, and Future: The Evolution of Tricky User Interfaces}.
\newblock \bibinfo{journal}{\emph{Queue}} \bibinfo{volume}{18}, \bibinfo{number}{2} (\bibinfo{date}{apr} \bibinfo{year}{2020}), \bibinfo{pages}{67–92}.
\newblock
\showISSN{1542-7730}
\urldef\tempurl%
\url{https://doi.org/10.1145/3400899.3400901}
\showDOI{\tempurl}


\bibitem[Nebeling and Dey(2016)]%
        {nebeling_xdbrowser_2016}
\bibfield{author}{\bibinfo{person}{Michael Nebeling} {and} \bibinfo{person}{Anind~K. Dey}.} \bibinfo{year}{2016}\natexlab{}.
\newblock \showarticletitle{{XDBrowser}: {User}-{Defined} {Cross}-{Device} {Web} {Page} {Designs}}. In \bibinfo{booktitle}{\emph{Proceedings of the 2016 {CHI} {Conference} on {Human} {Factors} in {Computing} {Systems}}} \emph{(\bibinfo{series}{{CHI} '16})}. \bibinfo{publisher}{Association for Computing Machinery}, \bibinfo{address}{New York, NY, USA}, \bibinfo{pages}{5494--5505}.
\newblock
\showISBNx{978-1-4503-3362-7}
\urldef\tempurl%
\url{https://doi.org/10.1145/2858036.2858048}
\showDOI{\tempurl}


\bibitem[Nebeling et~al\mbox{.}(2013a)]%
        {nebeling_crowdadapt_2013}
\bibfield{author}{\bibinfo{person}{Michael Nebeling}, \bibinfo{person}{Maximilian Speicher}, {and} \bibinfo{person}{Moira~C. Norrie}.} \bibinfo{year}{2013}\natexlab{a}.
\newblock \showarticletitle{{CrowdAdapt}: enabling crowdsourced web page adaptation for individual viewing conditions and preferences}. In \bibinfo{booktitle}{\emph{Proceedings of the 5th {ACM} {SIGCHI} symposium on {Engineering} interactive computing systems}} \emph{(\bibinfo{series}{{EICS} '13})}. \bibinfo{publisher}{Association for Computing Machinery}, \bibinfo{address}{New York, NY, USA}, \bibinfo{pages}{23--32}.
\newblock
\showISBNx{978-1-4503-2138-9}
\urldef\tempurl%
\url{https://doi.org/10.1145/2494603.2480304}
\showDOI{\tempurl}


\bibitem[Nebeling et~al\mbox{.}(2013b)]%
        {nebeling2013crowdadapt}
\bibfield{author}{\bibinfo{person}{Michael Nebeling}, \bibinfo{person}{Maximilian Speicher}, {and} \bibinfo{person}{Moira~C Norrie}.} \bibinfo{year}{2013}\natexlab{b}.
\newblock \showarticletitle{CrowdAdapt: enabling crowdsourced web page adaptation for individual viewing conditions and preferences}. In \bibinfo{booktitle}{\emph{Proceedings of the 5th ACM SIGCHI symposium on Engineering interactive computing systems}}. \bibinfo{publisher}{Association for Computing Machinery}, \bibinfo{address}{New York, NY, USA}, \bibinfo{pages}{23--32}.
\newblock


\bibitem[Nichols and Myers(2009)]%
        {nichols_2009_description}
\bibfield{author}{\bibinfo{person}{Jeffrey Nichols} {and} \bibinfo{person}{Brad~A. Myers}.} \bibinfo{year}{2009}\natexlab{}.
\newblock \showarticletitle{Creating a Lightweight User Interface Description Language: An Overview and Analysis of the Personal Universal Controller Project}.
\newblock \bibinfo{journal}{\emph{ACM Trans. Comput.-Hum. Interact.}} \bibinfo{volume}{16}, \bibinfo{number}{4}, Article \bibinfo{articleno}{17} (\bibinfo{date}{nov} \bibinfo{year}{2009}), \bibinfo{numpages}{37}~pages.
\newblock
\showISSN{1073-0516}
\urldef\tempurl%
\url{https://doi.org/10.1145/1614390.1614392}
\showDOI{\tempurl}


\bibitem[Nichols et~al\mbox{.}(2002)]%
        {nichols_2002_remote}
\bibfield{author}{\bibinfo{person}{Jeffrey Nichols}, \bibinfo{person}{Brad~A. Myers}, \bibinfo{person}{Michael Higgins}, \bibinfo{person}{Joseph Hughes}, \bibinfo{person}{Thomas~K. Harris}, \bibinfo{person}{Roni Rosenfeld}, {and} \bibinfo{person}{Mathilde Pignol}.} \bibinfo{year}{2002}\natexlab{}.
\newblock \showarticletitle{Generating Remote Control Interfaces for Complex Appliances}. In \bibinfo{booktitle}{\emph{Proceedings of the 15th Annual ACM Symposium on User Interface Software and Technology}} (Paris, France) \emph{(\bibinfo{series}{UIST '02})}. \bibinfo{publisher}{Association for Computing Machinery}, \bibinfo{address}{New York, NY, USA}, \bibinfo{pages}{161–170}.
\newblock
\showISBNx{1581134886}
\urldef\tempurl%
\url{https://doi.org/10.1145/571985.572008}
\showDOI{\tempurl}


\bibitem[Nosek et~al\mbox{.}(2015)]%
        {nosek2015promoting}
\bibfield{author}{\bibinfo{person}{Brian~A Nosek}, \bibinfo{person}{George Alter}, \bibinfo{person}{George~C Banks}, \bibinfo{person}{Denny Borsboom}, \bibinfo{person}{Sara~D Bowman}, \bibinfo{person}{Steven~J Breckler}, \bibinfo{person}{Stuart Buck}, \bibinfo{person}{Christopher~D Chambers}, \bibinfo{person}{Gilbert Chin}, \bibinfo{person}{Garret Christensen}, {et~al\mbox{.}}} \bibinfo{year}{2015}\natexlab{}.
\newblock \showarticletitle{Promoting an open research culture}.
\newblock \bibinfo{journal}{\emph{Science}} \bibinfo{volume}{348}, \bibinfo{number}{6242} (\bibinfo{year}{2015}), \bibinfo{pages}{1422--1425}.
\newblock


\bibitem[Nosek et~al\mbox{.}(2018)]%
        {nosek2018preregistration}
\bibfield{author}{\bibinfo{person}{Brian~A Nosek}, \bibinfo{person}{Charles~R Ebersole}, \bibinfo{person}{Alexander~C DeHaven}, {and} \bibinfo{person}{David~T Mellor}.} \bibinfo{year}{2018}\natexlab{}.
\newblock \showarticletitle{The preregistration revolution}.
\newblock \bibinfo{journal}{\emph{Proceedings of the National Academy of Sciences}} \bibinfo{volume}{115}, \bibinfo{number}{11} (\bibinfo{year}{2018}), \bibinfo{pages}{2600--2606}.
\newblock


\bibitem[Nouwens et~al\mbox{.}(2022)]%
        {nouwensConsentOMaticAutomaticallyAnswering2022a}
\bibfield{author}{\bibinfo{person}{Midas Nouwens}, \bibinfo{person}{Rolf Bagge}, \bibinfo{person}{Janus~Bager Kristensen}, {and} \bibinfo{person}{Clemens~Nylandsted Klokmose}.} \bibinfo{year}{2022}\natexlab{}.
\newblock \showarticletitle{Consent-O-Matic: Automatically Answering Consent Pop-Ups Using Adversarial Interoperability}. In \bibinfo{booktitle}{\emph{Extended Abstracts of the 2022 CHI Conference on Human Factors in Computing Systems}} \emph{(\bibinfo{series}{CHI EA '22})}. \bibinfo{publisher}{Association for Computing Machinery}, \bibinfo{address}{New York, NY, USA}, \bibinfo{pages}{1--7}.
\newblock
\showISBNx{978-1-4503-9156-6}
\urldef\tempurl%
\url{https://doi.org/10.1145/3491101.3519683}
\showDOI{\tempurl}


\bibitem[Nouwens et~al\mbox{.}(2020)]%
        {nouwens_dark_2020}
\bibfield{author}{\bibinfo{person}{Midas Nouwens}, \bibinfo{person}{Ilaria Liccardi}, \bibinfo{person}{Michael Veale}, \bibinfo{person}{David Karger}, {and} \bibinfo{person}{Lalana Kagal}.} \bibinfo{year}{2020}\natexlab{}.
\newblock \bibinfo{booktitle}{\emph{Dark Patterns after the GDPR: Scraping Consent Pop-Ups and Demonstrating Their Influence}}.
\newblock \bibinfo{publisher}{Association for Computing Machinery}, \bibinfo{address}{New York, NY, USA}, \bibinfo{pages}{1–13}.
\newblock
\showISBNx{9781450367080}
\urldef\tempurl%
\url{https://doi.org/10.1145/3313831.3376321}
\showURL{%
\tempurl}


\bibitem[Oppenlaender et~al\mbox{.}(2020a)]%
        {oppenlaender_crowdui_2020}
\bibfield{author}{\bibinfo{person}{Jonas Oppenlaender}, \bibinfo{person}{Thanassis Tiropanis}, {and} \bibinfo{person}{Simo Hosio}.} \bibinfo{year}{2020}\natexlab{a}.
\newblock \showarticletitle{{CrowdUI}: {Supporting} {Web} {Design} with the {Crowd}}.
\newblock \bibinfo{journal}{\emph{Proceedings of the ACM on Human-Computer Interaction}} \bibinfo{volume}{4}, \bibinfo{number}{EICS} (\bibinfo{date}{June} \bibinfo{year}{2020}), \bibinfo{pages}{76:1--76:28}.
\newblock
\urldef\tempurl%
\url{https://doi.org/10.1145/3394978}
\showDOI{\tempurl}


\bibitem[Oppenlaender et~al\mbox{.}(2020b)]%
        {oppenlaender2020crowdui}
\bibfield{author}{\bibinfo{person}{Jonas Oppenlaender}, \bibinfo{person}{Thanassis Tiropanis}, {and} \bibinfo{person}{Simo Hosio}.} \bibinfo{year}{2020}\natexlab{b}.
\newblock \showarticletitle{CrowdUI: Supporting Web Design with the Crowd}.
\newblock \bibinfo{journal}{\emph{Proceedings of the ACM on Human-Computer Interaction}} \bibinfo{volume}{4}, \bibinfo{number}{EICS} (\bibinfo{year}{2020}), \bibinfo{pages}{1--28}.
\newblock


\bibitem[Park et~al\mbox{.}(2013)]%
        {park_openhtml_2013}
\bibfield{author}{\bibinfo{person}{Thomas~H. Park}, \bibinfo{person}{Ankur Saxena}, \bibinfo{person}{Swathi Jagannath}, \bibinfo{person}{Susan Wiedenbeck}, {and} \bibinfo{person}{Andrea Forte}.} \bibinfo{year}{2013}\natexlab{}.
\newblock \showarticletitle{{OpenHTML}: designing a transitional web editor for novices}. In \bibinfo{booktitle}{\emph{{CHI} '13 {Extended} {Abstracts} on {Human} {Factors} in {Computing} {Systems}}} \emph{(\bibinfo{series}{{CHI} {EA} '13})}. \bibinfo{publisher}{Association for Computing Machinery}, \bibinfo{address}{New York, NY, USA}, \bibinfo{pages}{1863--1868}.
\newblock
\showISBNx{978-1-4503-1952-2}
\urldef\tempurl%
\url{https://doi.org/10.1145/2468356.2468690}
\showDOI{\tempurl}


\bibitem[Prange et~al\mbox{.}(2022)]%
        {prange2022secure}
\bibfield{author}{\bibinfo{person}{Sarah Prange}, \bibinfo{person}{Niklas Thiem}, \bibinfo{person}{Michael Fr{\"o}hlich}, {and} \bibinfo{person}{Florian Alt}.} \bibinfo{year}{2022}\natexlab{}.
\newblock \showarticletitle{“Secure settings are quick and easy!”--Motivating End-Users to Choose Secure Smart Home Configurations}. In \bibinfo{booktitle}{\emph{Proceedings of the 2022 International Conference on Advanced Visual Interfaces}}. \bibinfo{publisher}{John Wiley \& Sons}, \bibinfo{address}{Hoboken, New Jersey}, \bibinfo{pages}{1--9}.
\newblock


\bibitem[Rogers(1975)]%
        {rogers_protection_1975}
\bibfield{author}{\bibinfo{person}{Ronald~W. Rogers}.} \bibinfo{year}{1975}\natexlab{}.
\newblock \showarticletitle{A protection motivation theory of fear appeals and attitude change}.
\newblock \bibinfo{journal}{\emph{The journal of psychology}} \bibinfo{volume}{91}, \bibinfo{number}{1} (\bibinfo{year}{1975}), \bibinfo{pages}{93--114}.
\newblock
\newblock
\shownote{Publisher: Taylor \& Francis}.


\bibitem[Schaffner et~al\mbox{.}(2022)]%
        {schaffnerUnderstandingAccountDeletion2022}
\bibfield{author}{\bibinfo{person}{Brennan Schaffner}, \bibinfo{person}{Neha~A. Lingareddy}, {and} \bibinfo{person}{Marshini Chetty}.} \bibinfo{year}{2022}\natexlab{}.
\newblock \showarticletitle{Understanding Account Deletion and Relevant Dark Patterns on Social Media}.
\newblock \bibinfo{journal}{\emph{Proceedings of the ACM on Human-Computer Interaction}} \bibinfo{volume}{6}, \bibinfo{number}{CSCW2} (\bibinfo{date}{Nov.} \bibinfo{year}{2022}), \bibinfo{pages}{417:1--417:43}.
\newblock
\urldef\tempurl%
\url{https://doi.org/10.1145/3555142}
\showDOI{\tempurl}


\bibitem[Seymour et~al\mbox{.}(2020)]%
        {seymour_informing_2020}
\bibfield{author}{\bibinfo{person}{William Seymour}, \bibinfo{person}{Martin~J. Kraemer}, \bibinfo{person}{Reuben Binns}, {and} \bibinfo{person}{Max Van~Kleek}.} \bibinfo{year}{2020}\natexlab{}.
\newblock \showarticletitle{Informing the {Design} of {Privacy}-{Empowering} {Tools} for the {Connected} {Home}}. In \bibinfo{booktitle}{\emph{Proceedings of the 2020 {CHI} {Conference} on {Human} {Factors} in {Computing} {Systems}}} \emph{(\bibinfo{series}{{CHI} '20})}. \bibinfo{publisher}{Association for Computing Machinery}, \bibinfo{address}{New York, NY, USA}, \bibinfo{pages}{1--14}.
\newblock
\showISBNx{978-1-4503-6708-0}
\urldef\tempurl%
\url{https://doi.org/10.1145/3313831.3376264}
\showDOI{\tempurl}


\bibitem[Shneiderman(2022)]%
        {shneiderman2022human}
\bibfield{author}{\bibinfo{person}{Ben Shneiderman}.} \bibinfo{year}{2022}\natexlab{}.
\newblock \bibinfo{booktitle}{\emph{Human-centered AI}}.
\newblock \bibinfo{publisher}{Oxford University Press}.
\newblock


\bibitem[Sin et~al\mbox{.}(2022)]%
        {sin2022dark}
\bibfield{author}{\bibinfo{person}{Ray Sin}, \bibinfo{person}{Ted Harris}, \bibinfo{person}{Simon Nilsson}, {and} \bibinfo{person}{Talia Beck}.} \bibinfo{year}{2022}\natexlab{}.
\newblock \showarticletitle{Dark patterns in online shopping: do they work and can nudges help mitigate impulse buying?}
\newblock \bibinfo{journal}{\emph{Behavioural Public Policy}} \bibinfo{volume}{1}, \bibinfo{number}{1} (\bibinfo{year}{2022}), \bibinfo{pages}{1--27}.
\newblock


\bibitem[Siroker and Koomen(2015)]%
        {siroker2015b}
\bibfield{author}{\bibinfo{person}{Dan Siroker} {and} \bibinfo{person}{Pete Koomen}.} \bibinfo{year}{2015}\natexlab{}.
\newblock \bibinfo{booktitle}{\emph{A/B testing: The most powerful way to turn clicks into customers}}.
\newblock \bibinfo{publisher}{John Wiley \& Sons}, \bibinfo{address}{Hoboken, New Jersey}.
\newblock


\bibitem[Smullen et~al\mbox{.}(2021)]%
        {smullen_managing_nodate}
\bibfield{author}{\bibinfo{person}{Daniel Smullen}, \bibinfo{person}{Yaxing Yao}, \bibinfo{person}{Yuanyuan Feng}, \bibinfo{person}{Norman Sadeh}, \bibinfo{person}{Arthur Edelstein}, {and} \bibinfo{person}{Rebecca Weiss}.} \bibinfo{year}{2021}\natexlab{}.
\newblock \showarticletitle{Managing {Potentially} {Intrusive} {Practices} in the {Browser}: {A} {User}-{Centered} {Perspective}}. In \bibinfo{booktitle}{\emph{Proceedings on {Privacy} {Enhancing} {Technologies}}} \emph{(\bibinfo{series}{{PETS} '21}, Vol.~\bibinfo{volume}{2021})}. \bibinfo{publisher}{Sciendo}, \bibinfo{address}{The Internet}, \bibinfo{pages}{500--527}.
\newblock
\urldef\tempurl%
\url{https://doi.org/10.2478/popets-2021-0082}
\showDOI{\tempurl}


\bibitem[Snyder(2003)]%
        {snyder2003paper}
\bibfield{author}{\bibinfo{person}{Carolyn Snyder}.} \bibinfo{year}{2003}\natexlab{}.
\newblock \bibinfo{booktitle}{\emph{Paper prototyping: The fast and easy way to design and refine user interfaces}}.
\newblock \bibinfo{publisher}{Morgan Kaufmann}, \bibinfo{address}{San Franciso, CA}.
\newblock


\bibitem[Soden et~al\mbox{.}(2019)]%
        {soden_chi4evil_2019}
\bibfield{author}{\bibinfo{person}{Robert Soden}, \bibinfo{person}{Michael Skirpan}, \bibinfo{person}{Casey Fiesler}, \bibinfo{person}{Zahra Ashktorab}, \bibinfo{person}{Eric P.~S. Baumer}, \bibinfo{person}{Mark Blythe}, {and} \bibinfo{person}{Jasmine Jones}.} \bibinfo{year}{2019}\natexlab{}.
\newblock \showarticletitle{{CHI4EVIL}: {Creative} {Speculation} on the {Negative} {Impacts} of {HCI} {Research}}. In \bibinfo{booktitle}{\emph{Extended {Abstracts} of the 2019 {CHI} {Conference} on {Human} {Factors} in {Computing} {Systems}}} \emph{(\bibinfo{series}{{CHI} {EA} '19})}. \bibinfo{publisher}{Association for Computing Machinery}, \bibinfo{address}{New York, NY, USA}, \bibinfo{pages}{1--8}.
\newblock
\showISBNx{978-1-4503-5971-9}
\urldef\tempurl%
\url{https://doi.org/10.1145/3290607.3299033}
\showDOI{\tempurl}


\bibitem[Soe et~al\mbox{.}(2022)]%
        {soe2022automated}
\bibfield{author}{\bibinfo{person}{Than~Htut Soe}, \bibinfo{person}{Cristiana~Teixeira Santos}, {and} \bibinfo{person}{Marija Slavkovik}.} \bibinfo{year}{2022}\natexlab{}.
\newblock \bibinfo{title}{Automated Detection of Dark Patterns in Cookie Banners: How to Do It Poorly and Why It Is Hard to Do It Any Other Way}.
\newblock
\newblock
\urldef\tempurl%
\url{https://doi.org/10.48550/arXiv.2204.11836}
\showDOI{\tempurl}
\showeprint[arxiv]{2204.11836}~[cs]


\bibitem[Story(2021)]%
        {story2021design}
\bibfield{author}{\bibinfo{person}{Peter Story}.} \bibinfo{year}{2021}\natexlab{}.
\newblock \emph{\bibinfo{title}{Design and Evaluation of Security and Privacy Nudges: From Protection Motivation Theory to Implementation Intentions}}.
\newblock \bibinfo{thesistype}{Ph.\,D. Dissertation}. \bibinfo{school}{Carnegie Mellon University}.
\newblock


\bibitem[Story et~al\mbox{.}(2022)]%
        {story2022increasing}
\bibfield{author}{\bibinfo{person}{Peter Story}, \bibinfo{person}{Daniel Smullen}, \bibinfo{person}{Rex Chen}, \bibinfo{person}{Yaxing Yao}, \bibinfo{person}{Alessandro Acquisti}, \bibinfo{person}{Lorrie~Faith Cranor}, \bibinfo{person}{Norman Sadeh}, {and} \bibinfo{person}{Florian Schaub}.} \bibinfo{year}{2022}\natexlab{}.
\newblock \showarticletitle{Increasing adoption of tor browser using informational and planning nudges}.
\newblock \bibinfo{journal}{\emph{UMBC Faculty Collection}} (\bibinfo{year}{2022}).
\newblock


\bibitem[Stuerzlinger et~al\mbox{.}(2006)]%
        {stuerzlinger_user_2006}
\bibfield{author}{\bibinfo{person}{Wolfgang Stuerzlinger}, \bibinfo{person}{Olivier Chapuis}, \bibinfo{person}{Dusty Phillips}, {and} \bibinfo{person}{Nicolas Roussel}.} \bibinfo{year}{2006}\natexlab{}.
\newblock \showarticletitle{User interface façades: towards fully adaptable user interfaces}. In \bibinfo{booktitle}{\emph{Proceedings of the 19th annual {ACM} symposium on {User} interface software and technology}} \emph{(\bibinfo{series}{{UIST} '06})}. \bibinfo{publisher}{Association for Computing Machinery}, \bibinfo{address}{New York, NY, USA}, \bibinfo{pages}{309--318}.
\newblock
\showISBNx{978-1-59593-313-3}
\urldef\tempurl%
\url{https://doi.org/10.1145/1166253.1166301}
\showDOI{\tempurl}


\bibitem[Sturdee and Lindley(2019)]%
        {sturdee2019sketching}
\bibfield{author}{\bibinfo{person}{Miriam Sturdee} {and} \bibinfo{person}{Joseph Lindley}.} \bibinfo{year}{2019}\natexlab{}.
\newblock \showarticletitle{Sketching \& {Drawing} as {Future} {Inquiry} in {HCI}}. In \bibinfo{booktitle}{\emph{Proceedings of the {Halfway} to the {Future} {Symposium} 2019}} \emph{(\bibinfo{series}{{HTTF} 2019})}. \bibinfo{publisher}{Association for Computing Machinery}, \bibinfo{address}{New York, NY, USA}, \bibinfo{pages}{1--10}.
\newblock
\showISBNx{978-1-4503-7203-9}
\urldef\tempurl%
\url{https://doi.org/10.1145/3363384.3363402}
\showDOI{\tempurl}


\bibitem[Sung et~al\mbox{.}(2009)]%
        {sung2009robots}
\bibfield{author}{\bibinfo{person}{JaYoung Sung}, \bibinfo{person}{Henrik~I Christensen}, {and} \bibinfo{person}{Rebecca~E Grinter}.} \bibinfo{year}{2009}\natexlab{}.
\newblock \showarticletitle{Robots in the wild: understanding long-term use}. In \bibinfo{booktitle}{\emph{Proceedings of the 4th ACM/IEEE international conference on Human robot interaction}}. \bibinfo{publisher}{Association for Computing Machinery}, \bibinfo{address}{New York, NY, USA}, \bibinfo{pages}{45--52}.
\newblock


\bibitem[Toth et~al\mbox{.}(2022)]%
        {toth_dark_2022}
\bibfield{author}{\bibinfo{person}{Michael Toth}, \bibinfo{person}{Nataliia Bielova}, {and} \bibinfo{person}{Vincent Roca}.} \bibinfo{year}{2022}\natexlab{}.
\newblock \showarticletitle{On dark patterns and manipulation of website publishers by {CMPs}}.
\newblock \bibinfo{journal}{\emph{Proceedings on Privacy Enhancing Technologies}} \bibinfo{volume}{2022}, \bibinfo{number}{3} (\bibinfo{date}{July} \bibinfo{year}{2022}), \bibinfo{pages}{478--497}.
\newblock
\showISSN{2299-0984}
\urldef\tempurl%
\url{https://doi.org/10.56553/popets-2022-0082}
\showDOI{\tempurl}


\bibitem[Truong et~al\mbox{.}(2006)]%
        {truong2006storyboarding}
\bibfield{author}{\bibinfo{person}{Khai~N. Truong}, \bibinfo{person}{Gillian~R. Hayes}, {and} \bibinfo{person}{Gregory~D. Abowd}.} \bibinfo{year}{2006}\natexlab{}.
\newblock \showarticletitle{Storyboarding: an empirical determination of best practices and effective guidelines}. In \bibinfo{booktitle}{\emph{Proceedings of the 6th conference on {Designing} {Interactive} systems}} \emph{(\bibinfo{series}{{DIS} '06})}. \bibinfo{publisher}{Association for Computing Machinery}, \bibinfo{address}{New York, NY, USA}, \bibinfo{pages}{12--21}.
\newblock
\showISBNx{978-1-59593-367-6}
\urldef\tempurl%
\url{https://doi.org/10.1145/1142405.1142410}
\showDOI{\tempurl}


\bibitem[Turatto et~al\mbox{.}(2007)]%
        {turatto_moving_attention_2007}
\bibfield{author}{\bibinfo{person}{Massimo Turatto}, \bibinfo{person}{Massimo Vescovi}, {and} \bibinfo{person}{Matteo Valsecchi}.} \bibinfo{year}{2007}\natexlab{}.
\newblock \showarticletitle{Attention makes moving objects be perceived to move faster}.
\newblock \bibinfo{journal}{\emph{Vision Research}} \bibinfo{volume}{47}, \bibinfo{number}{2} (\bibinfo{year}{2007}), \bibinfo{pages}{166--178}.
\newblock
\showISSN{0042-6989}
\urldef\tempurl%
\url{https://doi.org/10.1016/j.visres.2006.10.002}
\showDOI{\tempurl}


\bibitem[Ur et~al\mbox{.}(2012)]%
        {ur_smart_2012}
\bibfield{author}{\bibinfo{person}{Blase Ur}, \bibinfo{person}{Pedro~Giovanni Leon}, \bibinfo{person}{Lorrie~Faith Cranor}, \bibinfo{person}{Richard Shay}, {and} \bibinfo{person}{Yang Wang}.} \bibinfo{year}{2012}\natexlab{}.
\newblock \showarticletitle{Smart, useful, scary, creepy: perceptions of online behavioral advertising}. In \bibinfo{booktitle}{\emph{Proceedings of the {Eighth} {Symposium} on {Usable} {Privacy} and {Security}}} \emph{(\bibinfo{series}{{SOUPS} '12})}. \bibinfo{publisher}{Association for Computing Machinery}, \bibinfo{address}{New York, NY, USA}, \bibinfo{pages}{1--15}.
\newblock
\showISBNx{978-1-4503-1532-6}
\urldef\tempurl%
\url{https://doi.org/10.1145/2335356.2335362}
\showDOI{\tempurl}


\bibitem[Van~Bavel et~al\mbox{.}(2019)]%
        {van2019using}
\bibfield{author}{\bibinfo{person}{Ren{\'e} Van~Bavel}, \bibinfo{person}{Nuria Rodr{\'\i}guez-Priego}, \bibinfo{person}{Jos{\'e} Vila}, {and} \bibinfo{person}{Pam Briggs}.} \bibinfo{year}{2019}\natexlab{}.
\newblock \showarticletitle{Using protection motivation theory in the design of nudges to improve online security behavior}.
\newblock \bibinfo{journal}{\emph{International Journal of Human-Computer Studies}}  \bibinfo{volume}{123} (\bibinfo{year}{2019}), \bibinfo{pages}{29--39}.
\newblock


\bibitem[Wang et~al\mbox{.}(2014)]%
        {wang_field_2014}
\bibfield{author}{\bibinfo{person}{Yang Wang}, \bibinfo{person}{Pedro~Giovanni Leon}, \bibinfo{person}{Alessandro Acquisti}, \bibinfo{person}{Lorrie~Faith Cranor}, \bibinfo{person}{Alain Forget}, {and} \bibinfo{person}{Norman Sadeh}.} \bibinfo{year}{2014}\natexlab{}.
\newblock \showarticletitle{A field trial of privacy nudges for facebook}. In \bibinfo{booktitle}{\emph{Proceedings of the {SIGCHI} {Conference} on {Human} {Factors} in {Computing} {Systems}}}. \bibinfo{publisher}{ACM}, \bibinfo{address}{Toronto Ontario Canada}, \bibinfo{pages}{2367--2376}.
\newblock
\showISBNx{978-1-4503-2473-1}
\urldef\tempurl%
\url{https://doi.org/10.1145/2556288.2557413}
\showDOI{\tempurl}


\bibitem[Warner(2021)]%
        {warner_lawmakers_2021}
\bibfield{author}{\bibinfo{person}{Mark Warner}.} \bibinfo{year}{2021}\natexlab{}.
\newblock \bibinfo{title}{Lawmakers {Reintroduce} {Bipartisan} {Bicameral} {Legislation} to {Ban} {Manipulative} '{Dark} {Patterns}'}.
\newblock
\newblock
\urldef\tempurl%
\url{https://www.warner.senate.gov/public/index.cfm/2021/12/lawmakers-reintroduce-bipartisan-bicameral-legislation-to-ban-manipulative-dark-patterns}
\showURL{%
\tempurl}


\bibitem[Wen et~al\mbox{.}(2023)]%
        {wen2023empowering}
\bibfield{author}{\bibinfo{person}{Hao Wen}, \bibinfo{person}{Yuanchun Li}, \bibinfo{person}{Guohong Liu}, \bibinfo{person}{Shanhui Zhao}, \bibinfo{person}{Tao Yu}, \bibinfo{person}{Toby Jia-Jun Li}, \bibinfo{person}{Shiqi Jiang}, \bibinfo{person}{Yunhao Liu}, \bibinfo{person}{Yaqin Zhang}, {and} \bibinfo{person}{Yunxin Liu}.} \bibinfo{year}{2023}\natexlab{}.
\newblock \bibinfo{title}{Empowering LLM to use Smartphone for Intelligent Task Automation}.
\newblock
\newblock
\showeprint[arxiv]{2308.15272}~[cs.AI]


\bibitem[Wikström and Verganti(2013)]%
        {wikstrom_exploring_2013}
\bibfield{author}{\bibinfo{person}{Anders Wikström} {and} \bibinfo{person}{Roberto Verganti}.} \bibinfo{year}{2013}\natexlab{}.
\newblock \showarticletitle{Exploring storyboarding in pre-brief activities}. In \bibinfo{booktitle}{\emph{Proceedings of the 19th {International} {Conference} on {Engineering} {Design}}} \emph{(\bibinfo{series}{{ICED} '13}, Vol.~\bibinfo{volume}{7})}. \bibinfo{publisher}{Design Society}, \bibinfo{address}{Seoul, Korea}, \bibinfo{pages}{011--020}.
\newblock
\showISBNx{978-1-904670-50-6}
\urldef\tempurl%
\url{https://www.designsociety.org/publication/34565/Exploring+storyboarding+in+pre-brief+activities}
\showURL{%
\tempurl}
\newblock
\shownote{ISBN: 9781904670506}.


\bibitem[Wilson et~al\mbox{.}(2022)]%
        {wilson2022dark}
\bibfield{author}{\bibinfo{person}{Richard Wilson} {et~al\mbox{.}}} \bibinfo{year}{2022}\natexlab{}.
\newblock \showarticletitle{Dark Patterns and Epistemic Ignorance: An Educational Crisis}. In \bibinfo{booktitle}{\emph{European Conference on the Impact of Artificial Intelligence and Robotics}}, Vol.~\bibinfo{volume}{4}. \bibinfo{publisher}{Academic Conferences International}, \bibinfo{address}{Curtis Farm, United Kingdom}, \bibinfo{pages}{78--84}.
\newblock


\bibitem[Yin et~al\mbox{.}(2014)]%
        {yin_temporal_2014}
\bibfield{author}{\bibinfo{person}{Hongzhi Yin}, \bibinfo{person}{Bin Cui}, \bibinfo{person}{Ling Chen}, \bibinfo{person}{Zhiting Hu}, {and} \bibinfo{person}{Zi Huang}.} \bibinfo{year}{2014}\natexlab{}.
\newblock \showarticletitle{A temporal context-aware model for user behavior modeling in social media systems}. In \bibinfo{booktitle}{\emph{Proceedings of the 2014 {ACM} {SIGMOD} {International} {Conference} on {Management} of {Data}}} \emph{(\bibinfo{series}{{SIGMOD} '14})}. \bibinfo{publisher}{Association for Computing Machinery}, \bibinfo{address}{New York, NY, USA}, \bibinfo{pages}{1543--1554}.
\newblock
\showISBNx{978-1-4503-2376-5}
\urldef\tempurl%
\url{https://doi.org/10.1145/2588555.2593685}
\showDOI{\tempurl}


\bibitem[Zagal et~al\mbox{.}(2013)]%
        {zagal_dark_2013}
\bibfield{author}{\bibinfo{person}{José~P. Zagal}, \bibinfo{person}{Staffan Björk}, {and} \bibinfo{person}{Chris Lewis}.} \bibinfo{year}{2013}\natexlab{}.
\newblock \showarticletitle{Dark patterns in the design of games}. In \bibinfo{booktitle}{\emph{Foundations of {Digital} {Games} 2013}}. \bibinfo{publisher}{Society for the Advancement of the Science of Digital Games}, \bibinfo{address}{Chania, Greece}, \bibinfo{pages}{1--8}.
\newblock


\bibitem[Zhang and Guo(2018)]%
        {zhang_fusion_2018}
\bibfield{author}{\bibinfo{person}{Xiong Zhang} {and} \bibinfo{person}{Philip~J. Guo}.} \bibinfo{year}{2018}\natexlab{}.
\newblock \showarticletitle{Fusion: {Opportunistic} {Web} {Prototyping} with {UI} {Mashups}}. In \bibinfo{booktitle}{\emph{Proceedings of the 31st {Annual} {ACM} {Symposium} on {User} {Interface} {Software} and {Technology}}} \emph{(\bibinfo{series}{{UIST} '18})}. \bibinfo{publisher}{Association for Computing Machinery}, \bibinfo{address}{New York, NY, USA}, \bibinfo{pages}{951--962}.
\newblock
\showISBNx{978-1-4503-5948-1}
\urldef\tempurl%
\url{https://doi.org/10.1145/3242587.3242632}
\showDOI{\tempurl}


\bibitem[Zhang et~al\mbox{.}(2022)]%
        {zhang_storybuddy_2022}
\bibfield{author}{\bibinfo{person}{Zheng Zhang}, \bibinfo{person}{Ying Xu}, \bibinfo{person}{Yanhao Wang}, \bibinfo{person}{Bingsheng Yao}, \bibinfo{person}{Daniel Ritchie}, \bibinfo{person}{Tongshuang Wu}, \bibinfo{person}{Mo Yu}, \bibinfo{person}{Dakuo Wang}, {and} \bibinfo{person}{Toby Jia-Jun Li}.} \bibinfo{year}{2022}\natexlab{}.
\newblock \showarticletitle{{StoryBuddy}: {A} {Human}-{AI} {Collaborative} {Chatbot} for {Parent}-{Child} {Interactive} {Storytelling} with {Flexible} {Parental} {Involvement}}. In \bibinfo{booktitle}{\emph{{CHI} {Conference} on {Human} {Factors} in {Computing} {Systems}}} \emph{(\bibinfo{series}{{CHI} '22})}. \bibinfo{publisher}{Association for Computing Machinery}, \bibinfo{address}{New York, NY, USA}, \bibinfo{pages}{1--21}.
\newblock
\showISBNx{978-1-4503-9157-3}
\urldef\tempurl%
\url{https://doi.org/10.1145/3491102.3517479}
\showDOI{\tempurl}


\bibitem[Zhou et~al\mbox{.}(2018)]%
        {zhou_atrank_2018}
\bibfield{author}{\bibinfo{person}{Chang Zhou}, \bibinfo{person}{Jinze Bai}, \bibinfo{person}{Junshuai Song}, \bibinfo{person}{Xiaofei Liu}, \bibinfo{person}{Zhengchao Zhao}, \bibinfo{person}{Xiusi Chen}, {and} \bibinfo{person}{Jun Gao}.} \bibinfo{year}{2018}\natexlab{}.
\newblock \showarticletitle{{ATRank}: {An} {Attention}-{Based} {User} {Behavior} {Modeling} {Framework} for {Recommendation}}. In \bibinfo{booktitle}{\emph{Proceedings of the {AAAI} {Conference} on {Artificial} {Intelligence}}} \emph{(\bibinfo{series}{{AAAI} '18}, Vol.~\bibinfo{volume}{32})}. \bibinfo{publisher}{AAAI}, \bibinfo{address}{New Orleans, Louisiana, USA}, \bibinfo{pages}{4564--4571}.
\newblock
\urldef\tempurl%
\url{https://doi.org/10.1609/aaai.v32i1.11618}
\showDOI{\tempurl}


\bibitem[Zou et~al\mbox{.}(2020)]%
        {zou_examining_2020}
\bibfield{author}{\bibinfo{person}{Yixin Zou}, \bibinfo{person}{Kevin Roundy}, \bibinfo{person}{Acar Tamersoy}, \bibinfo{person}{Saurabh Shintre}, \bibinfo{person}{Johann Roturier}, {and} \bibinfo{person}{Florian Schaub}.} \bibinfo{year}{2020}\natexlab{}.
\newblock \showarticletitle{Examining the {Adoption} and {Abandonment} of {Security}, {Privacy}, and {Identity} {Theft} {Protection} {Practices}}. In \bibinfo{booktitle}{\emph{Proceedings of the 2020 {CHI} {Conference} on {Human} {Factors} in {Computing} {Systems}}}. \bibinfo{publisher}{Association for Computing Machinery}, \bibinfo{address}{New York, NY, USA}, \bibinfo{pages}{1--15}.
\newblock
\showISBNx{978-1-4503-6708-0}
\urldef\tempurl%
\url{https://doi.org/10.1145/3313831.3376570}
\showURL{%
\tempurl}


\end{thebibliography}

\clearpage
\appendix
\section{Appendix}
\label{sec:appendix}

\subsection{Co-Design Workshop Participant Demographics}
\label{appendix:workshop_demographics}
\begin{table}[H]
    \small
  \caption{Demographics of the co-design workshop participants. This table includes participants' ID, gender, age, education level, occupational domain, daily Internet usage, and whether they had heard of the concept ``UX Dark Patterns''.}
  \label{table:workshop_demographics}
  \begin{tabularx}{\textwidth}{llllYlr}
    \toprule
    \textbf{ID} & \textbf{Gender} & \textbf{Age} & \textbf{Education} & \textbf{Industry} & \textbf{Internet Use} & \textbf{Know DP?} \\ 
    \midrule
    PA1  & Female & 21--25 & Bachelor's & Software and Information Services & 2--5 hrs & Yes \\
    PA2  & Male & 21--25 & Bachelor's & Software and Information Services & More than 8 hrs & Yes \\
    PA3  & Male & 21--25 & Some college & Software and Information Services & 5--8 hrs & No \\
    PA4  & Female & 21--25 & Bachelor's & Health Care and Social Assistance & 2--5 hrs & No \\
    PA5  & Female & 21--25 & Bachelor's & Health Care and Social Assistance & More than 8 hrs & No \\
    PA6  & Female & 26--34 & Bachelor's & Software and Information Services & 5--8 hrs & Yes \\
    PA7  & Female & 26--34 & Master's & Software and Information Services & 5--8 hrs & Yes \\
    PA8  & Male & Not reported & Master's & Education & 5--8 hrs & Yes \\
    PA9  & Female & 35--50 & Bachelor's & Education & More than 8 hrs & No \\
    PA10 & Male & 26--34 & Master's & Education & 5--8 hrs & No \\
    PA11 & Female & Above 50 & Master's & Education & 2--5 hrs & Yes \\
    PA12 & Male & 26--34 & Master's & Education & 5--8 hrs & Yes \\
    \bottomrule
\end{tabularx}
\end{table}

\subsection{\ylhighlight{Themes Emerged from the Co-Design Workshops}}
\label{sec:workshop_analysis_themes}

{
\renewcommand{\arraystretch}{1.2}
\centering
\small
\begin{xltabular}{\linewidth}{L{0.2\textwidth}X}
    \caption{The top two levels of themes generated from our qualitative analysis for co-design workshops. The level-3 codes are not included due to the large quantity.}
    \label{tab:co-design_qualitative_themes}\\
    
    \toprule
    \textbf{Level-1 Theme} & \textbf{Level-2 Theme}\\ 
    \midrule
    \endfirsthead
    
    \multicolumn{2}{r}{continued from the previous page}\\
    \toprule
    \textbf{Level-1 Theme} & \textbf{Level-2 Theme}\\
    \midrule
    \endhead

    \multicolumn{2}{r}{to be continued}\\
    \endfoot
    
    \hline
    \endlastfoot
\multirow{5}{*}{\begin{tabular}[x]{@{}l@{}}Past Experience \\With Dark Patterns\end{tabular}} & Users experiences and perceptions of dark patterns are individualized for specific DP instances. \\
\cline{2-2}
& Users are frustrated and concerned about dark patterns, but even when they are aware and intend to change, they still feel manipulated. \\
\cline{2-2}
& Even when users are aware of and concerned about dark patterns, they might have to put up with it because the platform service is essential to them.\\
\hline
\multirow{6}{*}{\begin{tabular}[x]{@{}l@{}}Factors Making Dark\\ Patterns Annoying\end{tabular}} & Users perceive dark patterns that are not expected, involve deception, potentially cause financial loss, and are hard to solve / make them lose autonomy more annoying. \\
\cline{2-2}
& Users are more directly affected by individual dark pattern examples, their understanding of dark pattern come from specific interaction with these instances \\
\cline{2-2}
& Users subconsciously evaluate the benefits/loss caused by each dark pattern instance. Such evaluation later is translated into individual coping methods. \\ 
\hline
\multirow{8}{*}{\begin{tabular}[x]{@{}l@{}}Desired Support \\From Intervention\end{tabular}} & Users desire impact and severity of the impact information from our extension to support their subconscious process of benefit / loss calculation to develop coping strategy for dark patterns. \\
\cline{2-2}
& Users proposed a variety of strategies to make changes to specific interface components with dark patterns. \\
\cline{2-2}
& Users want to change the interface layout to make relevant and useful information prominent. \\ 
\cline{2-2}
& Users want to change the user flow of the service against dark patterns.\\
\cline{2-2}
& Users want to have autonomy over general extension use experience.\\
\bottomrule
\end{xltabular}
}

\subsection{Dark Pattern Instance Details}
\label{appendix:dark_pattern_instance}

{
    \centering
    \small
\begin{xltabular}{\linewidth}{lp{3cm}X}
    \caption{Dark pattern instances with types~\cite{bongard-blanchy_i_2021} and descriptions.}
    \label{instance_type_1}\\
    
    \toprule
    \textbf{Name} & \textbf{Type} & \textbf{Description}\\ \midrule
    \endfirsthead
    
    \multicolumn{3}{r}{continued from the previous page}\\
    \toprule
    \textbf{Name} & \textbf{Type} & \textbf{Description}\\ 
    \endhead

    \multicolumn{3}{r}{to be continued}\\
    \endfoot
    
    \hline
    \endlastfoot

Prominent ``Buy Now'' Button & False Hierarchy & The ``Buy Now'' button \yuwen{on the product page is designed in a more prominent orange color, making it easier to click on than the safer ``Add to Cart'' button, although they serve similar purposes. ``Buy Now'' provides} a frictionless experience that accelerates customers' checkout process and encourages purchase~\cite{kuang2019user}. It improves Amazon's conversion rate but can potentially make users buy unnecessary items they regret later.  \\
\hline
Disguised Ads & Hidden Information / False Hierarchy & The interface design put the ``sponsored'' tag at the bottom corner of the ad in \yuwen{a small and gray font. It's easy for users to miss the tag and not realize the content is sponsored. In some interfaces the tag can also be positioned to the top right of the item. This lack of position consistency makes it hard for users to catch the tag every time.} \\
\hline
Fake Discounts & Hidden Information & Discounts information may be exaggerated or shown in a misleading way, with a little disguised information button leading to another webpage describing the detials for different types of discount information~\cite{amazon_pricing_explanation}. It tricks users into thinking the items are on sale and buying it while the price might be the same according to 3rd party price tracking sources.\looseness=-1\\
\hline
Limited Time Recommendation & Limited-Time Message / False Hierarchy & The recommended items often appear on the homepage with a limited time offer information. The design also makes this part dominant and takes the whole top sections of the viewport. \\
\hline
Video Autoplay & Autoplay & When users hover over recommended videos on the YouTube homepage, it starts to play automatically. It exploits the psychological fact that users are more attracted to moving things~\cite{turatto_moving_attention_2007} and tries to intrigue users so they would start watching the videos. \\
\hline
Hiding Dislike Count & Hidden Information & YouTube only displays the like count but not the dislike counts. This may cause bias, as a video's collective rating \yuwen{helps users choose videos to watch and affect their evaluations of the content}. Hiding the other half of the information may inflate users' positive perception of all videos across the platform.  \\
\hline
Auto Recommendations & Autoplay & Video recommendations on the sidebar after a video plays are individualized to cater to users' watching preferences, and autoplay on hover makes users even more likely to watch them. \\ \hline
Hiding Total Episode Time & Hidden Information & The timeline only shows the time remaining, not how long you have spent on the episode. This can prevent users from realizing how long they have spent watching.  \\
\hline
Automatic Preview & Autoplay / False Hierarchy & Netflix automatically plays the featured trailer for you upon arriving on the site. It is also displayed disproportionately to other videos, which are essentially the same to users.\\
\hline
Fake Trending Content & High-Demand Message / Pre-Selection & \yuwen{In the ``trending'' section, Twitter personalizes the content for individual users but makes it seem to base on just the content's popularity. Many users are not aware of this and do not like this deceptive feature.} \\
\hline
Disguised Suggested Tweets & Hidden Information / High-Demand Message & This is suggested content by the Twitter algorithm, but it does not explicitly label itself as ``sponsored'' or ``suggested''. Instead, it uses confusing labels such as ``Popular videos'' which users often miss. \\
\hline
Sneaking Short Videos Into Feed & High-Demand Message / Hidden Information & \yuwen{Facebook often sneaks short video contents users don't follow into their feeds in the form of a widget named ``Reels'' to promote short video contents on the platform. It is similar to the dark pattern ``sneak into basket''~\cite{gray_dark_2018} for shopping websites, but instead of losing money for buying unwanted items, users lose time over content they do not originally plan to consume.\looseness=-1 } \\
\hline
Disguised Sponsorship & Hidden Information & Content prompted by the Facebook algorithm simply has a light-colored text label saying ``Sponsored'' or ``Suggested for you'' which many users miss. Otherwise, they appear identical to regular posts in a user's feed. \\ 
\bottomrule
\end{xltabular}
}

\begin{figure}[H]
\centering
\includegraphics[width=\linewidth]{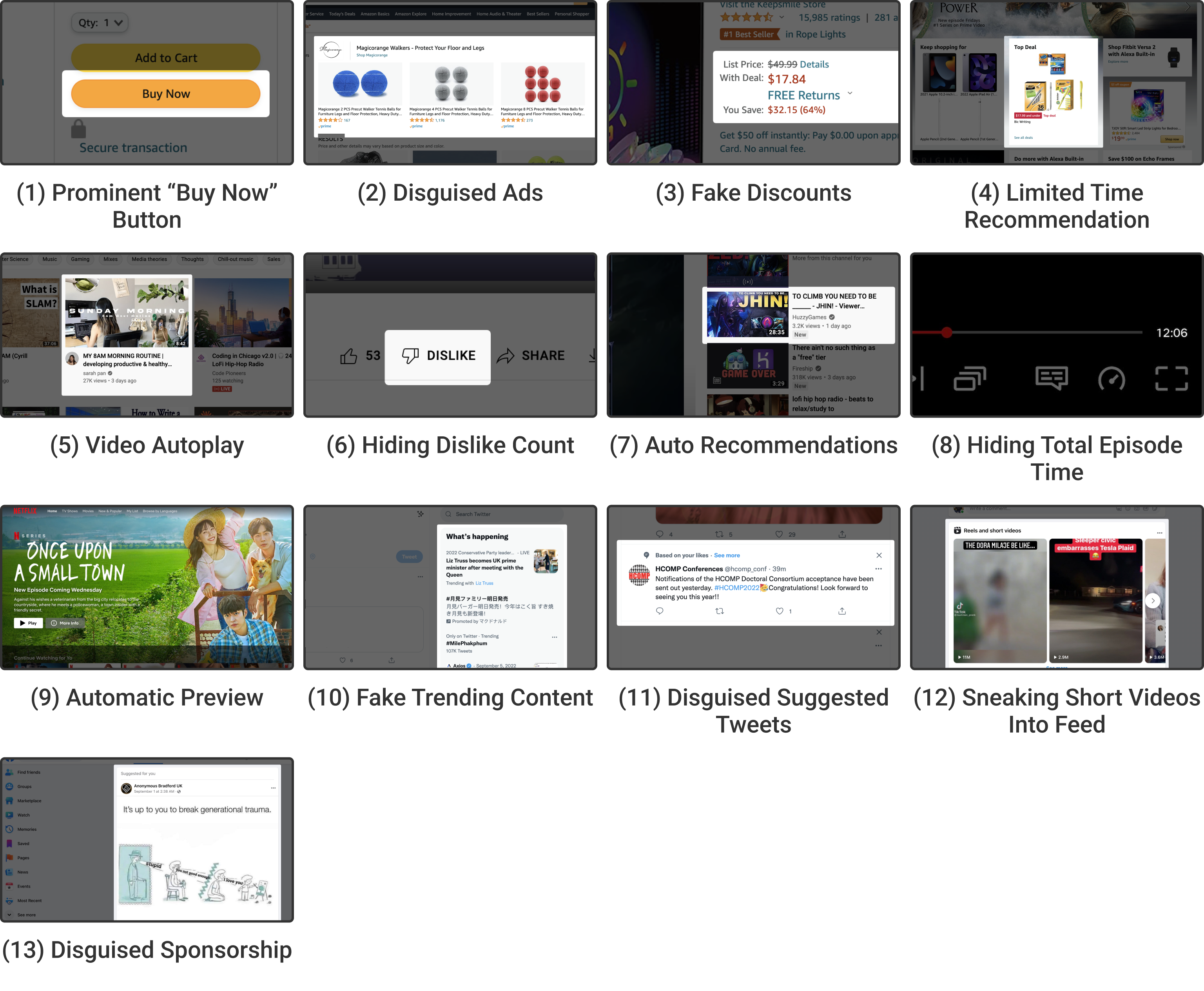}
\caption{Screenshots of sampled dark pattern instances.} \label{dark_pattern_instance}
\end{figure}

\subsection{UI Enhancement Details}
\label{appendix:ui_enhancement}

{
    \centering
    \small
\begin{xltabular}{\linewidth}{llX}
    \caption{UI enhancements with targeted dark patterns, intervention strategies, and descriptions.}
    \label{ui_enhancements_1}\\
    
    \toprule
    \textbf{Dark Pattern} & \textbf{Intervention} & \textbf{Description}\\ 
    \midrule
    \endfirsthead
    
    \multicolumn{3}{r}{continued from the previous page}\\
    \toprule
    \textbf{Dark Pattern} & \textbf{Intervention} & \textbf{Description}\\ 
    \endhead

    \multicolumn{3}{r}{to be continued}\\
    \endfoot
    
    \hline
    \endlastfoot
\multirow{5}{*}{Prominent ``Buy Now'' Button} & Hiding & \textsc{Dark Pita} will make the ``Buy Now'' button disappear. Users can still purchase the item by adding it to the shopping cart and checking out there. \\
\cline{2-3}
& Fairness & \textsc{Dark Pita} will make the ``Buy Now'' button be in the same color as the regular ``Add to Cart'' option. \\
\cline{2-3}
& Friction & \textsc{Dark Pita} will add an overlay as friction before proceeding to purchase when users try to click buy now.\\
\hline
\multirow{6}{*}{Disguised Ads} & Hiding & \textsc{Dark Pita} will hide the disguised ads, which eliminates ads camouflaged as regular items. \\
\cline{2-3}
& Friction & \textsc{Dark Pita} will make the cursor invisible if it navigates through the ads area, which rouses users' attention. \\
\cline{2-3}
& Information Disclosure & \textsc{Dark Pita} will catch the content of the ambiguous ads and explicitly present their ads identity by adding extra text labels. \\ 
\cline{2-3}
& Counterfactual Thinking & \textsc{Dark Pita} will mark that the item(s) may be promoted because they paid Amazon. This may also help users avoid unnecessary browsing. \\
\hline
\multirow{8}{*}{Fake Discounts} & Hiding & \textsc{Dark Pita} can hide the discount information and prevent it from influencing users' decisions. \\
\cline{2-3}
& Information Disclosure & \textsc{Dark Pita} will help users understand the rationales behind the price and the marketing jargon. \\
\cline{2-3}
& Counterfactual Thinking & \textsc{Dark Pita} will add some visual effects and remind users to give a second thought before making their purchase decisions. \\ 
\cline{2-3}
& Action Guide & When users  hover over the discount price, \textsc{Dark Pita} will provide some suggestions on actions that users can take towards this item. Those actions are in accordance with users' long-term goals.\\
\hline
\multirow{5}{*}{Limited Time Recommendation} & Hiding & \textsc{Dark Pita} will hide ALL recommended-items sections on the homepage, which drastically eliminates potential distractions. \\
\cline{2-3}
& Counterfactual Thinking & \textsc{Dark Pita} will add visual effects and remind users to think twice before mindlessly browsing. \\ 
\cline{2-3}
& Reflection & \textsc{Dark Pita} will track and show users' extra cost on Amazon that is likely caused by dark patterns. \\
\hline
\multirow{6}{*}{Video Autoplay} & Hiding & \textsc{Dark Pita} will hide ALL recommended videos on the homepage. Users can still use the search bar on top; this is helpful when users visit YouTube with a specific purpose in mind. \\
\cline{2-3}
& Disabling & \textsc{Dark Pita} will disable the preview function. This may prevent users from being distracted by the motions. \\ 
\cline{2-3}
& Reflection & \textsc{Dark Pita} will track and show the extra time users spend on YouTube due to dark patterns. \\
\hline
Hiding Dislike Count & Information Disclosure & \textsc{Dark Pita} will show the hidden dislike counts. \\
\hline
\multirow{6}{*}{Auto Recommendations} & Hiding & \textsc{Dark Pita} will hide ALL recommended videos on the sidebar. This is helpful when users want to prevent themselves from binge-watching. \\
\cline{2-3}
& Disabling & \textsc{Dark Pita} will disable the preview function. This may prevent users from being distracted and spending too much time on YouTube. \\ 
\cline{2-3}
& Reflection & \textsc{Dark Pita} will track and show the time users spend on YouTube and let users know how much is through the dark patterns on YouTube interfaces. \\
\hline
Hiding Total Episode Time & Reflection & \textsc{Dark Pita} tracks and shows the time users spend on Netflix. This may prevent binge-watching. \\
\hline
Automatic Preview & Disabling & \textsc{Dark Pita} will disable background preview on Netflix homepage. This may help prevent users from being distracted by the featured content. \\
\hline
Fake Trending Content & Hiding & \textsc{Dark Pita} will hide this section to help you focus on your feed. Users can always use the ``Explore'' tab on the left sidebar. This may help users reduce Twitter consumption. \\
\hline
\multirow{3}{*}{Disguised Suggested Tweets} & Information Disclosure & \textsc{Dark Pita} will detect this type of tweets and explicitly mark them as promoted for users. \\
\cline{2-3}
& Friction & \textsc{Dark Pita} will detect this type of tweets and replace them with an overlay. If users still want to see the tweet, they can click on the reveal button. \\ 
\hline
\multirow{6}{*}{Sneaking Short Videos Into Feed} & Hiding & \textsc{Dark Pita} will hide the Reels. \\
\cline{2-3}
& Counterfactual Thinking & \textsc{Dark Pita} will prompt users to think about the mechanism behind the selected content. \\
\cline{2-3}
& Friction & \textsc{Dark Pita} will add an overlay to the Reels to prevent users from immediately being distracted by them. If users still want to view them, click on the reveal button. \\
\hline
\multirow{2}{*}{Disguised Sponsorship} & Hiding & \textsc{Dark Pita} will hide the selected suggested content for users. \\
\cline{2-3}
& Information Disclosure & \textsc{Dark Pita} will explicitly label it as suggested content. \\
\bottomrule
\end{xltabular}
}

\subsection{Deployment Participant Demographics}
\label{appendix:deployment_demographics}

\begin{table}[H]
    \small
  \caption{Demographics of the deployment study participants. This table includes participants' ID, gender identity, age, education level, occupational domain, daily Chrome usage, and whether heard of ``Dark Patterns in UX'' or not.}
  \label{table:deployment_demographics}
  \begin{tabularx}{\textwidth}{llllYlr}
    \toprule
    \textbf{ID} & \textbf{Gender} & \textbf{Age} & \textbf{Education} & \textbf{Industry} & \textbf{Daily Chrome Use} & \textbf{Know DP?} \\ 
    \midrule
    PB1  & Female & 18--20 & High school & College Student & Less than 1 hrs & No \\
    PB2  & Female & 26--34 & Master's & Media & 1--2 hrs & Yes \\
    PB3  & Female & 21--25 & Master's & Software and Information Services & More than 5 hrs & Yes \\
    PB4  & Male & 26--34 & Master's & Media & 2--5 hrs & No \\
    PB5  & Male & Above 50 & Some college & Retail & More than 5 hrs & No \\
    PB6  & Male & 35--50 & Bachelor's & Software and Information Services & More than 5 hrs & No \\
    PB7  & Female & 26--34 & Master's & Education & 2--5 hrs & Yes \\
    PB8  & Female & 21--25 & Some college & Software and Information Services & More than 5 hrs & Yes \\
    PB9  & Male & Above 50 & Doctoral & Education & 2--5 hrs & Yes \\
    PB10 & Female & 26--34 & Bachelor's & Arts, Entertainment, and Recreation & 2--5 hrs & Yes \\
    PB11 & Male & 21--25 & Some college & Telecommunications & More than 5 hrs & No \\
    PB12 & Male & 21--25 & Master's & Computer and Electronics & More than 5 hrs & Yes \\
    PB13 & Male & 26--34 & Master's & Software and Information Services & More than 5 hrs & No \\
    PB14 & Male & 26--34 & Bachelor's & Scientific or Technical Services & 2--5 hrs & No \\
    PB15 & Male & 21--25 & Bachelor's & Education & More than 5 hrs & Yes \\
    \bottomrule
\end{tabularx}
\end{table}

\subsection{Deployment Study Interview Protocols}
\label{appendix:deploymen_protocol}

We used the following protocols to conduct our three stages of interviews in our deployment studies. Follow-up questions were asked whenever the interviewer(s) saw fit.
\subsubsection{Entry Interview (1 hour)}
\begin{enumerate}
    \item Gather participants' perceptions and experiences with dark patterns in daily lives
    \begin{enumerate}
        \item Give a general introduction to dark patterns
        \item Ask participants about:
        \begin{enumerate}
            \item websites they usually go to and the dark patterns there (Netflix, YouTube; Amazon; Twitter, Facebook)
            \item the negative emotions, felt manipulation, self-autonomy loss, and the likelihood of being influenced when seeing these dark patterns
        \end{enumerate}
    \end{enumerate}
    \item Show participants some dark patterns and how to use \textsc{Dark Pita} to change them
    \begin{enumerate}
        \item Tell participants that in our study we encourage them to:
        \begin{enumerate}
            \item change a few dark patterns on the websites they use every day
            \item come up with more UI design enhancement strategies they desire
            \item send at least one diary note every 2 days
            \item come up with more dark patterns they encounter on websites we don't yet support 
        \end{enumerate}
        \item Give participants our manual, containing information and Q\&A on using our extension, and let them know how to contact us or ask questions
    \end{enumerate}
\end{enumerate}

\subsubsection{Check-in Interview (30 minutes)}
\begin{enumerate}
    \item Check in with participants on their questions \& issues during the past week using our extension
    \item Before the study, check any outlier in the participant's user log or diary notes, make clarifications if needed
    \item Ask participants about the dark patterns they used our extension to change and ask their thoughts on:
    \begin{enumerate}
        \item how do they think the change impacted their behavior on these websites?
        \item in the long term, which changes do you want to keep? Which ones do not? Why?
        \item Do you have any alternative enhancements for this dark pattern you desire?
    \end{enumerate}
    \item Did the participant find dark patterns on other websites?
    \item Remind participants of sending diary notes regularly
    \item Schedule a third interview session with the participant
\end{enumerate}

\subsubsection{Exit Interview (1 hour)}
\begin{enumerate}
    \item Understand and clarify any outliers in user log data and diary notes
    \item Have the participant talk about their experiences with \textsc{Dark Pita} during the past 2 weeks, with them referencing the websites for more contexts
    \item More specifically, what dark patterns did they use \textsc{Dark Pita} to change?
    \begin{enumerate}
        \item Why did the participant change it? How did the change impact their online experience?
        \item How did the change make the participant feel emotional?
        \item Which UI enhancement was the participant's favorite? Why?
    \end{enumerate}
    \item What dark patterns did the participant not use \textsc{Dark Pita} to change?
    \begin{enumerate}
        \item Why? Is it because of the dark pattern or the intervention?
        \item If it is the intervention, what intervention does the participant desire?
    \end{enumerate}
    \item Questions on educational values
    \begin{enumerate}
        \item Did the participant learn anything new about dark patterns? If so, what are they?
        \item Anything else the participant did to learn more about dark patterns besides our study?
        \begin{enumerate}
            \item Why? How did the participant get interested? What motivated them to learn more?
            \item what experience they had, or which feature in \textsc{Dark Pita} made them want to learn more about dark patterns?
            \item What new thoughts on dark patterns \& intervention techniques did the participant have?
        \end{enumerate}
    \end{enumerate}
    \item Would the participant like to continue using the tool in the long term? Why? 
    \item What change does the participant want to make in the future?
    \item Other general feedback?
\end{enumerate}

\subsection{\ylhighlight{Themes Emerged from the Deployment Study}}
\label{appendix: deployment qualitative analysis}

{
\renewcommand{\arraystretch}{1.2}
\centering
\small
\begin{xltabular}{\linewidth}{L{0.3\textwidth}X}
    \caption{Qualitative analysis themes for the deployment study interviews. Themes from our second round of labeling were already comprehensive and our third round of labeling yielded only limited improvement over the second round. As a result, we simply used our second-round themes as level-1.}
    \label{tab:deployment_qualitative_themes}\\
    
    \toprule
    \textbf{Level-1 Theme} & \textbf{Level-2 Theme}\\ 
    \midrule
    \endfirsthead
    
    \multicolumn{2}{r}{continued from the previous page}\\
    \toprule
    \textbf{Level-1 Theme} & \textbf{Level-2 Theme}\\
    \midrule
    \endhead

    \multicolumn{2}{r}{to be continued}\\
    \endfoot
    
    \hline
    \endlastfoot
\multirow{8}{*}{\begin{tabular}[x]{@{}l@{}}Dark Pita was able to help users\\ \textbf{\textit{understand}} the concept of DP,\\ \textbf{\textit{discover}} DPs on the current\\ platform, and \textbf{\textit{transfer}} the\\ knowledge to other platforms\end{tabular}} & Our probe increased their knowledge about dark patterns \\
\cline{2-2}
& Dark Pita made the user realize the large number of dark patterns that exist on the website \\
\cline{2-2}
& Users are inspired to generalize Dark-Pattern-related knowledge to their daily usage of other platforms/interfaces \\
\cline{2-2}
& Dark Pita helped participants see dark patterns more explicitly, some of which they may have felt annoyed by before \\
\cline{2-2}
& Intervention provides a new perspective to online service. \\
\cline{2-2}
& Users want to use the probe further to other sites, services, and scenarios. \\
\hline
\multirow{5}{*}{\begin{tabular}[x]{@{}l@{}}Users are concerned if the\\ platform provides the ability to\\ change the interface, it will be\\ used to collect preference data\\ and still benefit the companies.\end{tabular}} & Websites may provide some options for users to change dark patterns. \\
\cline{2-2}
& While companies and users may aim at the same techniques, their goals could diverge and even be against each other. \\
\cline{2-2}
& Users think the options provided by the services are not good or even annoying because they are still out of the company's benefits -- to understand users' preferences. \\
\hline
\multirow{7}{*}{\begin{tabular}[x]{@{}l@{}}Users expect a neutral\\ community to provide such tools\\ other than the service providers,\\ because of misalignment between \\users' and companies' goals\end{tabular}} & Future exceptions about user empowerment. \\
\cline{2-2}
& Websites may provide some options for users to change dark patterns. \\
\cline{2-2}
& While companies and users may aim at the same techniques, their goals could diverge and even be against each other. \\
\cline{2-2}
& Users think the options provided by the services are not good or even annoying because they are still out of the company's benefits -- to understand users' preferences. \\
\hline
\multirow{4}{*}{\begin{tabular}[x]{@{}l@{}}Dark Pita enables users change\\ dark patterns, giving users posi-\\tive feelings and self-autonomy.\end{tabular}} & Users feel empowered with actions of our probe. \\
\cline{2-2}
& Good feelings about being empowered to change dark patterns. \\
\cline{2-2}
& Users think many other people would like to have the ability to change dark pattern interfaces \\
\hline
\multirow{6}{*}{\begin{tabular}[x]{@{}l@{}}Design Implications for interven-\\tion: less intrusive (visual + exp-\\erience), more controllable, more \\ straightforward, less FOMO.\end{tabular}} & Visually clear + less intrusive intervention gives better perception \\
\cline{2-2}
& Users' requests for more customizability and autonomy over their interface \\
\cline{2-2}
& Do NOT interrupt the normal experience. \\
\cline{2-2}
& Hiding elements on the website may give users FOMO and create backlash (users want to see more) \\
\bottomrule
\end{xltabular}
}

\end{document}